\documentclass[]{aa}  
\usepackage{amsmath}
\usepackage{graphicx}
\usepackage{txfonts}
\usepackage{natbib} 
\bibpunct{(}{)}{;}{a}{}{,} 
 
\usepackage{verbatim}
\usepackage{url}
\usepackage{epstopdf}

\usepackage{color}

\begin{document}

   \title{Asteroseismic age estimates of RGB stars in open clusters} 

   \subtitle{A statistical investigation of different estimation methods}

   \author{G. Valle \inst{1,2,3}, M. Dell'Omodarme \inst{3}, E. Tognelli\inst{1,2,3}, P.G. Prada Moroni
     \inst{2,3}, S. Degl'Innocenti \inst{2,3} 
          }
   \titlerunning{Statistical investigation on asteroseismic age estimates of clusters RGB stars}
   \authorrunning{Valle, G. et al.}

   \institute{
INAF - Osservatorio Astronomico di Collurania, Via Maggini, I-64100, Teramo, Italy 
\and
 INFN,
 Sezione di Pisa, Largo Pontecorvo 3, I-56127, Pisa, Italy
\and
Dipartimento di Fisica "Enrico Fermi'',
Universit\`a di Pisa, Largo Pontecorvo 3, I-56127, Pisa, Italy
 }

   \offprints{G. Valle, valle@df.unipi.it}

   \date{Received 23/07/2018; accepted 13/09/2018}

  \abstract
{Open clusters (OCs) provide a classical target to calibrate the age scale and other stellar parameters. Despite their wide use,  some issues remain to be explored in detail.}
{We performed a theoretical investigation focused on the age estimate of red giant branch (RGB) stars in OCs based on mixed classical surface ($T_{\rm eff}$ and [Fe/H]) and asteroseismic ($\Delta \nu$ and $\nu_{\rm max}$) parameters. We aimed to evaluate the performances of three widely adopted fitting procedures, that is, a pure geometrical fit, a maximum likelihood approach, and a single stars fit, 
 in recovering stellar parameters. }
{A dense grid of stellar models was computed, covering different chemical compositions and different values of the mixing-length parameter. Artificial OCs were generated from these data by means of a Monte Carlo procedure for two different ages (7.5 and 9.0 Gyr) and two different choices of the  number of stars in the RGB evolutionary phase (35 and 80). The cluster age and other fundamental parameters were then recovered by means of the three methods previously mentioned. A Monte Carlo Markov chain approach was adopted for estimating the posterior densities of probability of the estimated parameters.   }
{The geometrical approach overestimated the age by about 0.3 and 0.2 Gyr for true ages of 7.5 and 9.0 Gyr, respectively. The value of the initial helium content was recovered unbiased within  the large random errors on the estimates. The maximum likelihood approach provided similar biases (0.1 and 0.2 Gyr) but with a variance reduced by a factor of between two and four with respect to geometrical fit. The independent fit of single stars showed a very large variance owing to its neglect of the fact that the stars came from the same cluster. The age of the cluster was recovered with no biases for 7.5 Gyr true age and with a bias of $-0.4$ Gyr for 9.0 Gyr.
The most important difference between geometrical and maximum likelihood approaches was the robustness against observational errors.  
For the first fitting technique, we found that estimations starting from the same sample but with different Gaussian perturbations on the observables suffer from a variability in the recovered mean of about 0.3 Gyr from one Monte Carlo run to another. This value was as high as 45\% of the intrinsic variability due to observational errors. On the other hand, for the maximum likelihood fitting method, this value was about 65\%. This larger variability led most simulations -- up to 90\% -- to fail to include the true parameter values in their estimated $1 \sigma$ credible interval.
Finally, we compared the performance of the three fitting methods for single RGB-star age estimation. 
The variability owing to the choice of the fitting method was minor, being about 15\% of the variability caused by observational uncertainties.}
{ Each method has its own merits and drawbacks. The single star fit showed the lowest performances. The higher precision of the maximum likelihood estimates is partially negated by the lower protection that this technique shows against random fluctuations compared to the pure geometrical fit.  Ultimately, the choice of the fitting method has to be evaluated in light of the specific sample and evolutionary phases under investigation.}


   \maketitle

\section{Introduction}\label{sec:intro}

Unlike other key stellar parameters, such as mass and radius, stellar ages -- with the exception of the Sun -- cannot be obtained by any direct measurement. This has profound consequences on  
a wide range of astrophysical questions, from the study of the evolution of planetary systems to
the understanding of the dynamical and chemical evolution of galaxies.
Therefore, several techniques 
have been developed to obtain age estimates that are as accurate as possible for single, binary, and cluster stars \citep[see][for a
review]{Soderblom2010}. 
These methods rely on different approaches -- from local interpolation to maximum likelihood (ML) and Bayesian fitting -- generally attempting to match the observational quantities against a set of precomputed stellar models to identify the best fit and possibly an associated statistical error.

In this framework, open clusters (OCs) provide a classical target to calibrate the age scale, and potentially other stellar parameters, which are notoriously difficult to evaluate for single stars, for example, the initial helium abundance or the convective core overshooting efficiency.   
Classically, the age of an OC is determined by fitting the HR diagram or its observational counterpart -- the colour-magnitude (CM) diagram -- with a set of stellar isochrones. 
A review of the age-estimation methods for OCs can be found in \citet{Mermilliod2000}, for example.
This approach, coupled with robust statistical treatment of the fitting procedure -- mainly by ML or Bayesian techniques -- leads to a major advancement in the age determination of OCs.
However the method suffers from some systematic errors that are sometimes difficult to properly address. A review of the approaches existing in the literature for tackling these problems can be found in \citet{Gallart2005}, who also address the question of composite resolved stellar populations.

A different approach to OC age determination became available in the last decade. The space-based missions CoRoT \citep[see e.g.][]{Appourchaux2008,Michel2008,Baglin2009} 
and {\it Kepler} \citep[see e.g.][]{Borucki2010, Gilliland2010} made precise asteroseismic measurements available for a large sample of stars. 
Among the selected targets in the original {\it Kepler} field of view there were four open clusters, that is, NGC 6791, NGC 6811, NGC 6819 and NGC 6866.
Although individual oscillations were available for  only a subset of stars with the best signal-to-noise ratios, two global asteroseismic quantities -- the average large frequency spacing $\Delta \nu$ and the
frequency of maximum oscillation power $\nu_{\rm max}$ -- are generally well determined.    
These quantities, combined with traditional non-seismic observables, such as effective temperature $T_{\rm eff}$, and metallicity [Fe/H], allow a different way to establish stellar masses, radii, and ages.

In particular, the detection of solar-like oscillations in G and K giants  \citep[see e.g.][]{Hekker2009,Mosser2010,Kallinger2010} makes these stars ideal targets to exploit the huge amount of asteroseismic observational data \citep{Miglio2012MNRAS, Miglio2012b,Stello2015}. This interest lead to the development of fast and automated methods to obtain insight into the stellar fundamental parameters exploiting classical and asteroseismic constraints \citep[see, among many,][]{Stello2009,Quirion2010,Basu2012,Silva2012, scepter1, Lebreton2014, Casagrande2014, Metcalfe2014, eta,Casagrande2016}.
Several works in the literature exploited the detected oscillations for red giant branch (RGB) stars in clusters to provide an estimate of their age 
\citep[see e.g.][]{Hekker2011,Miglio2012MNRAS,Miglio2012b,Corsaro2012,Wu2014,Miglio2015,Lagarde2015,Casagrande2016}. These estimates have the advantage of being independent of distance and reddening and provide an invaluable alternative way to check the age obtained by classical photometric fits of the cluster colour-magnitude diagram (CMD). 
The power of an integrated approach, that is, one that exploits data from various sources such as photometry, asteroseismology, and even binary systems, has been already demonstrated in the literature \citep[e.g.][]{Sandquist2016}.

The availability of a large amount of high-quality data motivated the adoption of powerful statistical methods to derive stellar characteristics. 
Most of the studies presenting asteroseismic age estimates of OCs adopt a maximum likelihood or Bayesian technique using a mixture of classical surface parameters (usually effective temperature and metallicity) and asteroseismic constraints.  
Despite the growing popularity of these approaches, the question of the theoretical foundations of the OC fit when such  constraints are available is largely unexplored in the literature. Indeed, a comprehensive study of the biases and the random variability to be expected in the OC asteroseismic age estimate from fitting RGB stars is still lacking.

This deficiency is clearly caused by the tremendous computational effort needed to shed light on this topic, an effort that was out of reach until recent years. 
A similar investigation is needed to carefully consider the different sources of theoretical variability that influence the stellar evolution but  are not directly constrained by the observations. Failure to address some relevant sources of variability can lead to meaningless age calibration \citep[see e.g.][]{TZFor, BinTeo}. 
The problem is of particular relevance because the stellar evolution theory is still affected by non-negligible uncertainty \citep[see e.g.][]{incertezze1, Stancliffe2015, Dotter2017, Salaris2017, Tognelli2018}. 
Some relevant error sources are related to the treatment of convection  and hamper a firm prediction on the extension of convective cores beyond the classical Schwarzschild border -- often modelled by assuming a parametric overshooting -- and of the efficiency of the superadiabatic convection, generally treated following the mixing-length scheme.
To evaluate the effect of these uncertainties on the results grid of models with a wide range of masses, chemical compositions, overshooting, and external convection efficiencies are required. However, such a grid is not freely available in the literature and should be computed ad-hoc when a similar investigation is attempted.

In this  investigation we explore some aspects of the performances attainable  for cluster stars in the RGB phase. We focus on the fitting of stars whenever either classical surface or global asteroseismic observables are available. In the former group we consider the effective temperature and the metallicity [Fe/H], while in the latter we look at the average large frequency spacing $\Delta \nu$ and the
frequency of maximum oscillation power $\nu_{\rm max}$. 
Our aims are twofold. Firstly, we are interested in evaluating the minimal statistical biases and errors on the recovered age (and on other stellar parameters varied in the fit) in various scenarios, with different reference age and size of the studied samples. These quantities are required to plan a sensible investigation on real observed objects. Secondly, we are interested in a thorough statistical comparison of some archetypal techniques adopted in the literature for cluster fitting. To this purpose we assess the relative merits and drawbacks of three approaches widely adopted in the literature, and quantify the bias in the reconstructed age for each of them. 
  
The study is organised as follows. In Sect.~\ref{sec:method} we present the framework for the investigation, the adopted Monte Carlo approach, the grid of stellar models, and the fitting methods discussed in the text. In Sect.~\ref{sec:results} we present the results of the investigation and discuss the biases in the recovered parameters. The impact of the observational errors is further analysed in Sect.~\ref{sec:raneff}.
A comparison between two of the adopted fitting methods is performed in Sect.~\ref{sec:cfr} for the purpose of assessing the variability in RGB single star age estimates due to the adoption of different recovery techniques. Some conclusions are presented in Sect.~\ref{sec:conclusions}.

\section{Methods}\label{sec:method}

Our theoretical analysis has been performed on a mock data set of artificial stars computed at a selected chemical composition and external convection efficiencies to simulate open cluster data. 
The age of the synthetic cluster, along with other stellar parameters, are recovered through several fitting methods, relying on a grid of stellar models with different chemical input (i.e. metallicity $Z$ and initial helium abundance $Y$) and mixing-length parameter $\alpha_{\rm ml}$, to evaluate the relative performances of the procedures.

However, we did not take into account other known sources of variability. In particular, all the models were computed keeping the input physics fixed at the reference values and for a given  heavy-element solar mixture. This choice stems for the consideration that it is not possible to model these sources of uncertainty using only one scaling parameter. 
As an example, the variability expected on the mean Rosseland opacity tables can be evaluated by comparing the computations  by different groups. For a MS star, especially near the solar mass, mean differences of about 5\% have been pointed out in the literature \citep{rose2001,OP2005}. However, for different ranges of temperature and density, such as those considered in this work, these differences show larger variations. Therefore a single scaling cannot precisely account for the opacity uncertainties.    

The reference case chosen in this work approximatively mimics the old open cluster NGC 6791; in the following we assume the reference chemical composition $Z = 0.02674$, $Y = 0.302$, corresponding to a helium-to-metal enrichment ratio $\Delta Y/\Delta Z = 2.0$. This choice implies [Fe/H] = 0.3, assuming the solar heavy-element mixture by \citet{AGSS09}. We fix two different ages at which the recovery is performed, that is, 7.5 Gyr and 9.0 Gyr.

In the subsequent step we sampled $n$ stars from the reference isochrones in a way to systematically cover the whole RGB\footnote{The synthetic stars cover the range of $\Delta \nu$ [10.8, 1.2] $\mu$Hz, assuming a reference $\Delta \nu_{\sun} = 135$ $\mu$Hz}.  Gaussian perturbations were applied to classical and asteroseismic observables of these synthetic stars, mimicking the observational errors. The assumed uncertainties for this process were: 75 K in $T_{\rm eff}$, 0.1 dex in [Fe/H], 1\% in  $\Delta \nu$, and 2.5\% in $\nu_{\rm max}$. The errors on the surface parameters were set from the errors quoted in the APOKASC catalogue \citep{Pinsonneault2014} and in \citet{Tayar2017}, while the uncertainties on the asteroseismic quantities were chosen 
taking into account the values quoted in SAGA of 0.7\% and 1.7\%  on $\Delta \nu$ and 
$\nu_{\rm max}$ , respectively \citep{Casagrande2014}, and those in the APOKASC catalogue
of 2.2\% and 2.7\%, respectively.
To assess the importance of the sample size on the final results, two different values of $n$ were considered: $n = 35$ for a scarcely populated RGB, and $n = 80$ for a well-populated RGB. These choices are representative of real numbers of RGB stars adopted in studies to constrain open cluster parameters by adopting asteroseismic quantities  \citep[e.g.][]{Basu2011, Hekker2011}.

In summary, four scenarios have been considered: a sample of 35 stars at age 7.5 Gyr (S35 in the following), a sample of 80 stars at 7.5 Gyr (S80), a sample of 35 stars at 9.0 Gyr (S35-9) and finally a sample of 80 stars at 9.0 Gyr (S80-9). Moreover, starting from the same artificial sample, the perturbation step was repeated $N = 80$ times to assess the differences in the age estimate solely due to random errors in the measurement process. 

For the recovery step we needed to compute a dense grid of stellar models around the reference scenario. To obtain a sensible age error estimate, the stellar grid should span an appropriate range not only in mass (or age for isochrones) but also in other quantities that impact the stellar evolution, such as the initial chemical composition ($Z$ and $Y$), and in parameters that affect the position of the RGB, such as the value of the mixing-length parameter $\alpha_{\rm ml}$. The grid of stellar models is presented in Sect.~\ref{sec:grids}.
The actual recovery was performed by means of three different methods, fully described in Sect.~\ref{sec:fittingML}. 

\subsection{Stellar model grid}
\label{sec:grids}

The model grids were computed for  masses in the range [0.4, 1.3] $M_{\sun}$, with a step of 0.05 $M_{\sun}$.
The initial metallicity [Fe/H] was varied from 0.15 to 0.45 dex, with
a step of 0.05 dex. 
The solar heavy-element mixture by \citet{AGSS09} was adopted. 
Nine initial helium abundances were considered at fixed metallicity by adopting the commonly used
linear relation $Y = Y_p+\frac{\Delta Y}{\Delta Z} Z$
with the primordial abundance $Y_p = 0.2485$ from WMAP
\citep{peimbert07a,peimbert07b} and with a helium-to-metal enrichment ratio $\Delta Y/\Delta Z$
from 1 to 3 with a step of 0.25 \citep{gennaro10}. 

The FRANEC code \citep{scilla2008, Tognelli2011} was used to compute the stellar models, in the same
configuration as was adopted to compute the Pisa Stellar
Evolution Data Base\footnote{\url{http://astro.df.unipi.it/stellar-models/}} 
for low-mass stars \citep{database2012}. 
The models were computed
for five different values of mixing-length parameter $\alpha_{\rm ml}$ in the range [1.54, 1.94] with a step of 0.1, with 1.74 being the solar-scaled value\footnote{The calibration is performed repeating the Sun evolution by changing $Z$, $Y$ and $\alpha_{\rm ml}$. The iteration stops when, at the Sun age, the computed radius, luminosity, effective temperature, and photospheric [Fe/H] match the observed values with relative tolerance $10^{-4}$.}. Microscopic diffusion is considered according to \citet{thoul94}.
Convective core overshooting was not included because for the range of mass relevant for the analysis (the reference scenario mass in RGB is about 1.15 $M_{\sun}$) it does not contribute to the stellar evolution.
Further details on the stellar models are fully
described in \citet{cefeidi,eta,binary} and references therein.  

Isochrones, in the age range [5, 10] Gyr with a time step of 100 Myr,  were computed according to the procedure described in 
\citet{database2012, stellar}.

The average large frequency spacing $\Delta \nu$ and the frequency of maximum 
oscillation power $\nu_{\rm max}$ are obtained using the scaling relations from
the solar values \citep{Ulrich1986, Kjeldsen1995}: 
\begin{eqnarray}\label{eq:dni}
\frac{\Delta \nu}{\Delta \nu_{\sun}} & = &
\sqrt{\frac{M/M_{\sun}}{(R/R_{\sun})^3}} \quad ,\\  \frac{\nu_{\rm
                max}}{\nu_{\rm max, \sun}} & = & \frac{{M/M_{\sun}}}{ (R/R_{\sun})^2
        \sqrt{ T_{\rm eff}/T_{\rm eff, \sun}} }. \label{eq:nimax}
\end{eqnarray}
The validity of these scaling relations in the RGB phase has been questioned in recent years \citep{Epstein2014, Gaulme2016, Viani2017}, posing a serious problem  whenever adopted for a comparison with observational data. Fortunately,  this question is of minor importance for our aims, 
because both artificial data and the recovery grid are computed using the same scheme.

\subsection{Fitting procedures}\label{sec:fittingML}

For the age recovery phase we employed three different techniques: a pure geometrical isochrone fitting, a Bayesian ML approach, and an independent fit of single stars.
All these methods are commonly adopted in the literature \citep[see among many][]{Frayn2002, Pont2004, Jorgensen2005, vonHippel2006, Gai2011, Casagrande2016, Creevey2017}, and several aspects of their relative performances are well understood. 
A comprehensive review of their use for age fitting -- but for different observable constraints -- can be found in 
\citet{Valls2014} and references therein.
The comparisons of the techniques presented in the literature are frequently made using main sequence stars -- for which the reliability of the theoretical stellar model is the highest -- and often do not assess the relevance for the results of the different sources of uncertainties in the theoretical stellar models. Indeed, a general theoretical analysis of the age recovery for RGB stars with accurate asteroseismic observational data is still lacking in the literature. 

To this aim, the present analysis assesses the biases and the uncertainty  -- due solely to the observational
errors -- on the final age estimates for RGB stars, focussing on several key quantities governing the stellar evolution. In other words, we work in an ideal framework in which the synthetic stars are sampled from the same grid adopted in the recovery stage. In this way we evaluate the theoretical minimum uncertainty, and, most importantly, the unavoidable biases that characterise the different adopted methods. 
 
\subsubsection{Geometrical fit} \label{sec:sub-geo}
 
The first recovery technique is based on pure geometrical fitting of isochrones to the observed quantities. 
Let $\theta = (\alpha_{\rm ml}, \Delta Y/\Delta Z, Z, {\rm age})$ be the vector of isochrone parameters and $q_i \equiv \{T_{\rm eff}, {\rm [Fe/H]}, \Delta \nu, \nu_{\rm max}\}_i$ be the vector of observed quantities for each star $i$ in the sample ($i = 1, \ldots, N$). 
Although for numerical reasons the fit is performed on the $\Delta Y/\Delta Z$ quantity, this quantity is linked to the initial helium abundance by the linear relation given in Sect.~\ref{sec:grids}. The discussion of the results is therefore frequently focussed on the latter quantity.

We set  $\sigma_i$ to be the vector of observational uncertainties for the $i$th star.
For each point $j$ on a given isochrone we define $q_j(\theta)$ as the vector of theoretical values. Finally, we compute the geometrical distance $d_{ij}(\theta)$ between the observed star $i$ and the $j$th point on the isochrone, defined as
\begin{equation}
d_{ij}(\theta) = \left\lVert \frac{q_i - q_j(\theta)}{\sigma_i} \right\rVert. \label{eq:dist}
\end{equation} 
The statistic
\begin{equation}
\chi^2(\theta) = \sum_{i=1}^N \min_j d_{ij}^2(\theta)
\end{equation} 
can then be adopted to compute the probability $P(\theta)$ that the sample comes from the given isochrone. 
In fact, if stars are uniformly distributed along the isochrone (this is indeed not the case, but see the following discussion), and assuming an independent Gaussian error model, it follows that
\begin{equation}
P(\theta) \propto \exp(-\chi^2/2), \label{eq:prob-geom}
\end{equation} 
where we neglect the normalization constant, which only depends on the observational errors $\sigma$ and not on $\theta,$ and has no practical importance in the posterior probability estimation. In fact, as detailed below, the relative merits of two sets of parameters $\theta_1$ and $\theta_2$ are evaluated on the basis of the ratio $P(\theta_1)/P(\theta_2),$ meaning that the contribution  of the normalization constant cancels out. 
In other words, the method minimises the sum of the minimum distances from each point to the isochrone and computes the probability that the sample comes from the isochrone parametrized by $\theta$ as the product of probability that each star comes from the isochrone.

A key point to address in the computation is that the isochrone is represented by a sequence of points, whose position is dictated by the stellar evolution time step. Therefore some evolutionary phases are more densely populated than others, posing a possible problem in the evaluation of the minimum distance. To overcome this difficulty we determined the minimum distance of an observational point $q_i$ from an isochrone in two steps. First we found the minimum distance $d_1$ using Eq.~(\ref{eq:dist}),
\begin{equation}
d_1 = \min_j d_{ij}(\theta).\label{eq:dist2}
\end{equation}
We set $\tilde{\jmath}$ to be the index on the isochrone corresponding to $d_1$. 
Subsequently, the distances $d_2$ and $d_3$ from the straight line passing from the points $\tilde{\jmath} - 1$ and $\tilde{\jmath} + 1$, 
respectively, are evaluated. The distances $d_2$ and $d_3$ were considered valid if the projection of $q_i$ was found to lie on the segments connecting $q_{\tilde{\jmath}-1}$ to $q_{\tilde{\jmath}}$ and $q_{\tilde{\jmath}}$ to $q_{\tilde{\jmath}+1}$ , respectively. Then the minimum of the valid distances among $d_1$, $d_2$, and $d_3$ was adopted as the distance from the point $i$ to the isochrone. 

The method easily allows  one to specify a prior on the $\theta$ parameter, adding them as multiplicative factors in Eq.~(\ref{eq:prob-geom}). In the following we adopt flat priors for all the parameters.  

The results obtained in this way can be compared to a direct numerical integration of the likelihood function $P(\theta)$ over the whole grid. This procedure does not rely on isochrone interpolations but evaluates the likelihood for all the grid $\theta$ values and approximates the integral with a discrete sum. The marginalized distributions of each parameter are obtained by summing the likelihood function over all the other parameters. Then, the mean values and the standard deviations of the distributions are adopted as best estimates and their $1 \sigma$ errors.  
This approach works well when the grid is dense enough in the parameter space because it cannot explore combinations of parameters outside those adopted in the grid computation. An in-depth discussion of the relative performance of this approach with respect to others, such as Monte Carlo Markov chain (MCMC) simulations in the framework of a binary system, fit can be found in \citet{Bazot2012}.

\subsubsection{Maximum-likelihood fit} \label{sec:sub-ML} 
 
It is well known \citep[see][and references therein]{Valls2014} that pure geometrical fitting suffers from intrinsic degeneracy, and different sets of parameters can provide similar likelihoods. In particular, the age-metallicity degeneracy poses a significant problem, chiefly in the main sequence phase. Indeed the geometrical fit does not exploit all the information provided by the stellar evolution computation, neglecting the timescale of the evolution in the fit.
Therefore the second adopted recovery technique was a variation of the previous one, and accounts for the variation of the stellar evolution timescale climbing on the RGB. As  widely discussed in the literature, the inclusion of this information in the fitting procedure can lead to better estimates than those obtained with a pure geometrical fit  \citep[see][and references therein]{Valls2014}.
We define the likelihood that a star $i$ comes from a given isochrone as
\begin{equation}
L(\theta)_i \propto \sum_{j=1}^N  \exp(-d_{ij}^2/2) \; \Delta M_j \label{eq:likBayes}
,\end{equation}
where $\Delta M_j$ is the mass interval between the point $j$ and $j+1$ in the isochrone and accounts for the different evolutionary timescale along the isochrone. 
As in the previous case, the normalization constant can be safely neglected for our aims.

In other words the likelihood that a point comes from an isochrone can be viewed as the sum over all the isochrone points of the products of the probability that it comes from position $j$ in the isochrone ($\Delta M_j$) multiplied by the probability of a distance $d_{ij}$ of the observational point to the $j$th isochrone position (assuming a Gaussian error model). The distance $d_{ij}$ from observation to isochrone position was assessed with the same technique discussed above for the geometrical fit. 
The probability that the whole observed sample comes from the given isochrone is therefore simply the product of all the individual likelihoods:    
\begin{equation}
P(\theta) = \prod_{i=1}^N L(\theta)_i \label{eq:prob-Bayes}
.\end{equation} 
This method reduces to the previous one if all the contributions to $L(\theta)_i$ are neglected except the one from the point with minimum $d_{ij}$.
We note that in the formulation of the likelihood in Eq.~\ref{eq:likBayes},  a factor giving the contribution of the initial mass function (IMF) is often included; however its inclusion in the RGB phase is of negligible importance given the small differences in mass among the isochrone points (the mass range spanned by the theoretical points is 0.008 $M_{\sun}$). Therefore we prefer to not include a specific formulation of IMF, implicitly assuming a flat IMF in this mass region. This assumption is in agreement with the discussion in \citet{Valls2014} about the relevance of the IMF weight in different evolutionary phases. However, as a safety measure, we verified that assuming a Salpeter IMF alters the results in a negligible way. As in the case of the geometrical fit, flat priors were assumed on the parameters $\theta$.

As in the previous case, computational implementation is particularly important for the obtention of a reliable estimate from Eq.~(\ref{eq:likBayes}), which approximates the integral over the isochrone. In particular, the likelihood in the equation depends on the mass step between points. The distance in mass between consecutive points in the raw isochrone is not constant and typically too large to grant a fine approximation of the integral.
Therefore the isochrones entering in the likelihood evaluation were linearly interpolated in mass so that 600 points were located, equally spaced in mass, in the relevant RGB range (values of $\Delta \nu/\Delta \nu_{\sun} < 0.15$). Each distance was then evaluated and adopted for the computation of Eq.~(\ref{eq:likBayes}). The adopted choices are more than enough for granting stable results, as verified by a direct trial-and-error procedure.  

\subsubsection{SCEPtER fitting of individual stars} \label{sec:sub-scepter}

As a last method we relied on fitting individual stars, that is, each star was independently fitted relaxing the constraint that they should share the same chemical composition and superadiabatic convection efficiency. The fit was performed by means of the SCEPtER method \citep{eta, bulge, binary}, an  ML technique adopted in the past for single and binary stars. 
For the $i$ star, the technique computes a likelihood for each $j$ point of every given isochrone as
\begin{equation}
L(\theta)_j \propto \exp (-d_{ij}^2(\theta)/2)
\label{eq:lik-scepter}
,\end{equation}
with $d_{ij}(\theta)$ as in Eq.~(\ref{eq:dist}).
The likelihood function is evaluated for each grid point within $3 \sigma$ of
all the variables from $q_i$; we define $L_{\rm max}$ as the maximum value
obtained in this step. The estimated stellar quantities are obtained
by averaging the corresponding $\theta$ quantity of all the models with a likelihood
greater than $0.95 \times L_{\rm max}$.
The performances of this method are expected to be clearly inferior to the previous ones since it does not use the information that the stars come from a cluster. Indeed, it serves as a useful reference to compare the performance of the first two techniques and -- thanks to the large numbers law -- the uncertainty on the recovered mean age is expected to scale as the inverse of the square root of the sample size.

\subsection{Simulations scheme}\label{sec:schema}

To establish the accuracy and precision of the three techniques we relied on several Monte Carlo simulations. For all of four scenarios defined in Sect.~\ref{sec:method} (S35, S35-9,
S80, and S80-9), we performed $N = 80$ Monte Carlo different perturbations of the synthetic observational values. The number of simulations was assessed by means of a trial-and-error procedure, and is enough to reach the stability of the results.
Each synthetic sample was subjected to parameter estimation through the three techniques described in Sect.~\ref{sec:fittingML}. 
A suffix is added to the scenario names to identify the algorithm adopted in the fit. Therefore, as an example, scenario S35 refers to the geometrical fit of 35 artificial stars sampled at 7.5 Gyr, S35w refers to the same configuration fitted by the ML approach, and S35S to the fit of individual stars by means of SCEPtER.  

The SCEPtER estimation, which can be viewed as a local interpolation, promptly returns a set of parameters, $\theta,$ for each star. Therefore we obtained a dataset of $80 \times 35$ estimates for scenarios S35S and S35-9S and $80 \times 80$ estimates for scenarios S80S and S80-9S.  

The estimation for the first two methods is more complex than when fitting single stars and was performed adopting a MCMC process, using the likelihoods in Eqs.~(\ref{eq:prob-geom}) and~(\ref{eq:prob-Bayes}). This method is widely adopted in the literature, originating from the formulation of \citet{Metropolis1953} and \citet{Hastings1970}. 
The method identifies a good starting point and then explores the parameter space by means of random displacements, following a user-specified "jump function". Rules for accepting the sampled parameters are based on the comparison of the likelihood of the new set of parameters with that of the previous one. The initial part of the chain is considered non-stationary and is discarded (burning-in phase) from the analysis. The chain is then evolved for a length long enough to fully explore the posterior density space.     
Details on the procedure, the length of the burning-in and sampling chains, and on the convergence-enhancing algorithmic approach are provided in Appendix~\ref{app:MCMC}.

For the geometrical fitting, the evolution of eight parallel chains of 7\,000 repetitions for samples of 35 stars and 21\,000 repetitions for samples of 80 stars were adequate for convergence. For the ML approach, the burn-in chain length had to be increased to 9\,000.
The difference is due to the faster convergence of the pure geometrical fitting, whose likelihood function is smoother than that of the ML approach, which is sparse in the parameter space, causing the chain to have greater difficulty in accurately mapping the posterior probability density. 

\subsection{Differences between geometrical and ML approaches}\label{sec:diff-geo-ML}

The apparent similarity of the methods described in Sects.~\ref{sec:sub-geo} and \ref{sec:sub-ML} hides a significant difference in the underlying approaches and in the targeted best solution. The geometrical method is insensitive to any part of a fitting isochrone but the nearest point, so two isochrones that have an equal minimum distance but a different shape are indistinguishable. This shows that this method clearly does not exploit all the information coded in the isochrone. 

On the other hand, the ML fit can leverage not only the minimum distance, but also a portion of the isochrone in the neighbourhood of this point, as it considers all the points on the curve; see Eq.~(\ref{eq:likBayes}). The ML also accounts for the timescale of the evolution; this term acts as a prior in the fitting stage, always favouring slowly evolving phases over rapid ones.

It is therefore understandable that the best solution for the geometrical fit can be, and will  generally be, different from the ML one. Indeed, it is possible to show this behaviour even in a simple toy model (Appendix \ref{app:toy}). 
The Appendix considers a simple "isochrone", modelled by an arc of parabola, and one observational point in a two-dimensional (2D) space that can be shifted in one direction only (see Fig.~\ref{fig:lik-toy} of the appendix). The position of the point is then varied continuously and the likelihood from Eqs.~(\ref{eq:prob-geom}) and (\ref{eq:likBayes}) are computed.

It is obvious that the adoption of a pure geometrical approach leads to the best fitting isochrone being the one that exactly passes through the observational point. However, 
this is not the case for the ML approach, which favours a slightly biased value but allows for a greater portion of the curve to remain close to the observational point. 
Moreover, Appendix~\ref{app:toy} also shows that the variance of the ML solution is lower than that of the geometrical fit. This is logical because, as mentioned above, the ML approach has much more information to use for discriminating among isochrones and discarding solutions that are still acceptable for the  geometrical fitting.  
Therefore, as the sample size increases, it is expected that the ML error shrinks faster than the geometrical one. 

Although for real data the situation is much more complex, these general considerations still hold and are at the basis of the differences between the results obtained with the two methods  discussed in the following sections.

\section{Estimated parameters}
\label{sec:results}

\begin{figure*}
        \centering
        \includegraphics[height=6.0cm,angle=-90]{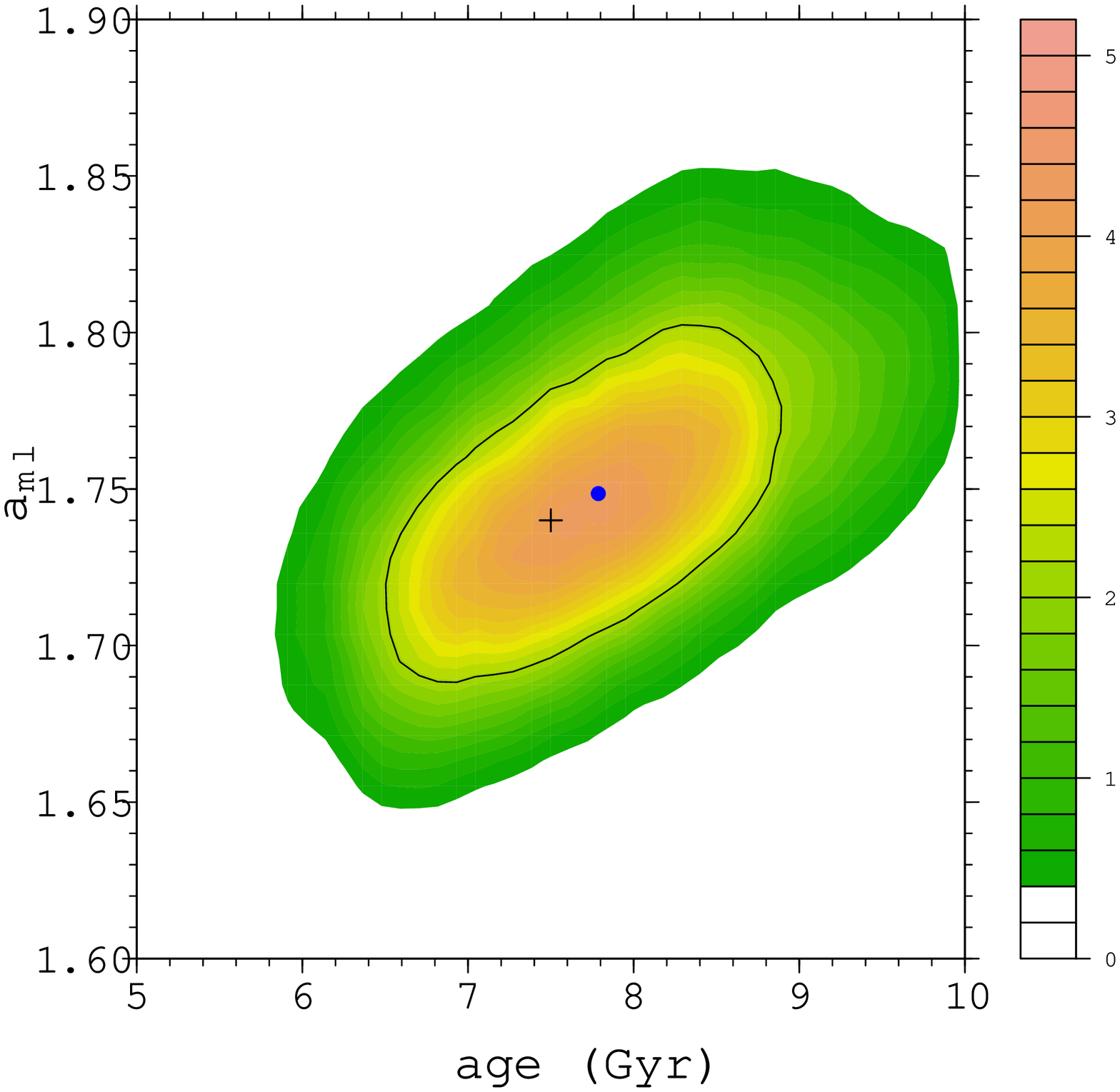}
        \includegraphics[height=6.0cm,angle=-90]{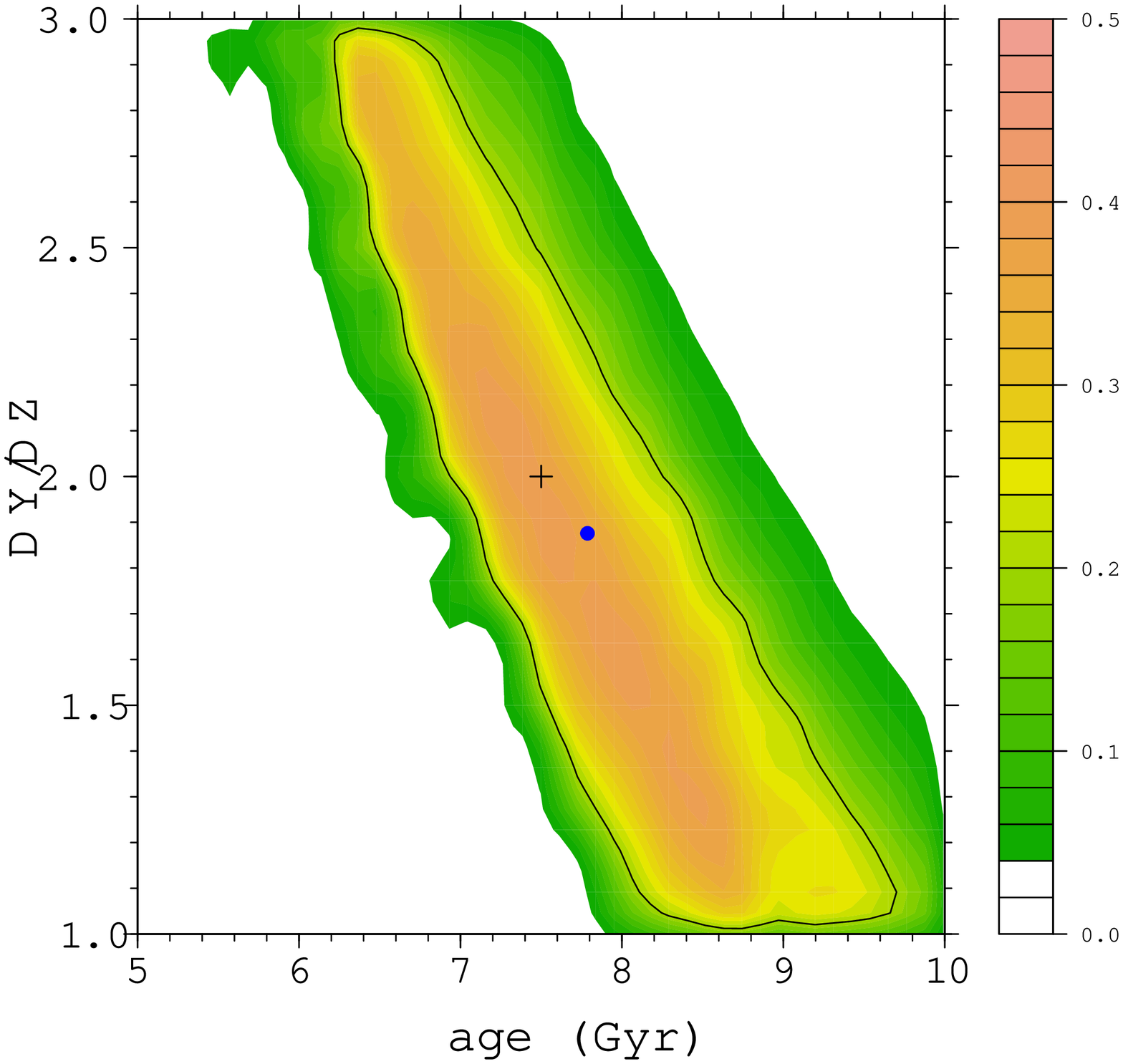}
        \includegraphics[height=6.0cm,angle=-90]{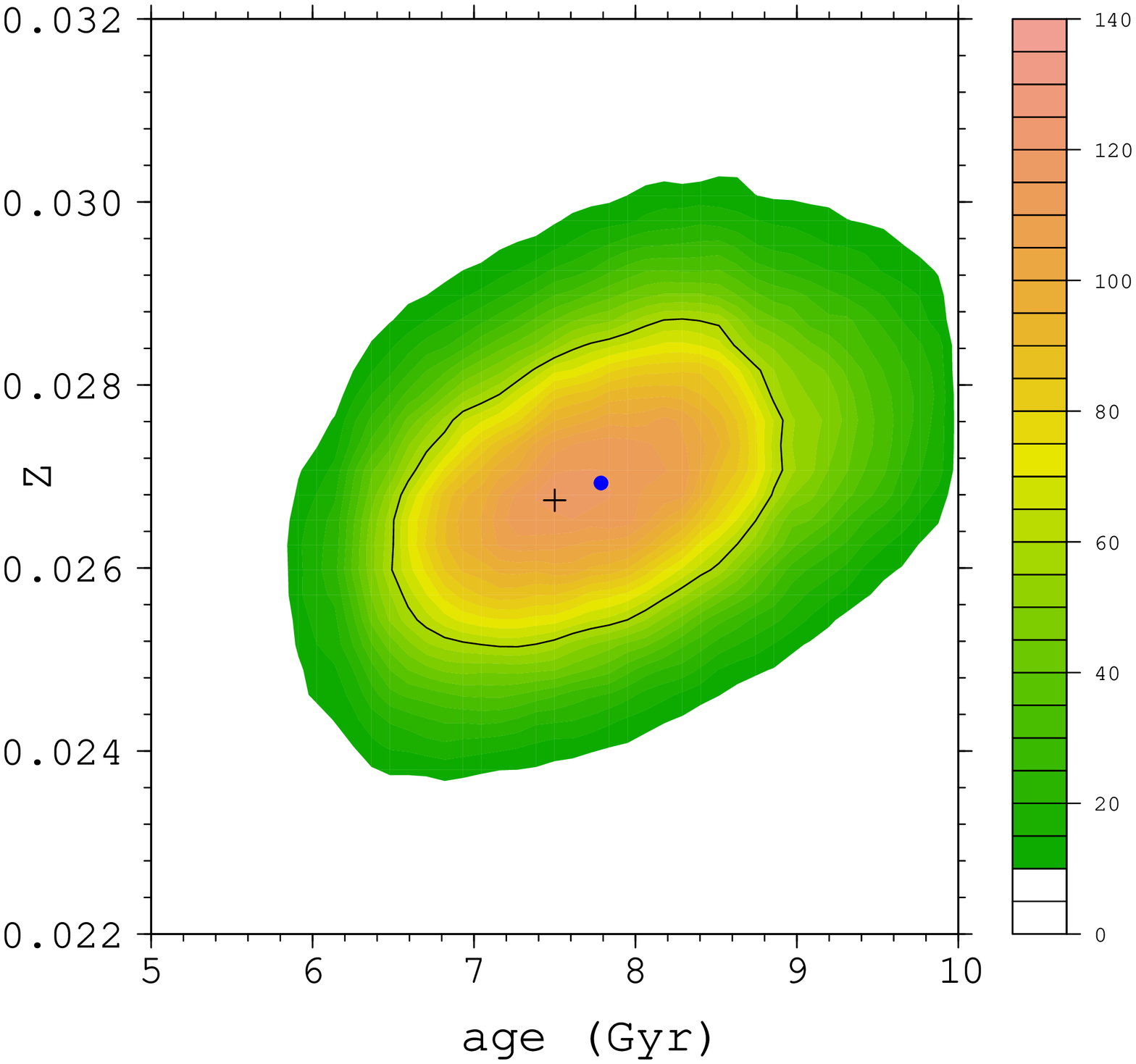}\\
        \includegraphics[height=6.0cm,angle=-90]{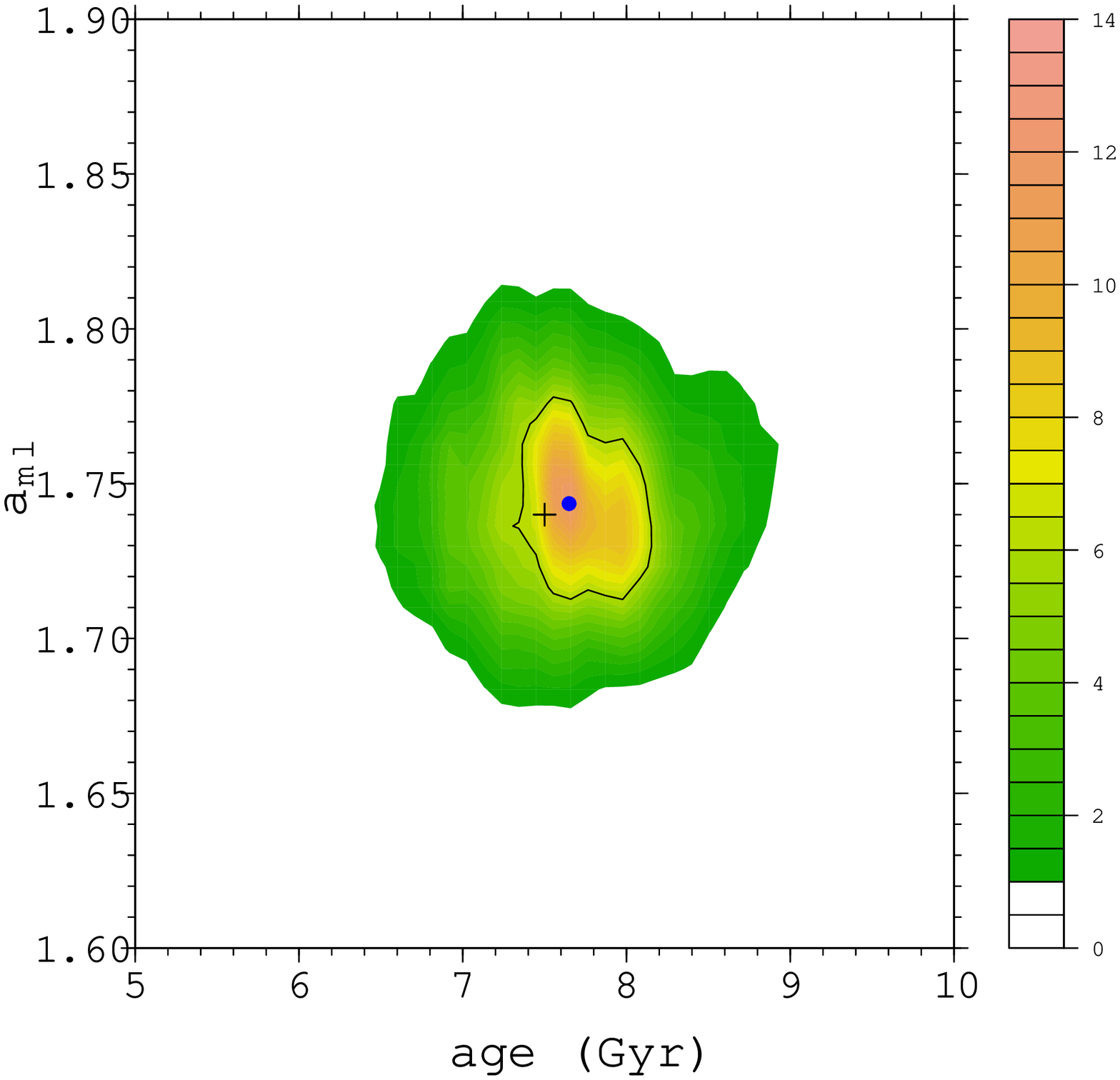}
        \includegraphics[height=6.0cm,angle=-90]{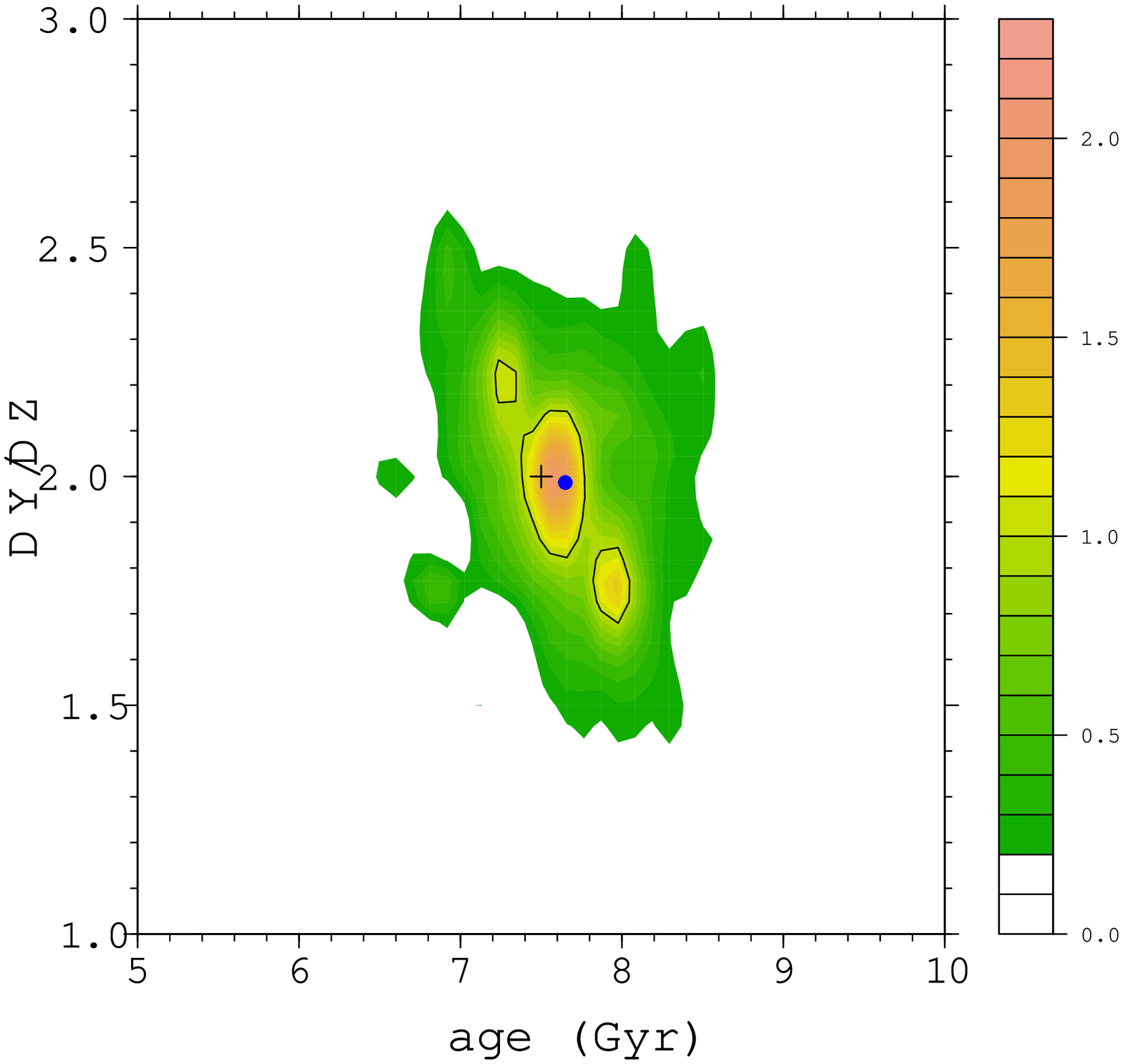}
        \includegraphics[height=6.0cm,angle=-90]{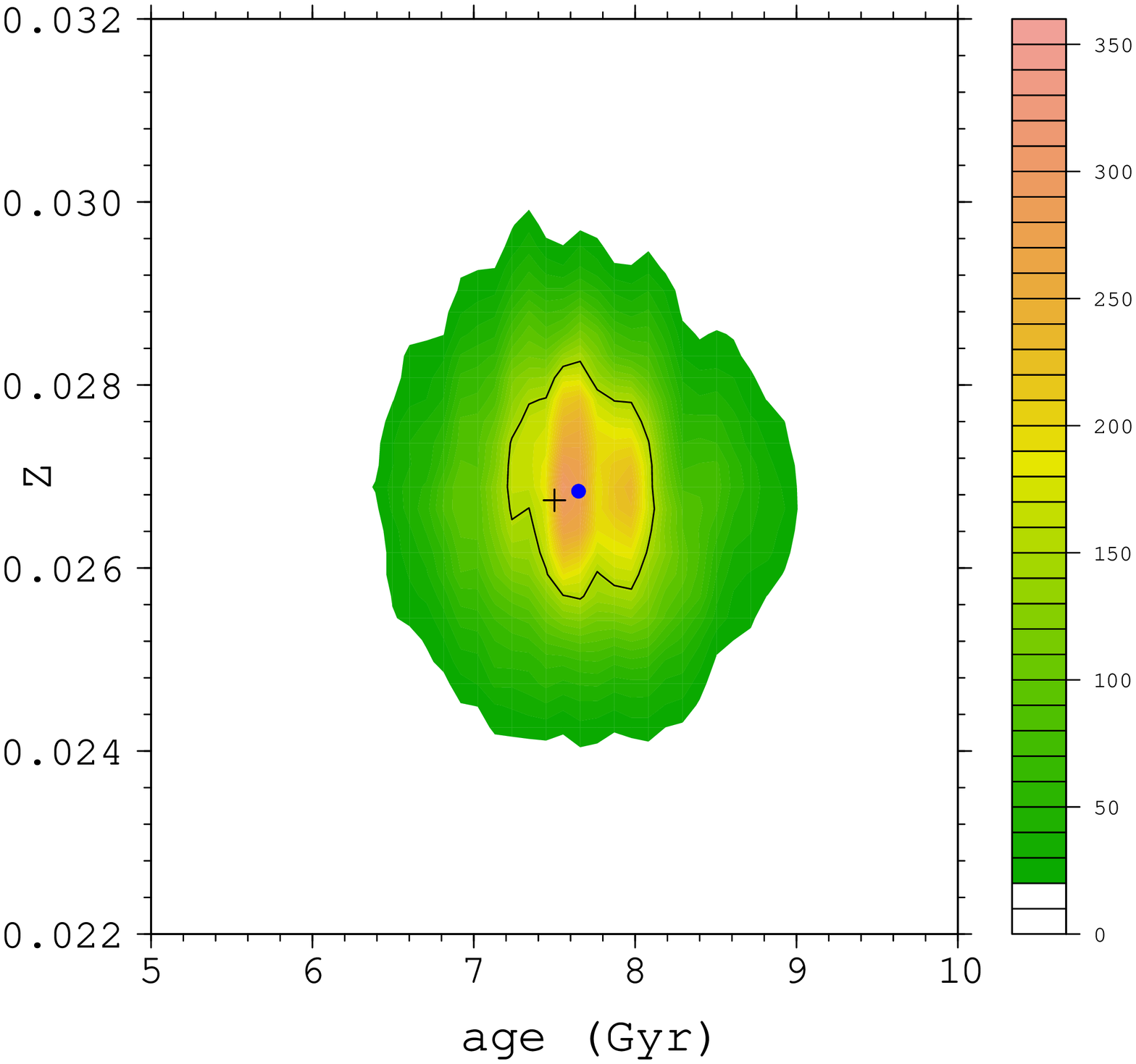}\\
        \includegraphics[height=6.0cm,angle=-90]{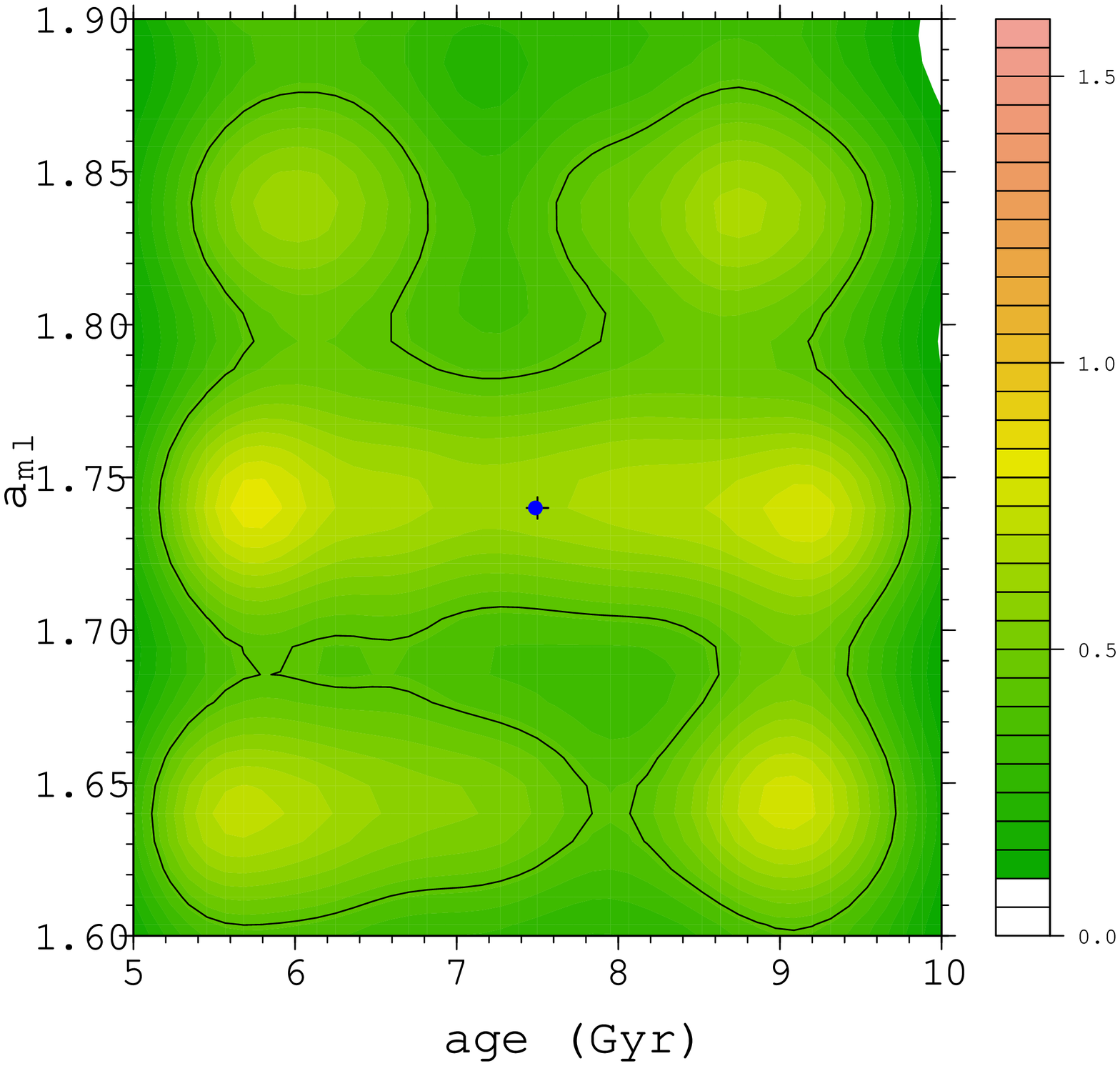}
        \includegraphics[height=6.0cm,angle=-90]{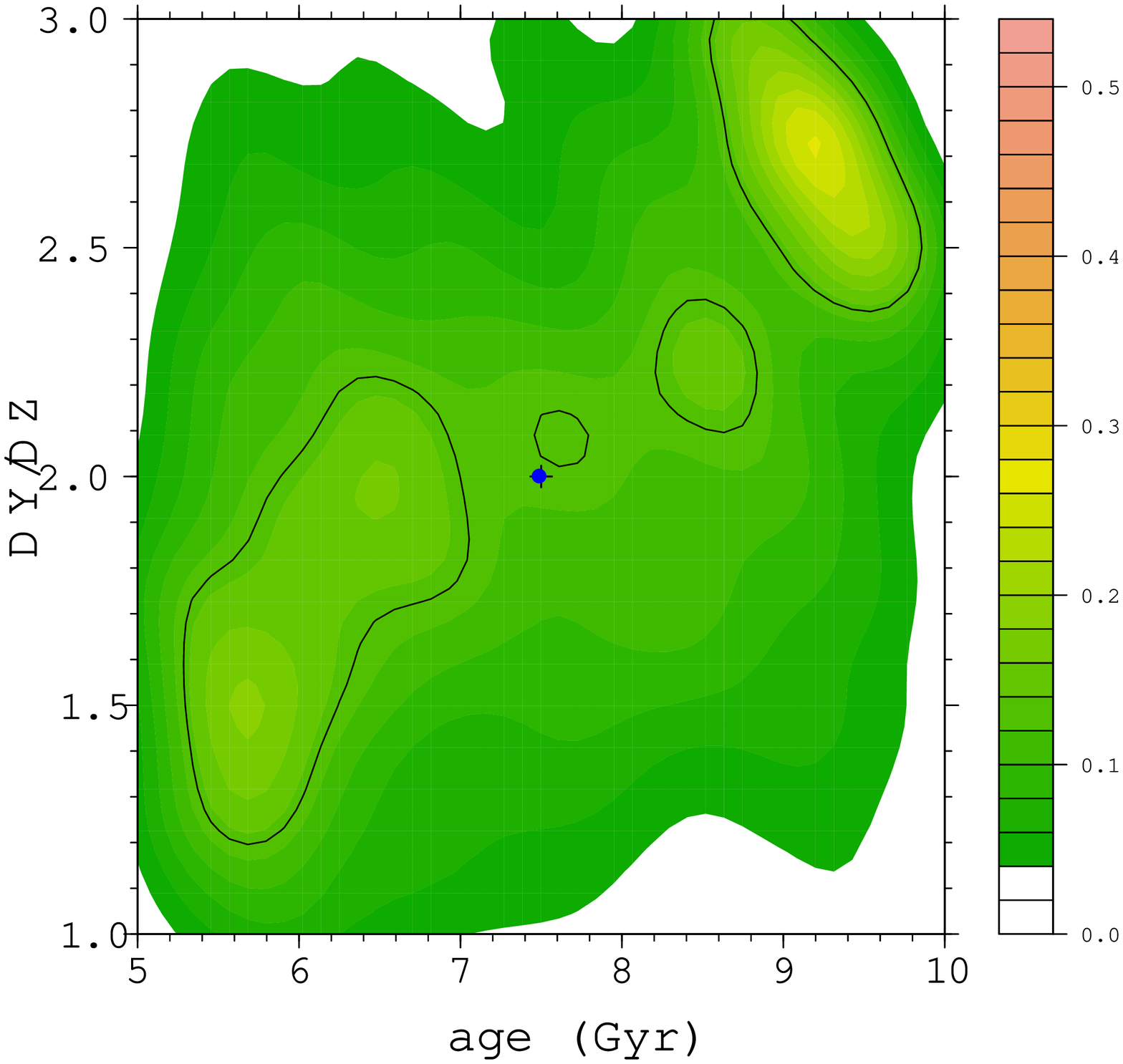}
        \includegraphics[height=6.0cm,angle=-90]{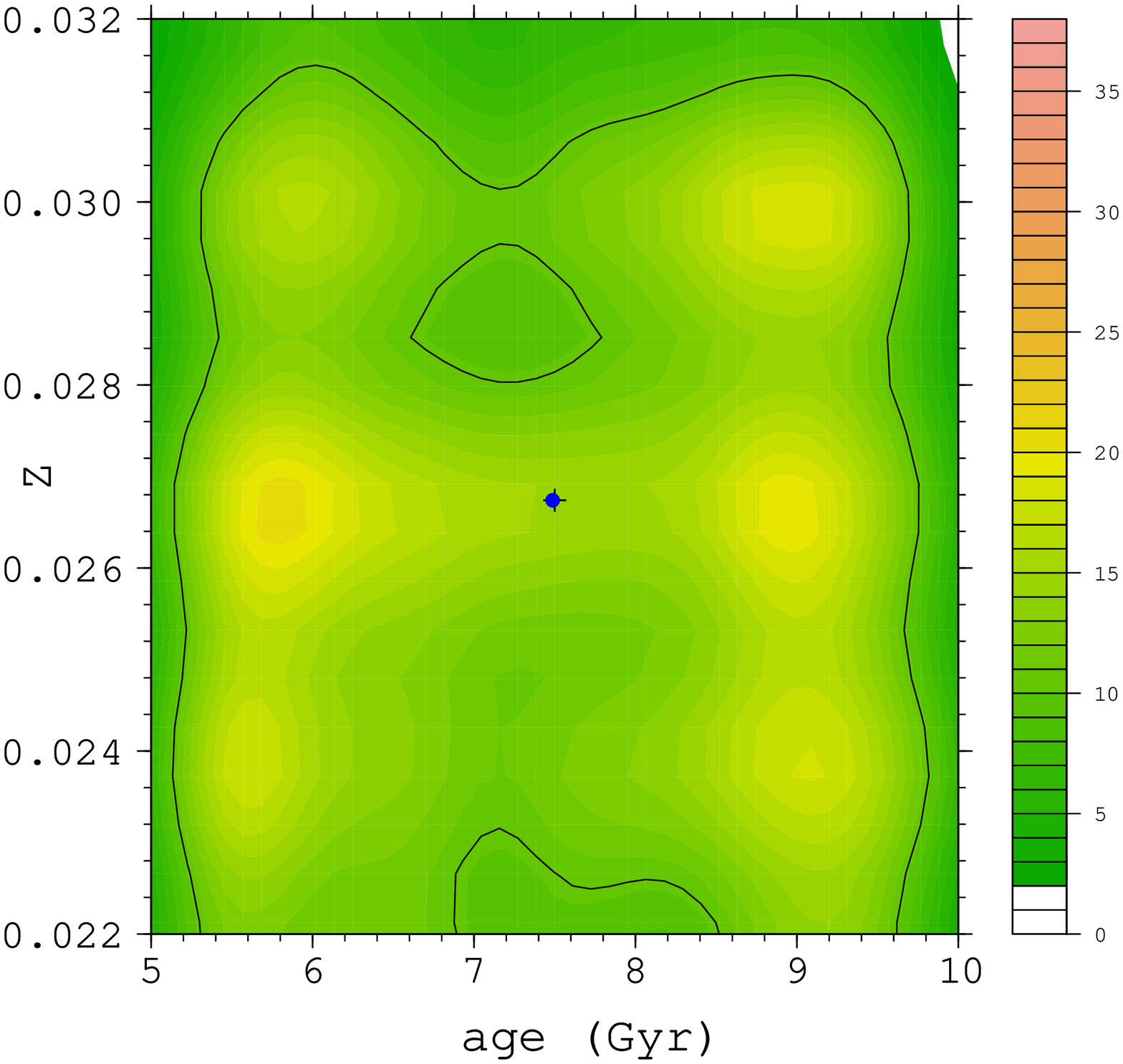}
        \caption{{\it Top row}, {\it left}: 2D density of probability in the age vs. $\alpha_{\rm ml}$ plane from samples of 35 RGB stars (true age of 7.5 Gyr). Parameters are recovered by means of the {\it pure geometrical} method (see text). The cross marks the position of the true values, while the filled circle marks the median of the recovered ones. {\it Middle}: same as in the {\it left} panel, but in the age vs. $\Delta Y/\Delta Z$ plane. {\it Right}: As in the {\it left} panel, but in the age vs. $Z$ plane. {\it Middle row}: As in the top row, but for maximum likelihood recovery. {\it Bottom row}: As in the top row, but for SCEPtER individual recovery.
        For ease of comparison, the corresponding panels in the three subsequent figures (Figs.~\ref{fig:res9.0Gyr35}, \ref{fig:res7.5Gyr80}, and \ref{fig:res9.0Gyr80}) share the same colour key.
        }
        \label{fig:res7.5Gyr35}
\end{figure*}

\begin{figure*}
        \centering
        \includegraphics[height=6.0cm,angle=-90]{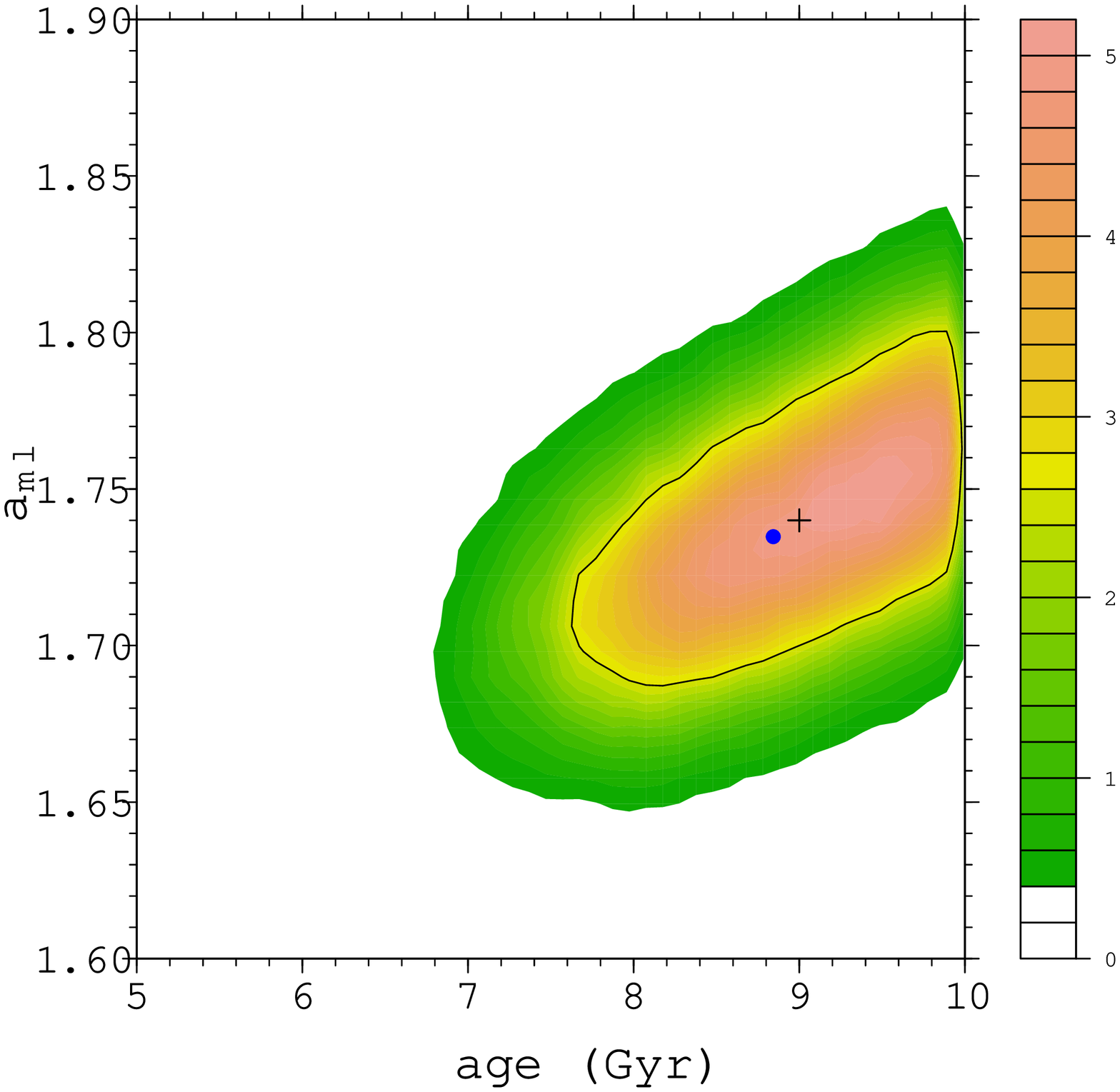}
        \includegraphics[height=6.0cm,angle=-90]{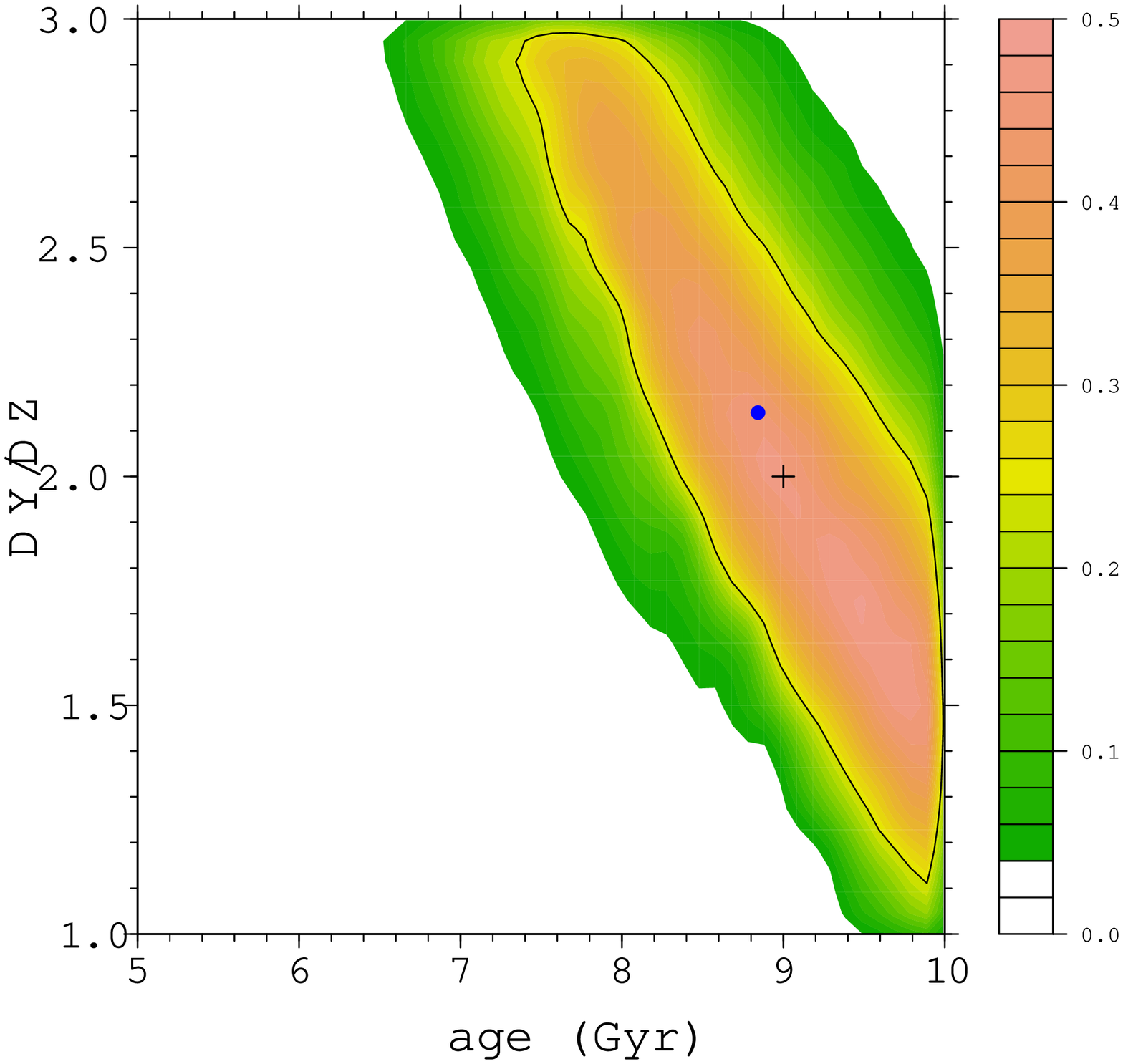}
        \includegraphics[height=6.0cm,angle=-90]{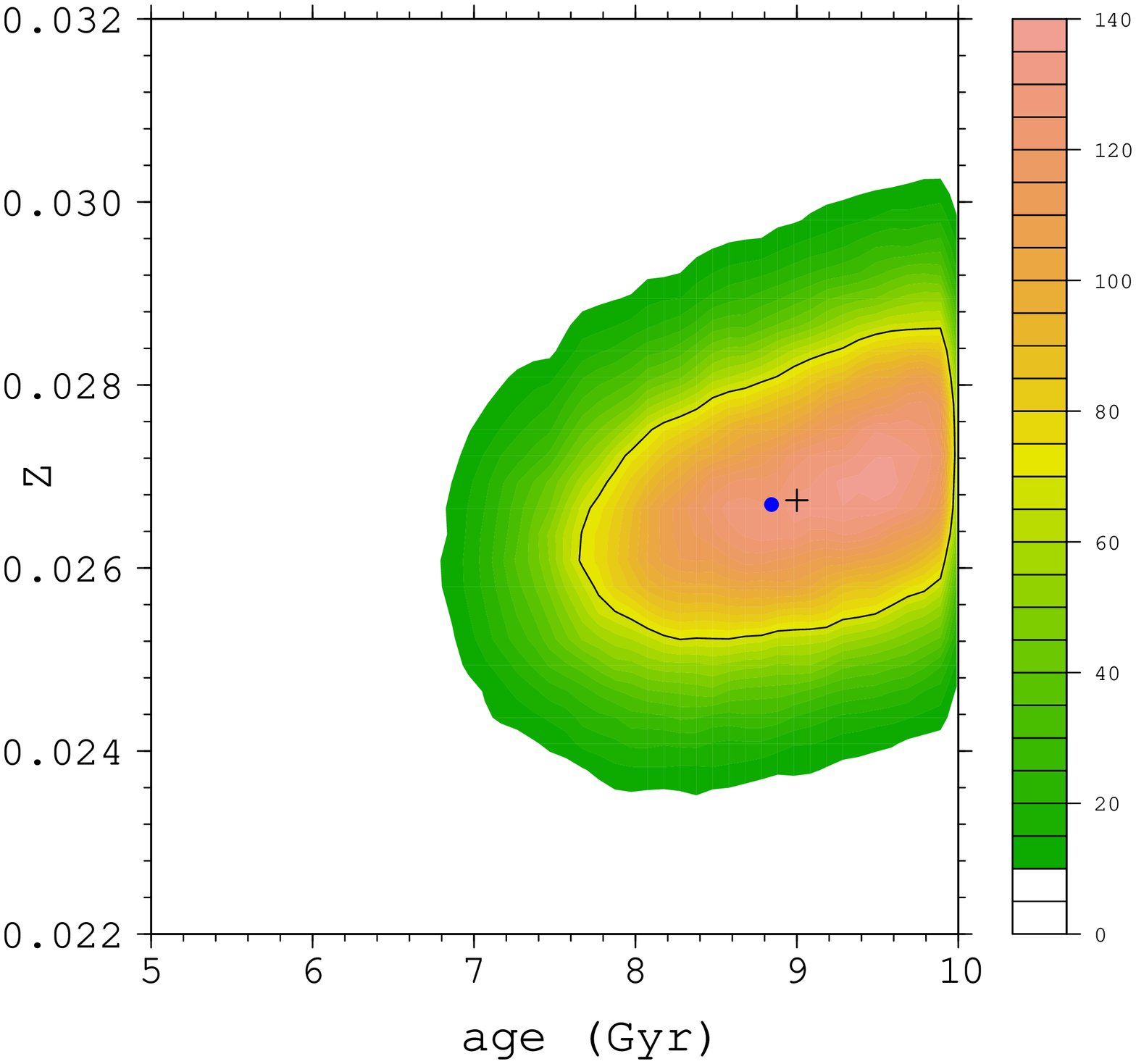}\\
        \includegraphics[height=6.0cm,angle=-90]{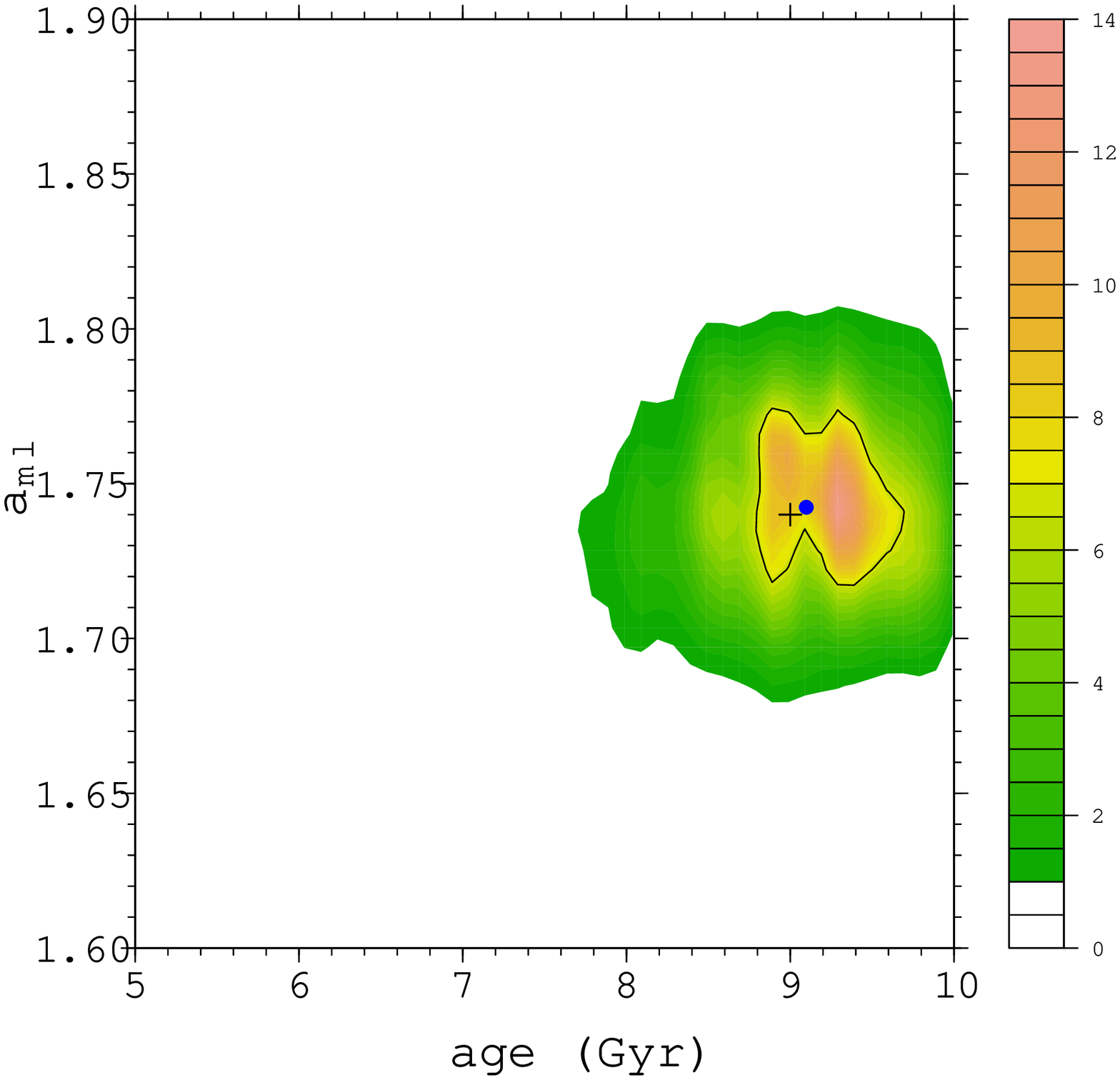}
        \includegraphics[height=6.0cm,angle=-90]{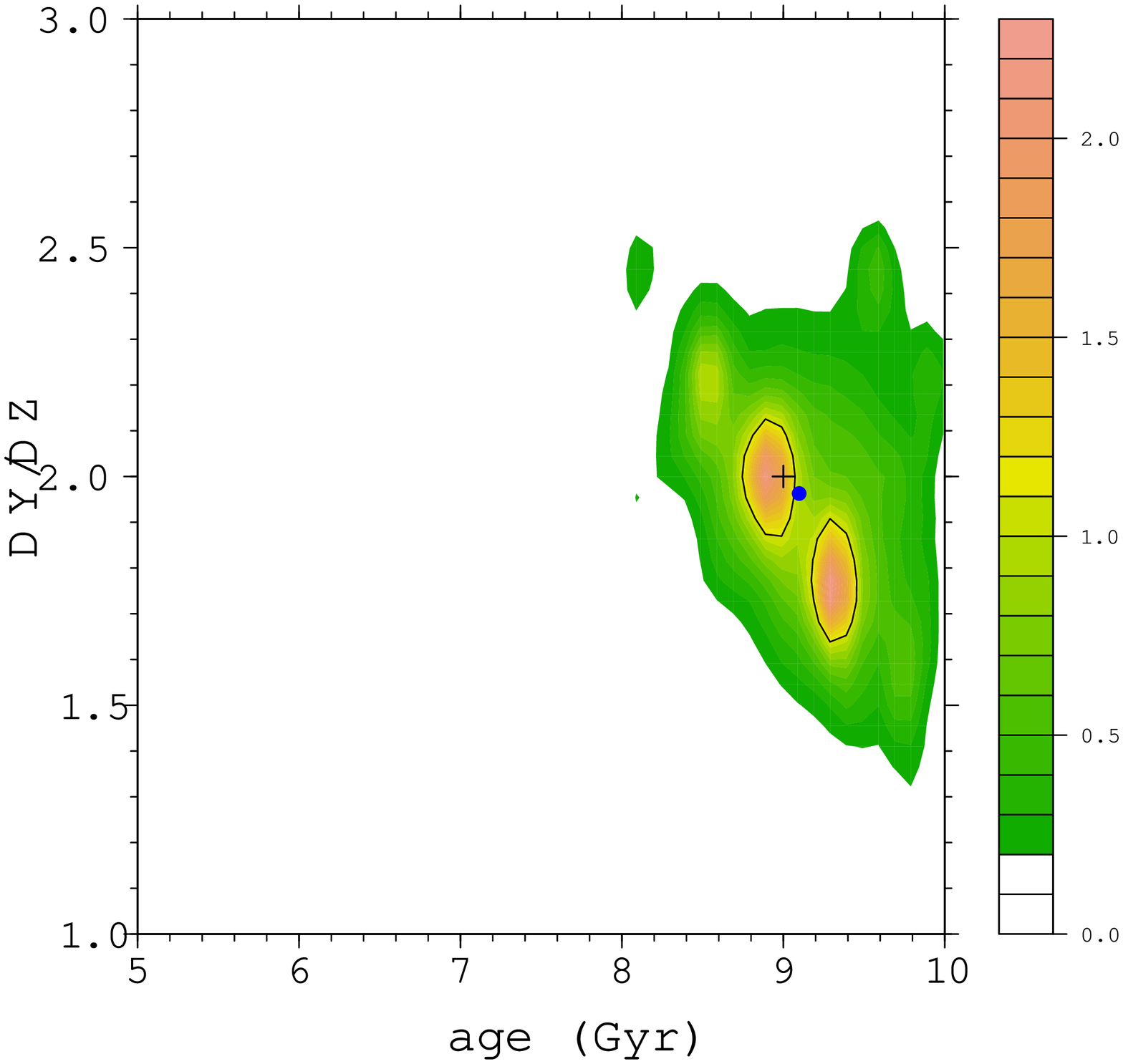}
        \includegraphics[height=6.0cm,angle=-90]{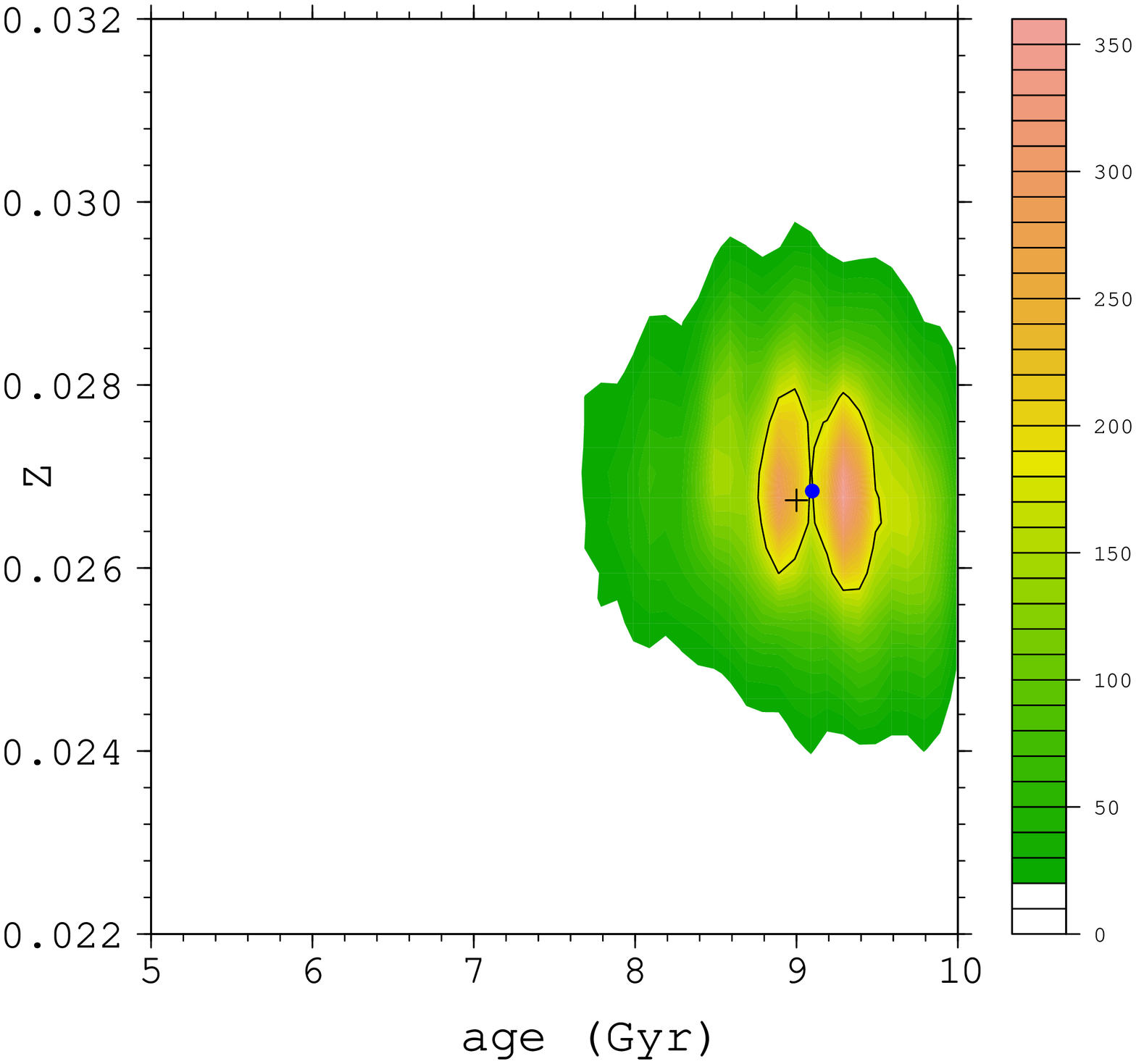}\\
        \includegraphics[height=6.0cm,angle=-90]{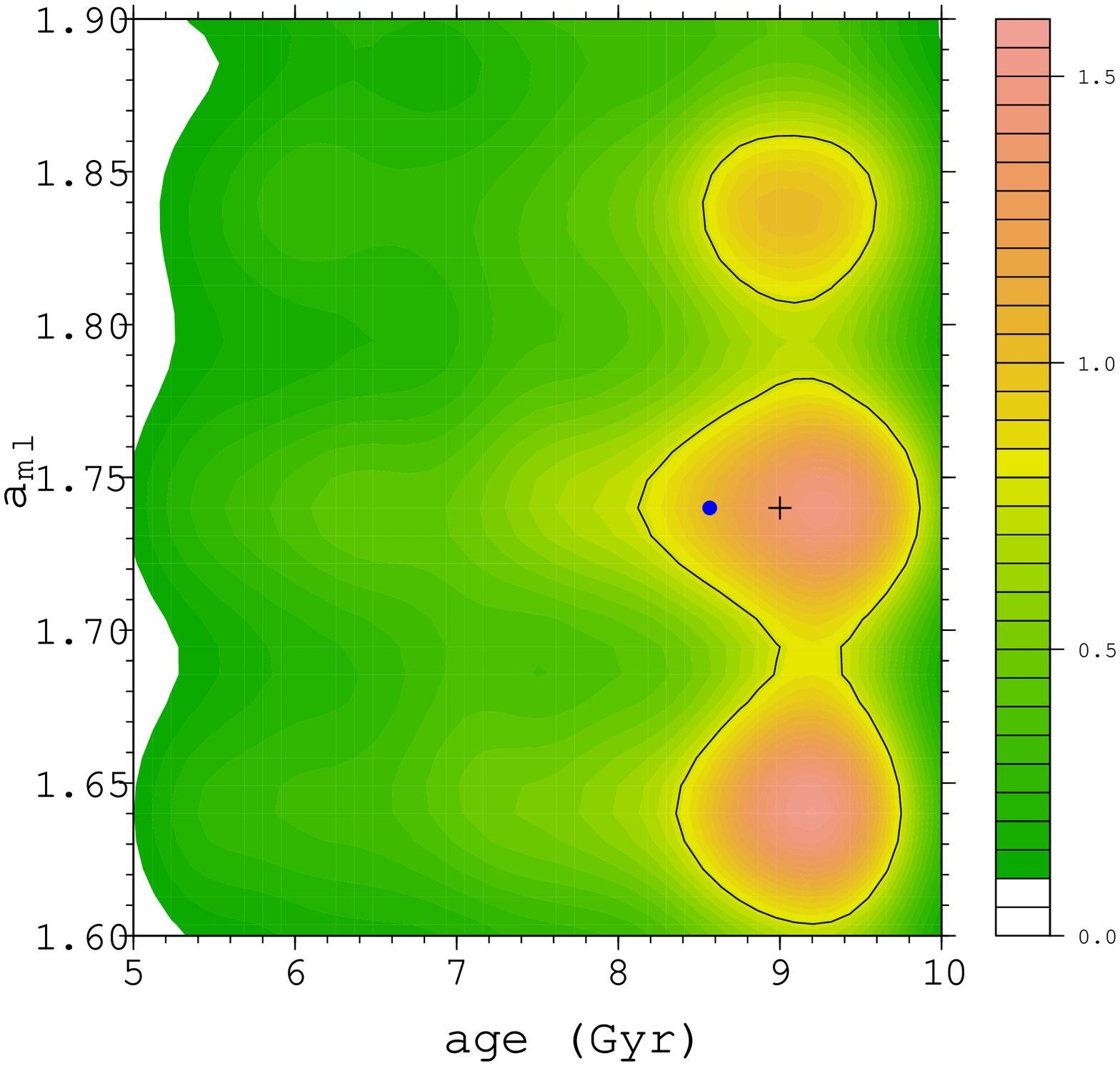}
        \includegraphics[height=6.0cm,angle=-90]{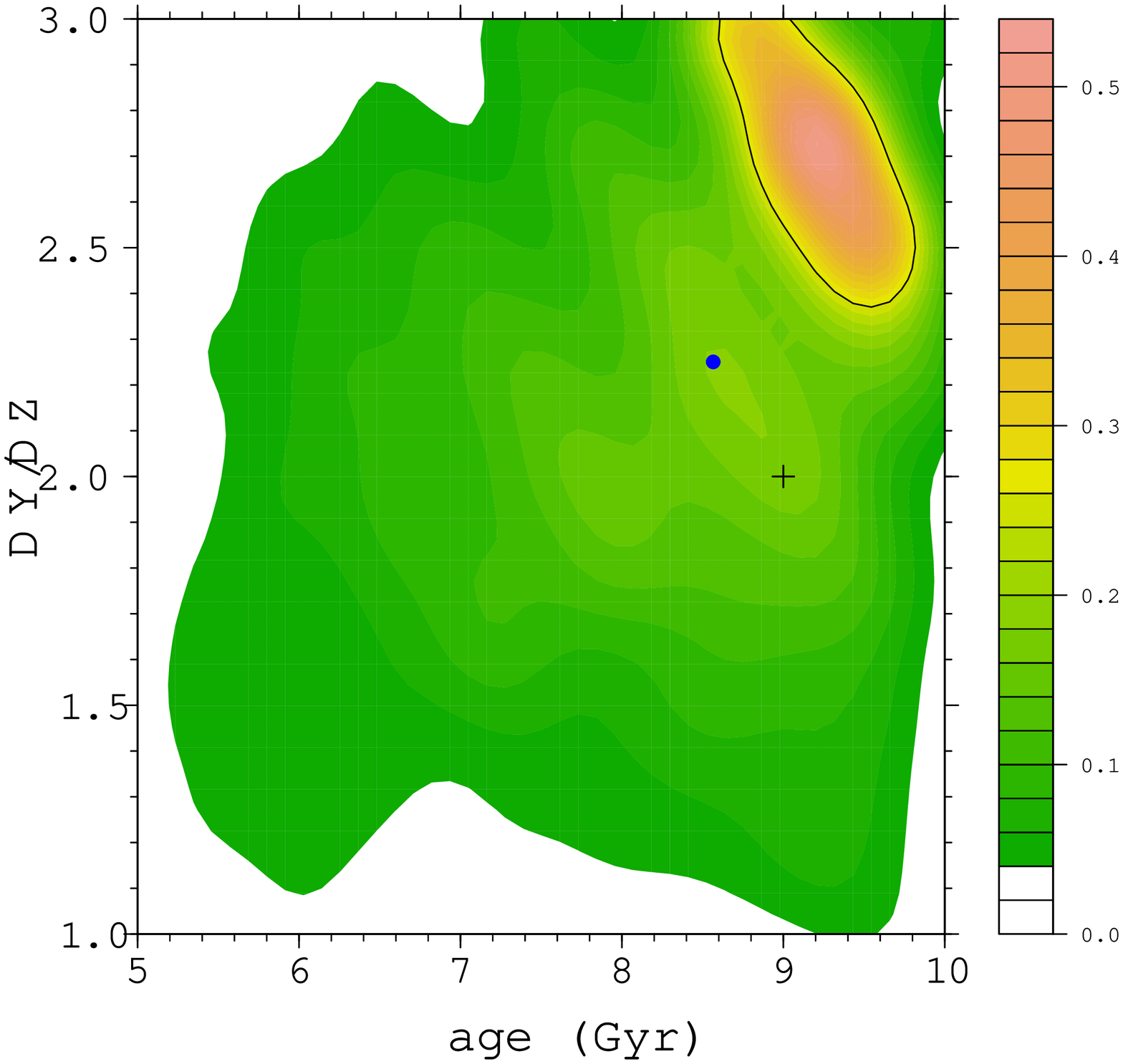}
        \includegraphics[height=6.0cm,angle=-90]{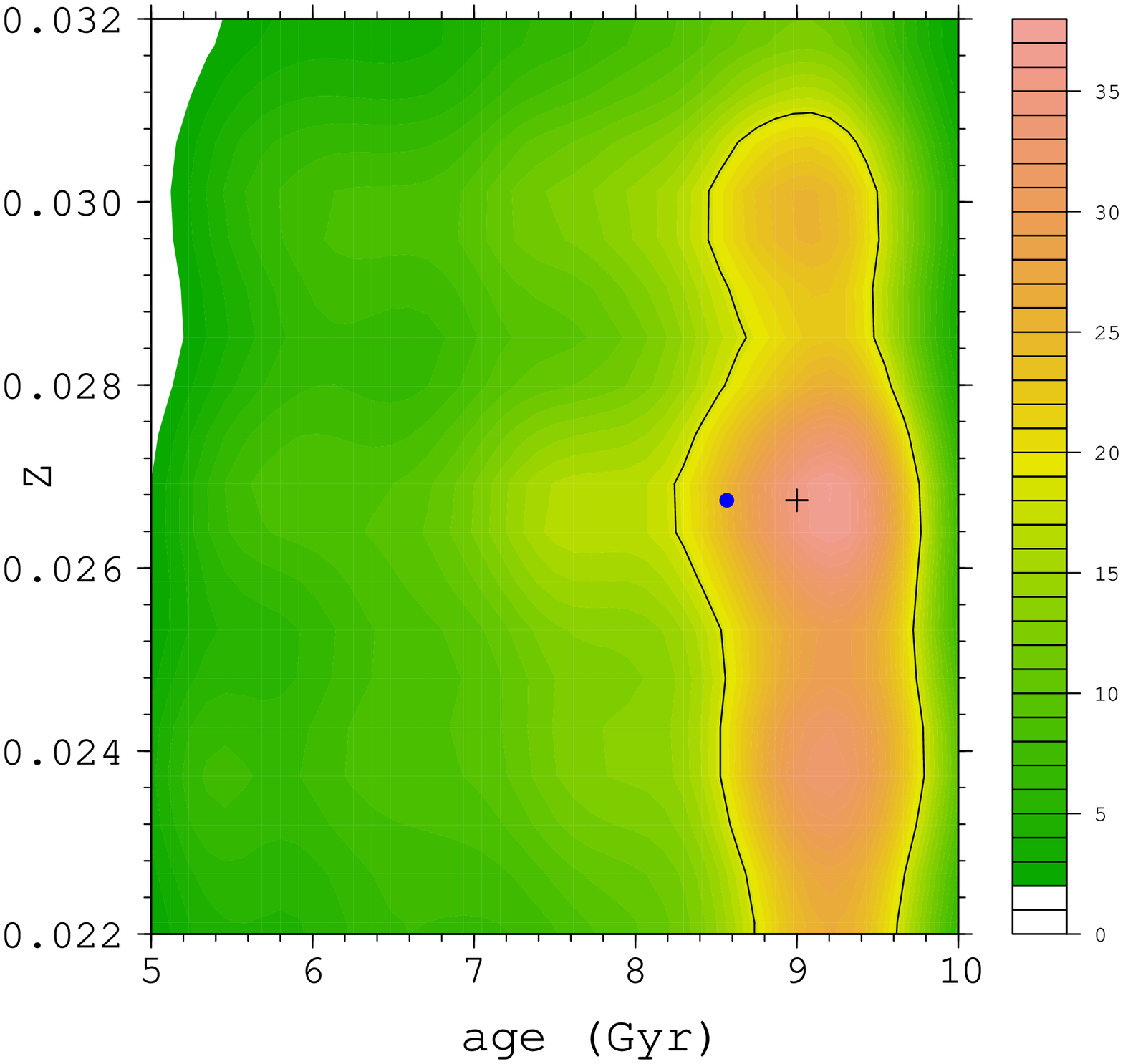}
        \caption{As in Fig.~\ref{fig:res7.5Gyr35}, but for a true age of 9.0 Gyr.  
        }
        \label{fig:res9.0Gyr35}
\end{figure*}

The three different estimation techniques described in Sect.~\ref{sec:fittingML} were applied to the four scenarios introduced in Sect.~\ref{sec:method}. Figures~\ref{fig:res7.5Gyr35} to \ref{fig:res9.0Gyr80} display the joint density of the obtained results in the  $\alpha_{\rm ml}$ versus age, $\Delta Y/\Delta Z$ versus age, and $Z$ versus age planes, while Table~\ref{tab:mainres} contains the median value of the marginalized parameters along with the 16th and 84th percentiles as  indicators of the distribution dispersion. 

It is apparent at first glance that the marginalized distributions have different characteristics depending on the method adopted for the fit. The cluster fit by the SCEPtER pipeline (Sect.~\ref{sec:sub-scepter}) produces diffuse distributions (third rows in Figs.~\ref{fig:res7.5Gyr35} to \ref{fig:res9.0Gyr80}) due to the possibility of each star to provide independently a best fit solution. The posterior probability density for the stellar parameters are nearly flat for S35S and S80S scenarios, and are only marginally peaked for S35-9S and S80-9S scenarios.
The posterior maps from the geometrical fit (Sect.~\ref{sec:sub-geo}, first rows in the figures) show clearly peaked unimodal distributions, with a somewhat large variance.
On the other hand, the marginalized distributions resulting from the ML method (Sect.~\ref{sec:sub-ML}, second rows in the figures) are heavily peaked and often multimodal. This explains the need in the MCMC construction to rely on a long chain to properly explore the posterior density.  

Apart from these differences, there is evidence that some parameters can be recovered well by each of the analysed methods. This is the case of both the  mixing-length value and the metallicity $Z$.
This general agreement can be easily understood because these two parameters heavily  affect the track position and morphology of RGB stars in the observable hyperspace. 

The helium-to-metal enrichment ratio $\Delta Y/\Delta Z,$  and therefore initial helium content through the relation  $Y = Y_p+\frac{\Delta Y}{\Delta Z} Z,$ was more difficult to estimate. The results for the geometrical and independent star fits show large uncertainties. This is expected owing to the well-known but weak  effect of the initial helium abundance on the position of the RGB. The ML approach performs much better in this case, providing a narrower error range that in all cases includes the unbiased value $\Delta Y/\Delta Z$ = 2.0. This is particularly true for scenarios S80w and S80-9w, where the larger sample size causes a reduction of the error range by a factor larger than three with respect to the pure geometrical approach. 
Although it is reasonable to obtain information on a cluster initial helium content if different evolutionary phases are taken into account, the  uncertainty obtained here (i.e. $\pm 0.03$ for the $Y$ value) is particularly small, being  only a little larger than  the precision of the helium content determination through the full CMD isochrone fitting.   

Regarding the age of the cluster, the methods can provide somewhat different estimates. 
For scenarios with a true age of 7.5 Gyr the differences between pure geometrical and ML fits are about 0.15 Gyr, the geometrical fit providing higher ages. For scenarios with a true age of 9.0 Gyr the difference increases to about 0.3 Gyr, but in this case the age estimate by the ML techniques is higher.

The following subsections are devoted to highlighting and discussing some specific results from the three techniques.

\subsection{Results from the pure geometrical fit}

The pure geometrical method (Sect.~\ref{sec:sub-geo}) showed a little bias, overestimating the age by about 0.3 and 0.2 Gyr in the S35 and S80 scenarios (relative errors of $4\%$ and $3\%$), respectively, and underestimating it by about 0.2 and 0.1 Gyr in the S35-9 and S80-9 scenarios ($-2\%$ and $-1\%$), respectively. The facts that the latter biases are smaller  than the former and that the signs of the relative differences change are linked to edge effects. In fact, approaching the grid edge, the age estimates are bounded by the upper value of the grid, that is, 10 Gyr. This obviously limits the possibility of  age overestimation. 
The distortion of the age estimate stems from the asymmetry of the stellar model grid in the hyperspace of parameters. As discussed in previous works \citep[see e.g.][]{bulge, overshooting, BinTeo} stellar track distances from a target point do not change symmetrically for symmetrical perturbations on the parameters adopted in their computation. In this particular case it happens that for scenario S35 and S80, isochrones with $\Delta Y/\Delta Z < 2$ tend to be a little closer to the synthetic data than those with $\Delta Y/\Delta Z > 2$. This leads to an underestimation of the initial 
helium abundance for the OC and consequently an overestimation of its age. For scenarios S35-9 and S80-9 the situation is complicated by the influence of edge effects that limit the possibility of the underestimation  of the initial helium content due to the impossibility to obtain an age outside the grid boundary.
Luckily, all the detected biases in the estimated ages are small, that is, about one third of the random errors due to observational uncertainties.  
The differences between the results for the scenarios with 35 and 80 artificial stars are minor, with a slight shrink of the error range and a much lower change in the estimated parameters than the random error component. 
Although it seems natural to anticipate that the estimated error shrinks with the square root of the sample size, this is only true when not accounting for the variance induced by the different artificial perturbations that occur among the $N$ Monte Carlo experiments. 
Therefore the total variance is composed of two terms: a term due to the uncertainty in the recovered cluster parameters for a given generation of synthetic cluster stars, which indeed scales as the inverse of the square root of the sample size, and a term that accounts for the dispersion of the $N$ recovered medians of each Monte Carlo simulation. 
This last component of the variance would not shrink with the sample size, leading to the weak dependence shown in Table~\ref{tab:mainres}.

As a consistency check, the results obtained by the geometrical method were compared to a direct integration of the likelihood function $P(\theta)$ over the whole grid, as described in Sect.~\ref{sec:sub-geo}.
Table~\ref{tab:directL} presents the results (identified by a suffix "D"), obtained combining the results from the $N = 80$ random artificial samples. These agree well with the MCMC results, as should be expected in light of two concurrent effects. First of all, the fitted values returned by the MCMC simulation are very close to points provided by the grid itself, so the grid resolution does not hamper the achievement of good accuracy. Second, the errors shown in Table~\ref{tab:mainres} on the recovered $\Delta Y/\Delta Z$ and age are large with respect to the grid steps. Therefore the coarseness of the grid does not limit the accuracy attainable using a direct approach. 

\begin{table*}[ht]
        \centering
        \caption{Estimated median (-est.), 16th ($\sigma-$), and 84th ($\sigma+$) percentiles of the stellar parameters ($\alpha_{\rm ml}$, $\Delta Y /\Delta Z$, $Z$, and age) recovered in the twelve considered scenarios.}
        \label{tab:mainres}     
        \begin{tabular}{lcccccccccccc}
                \hline\hline
                & \multicolumn{3}{c}{$\alpha_{\rm ml}$} &  \multicolumn{3}{c}{$\Delta Y /\Delta Z$} &  \multicolumn{3}{c}{$Z$} & \multicolumn{3}{c}{age (Gyr)}\\
                & est. & $\sigma-$ & $\sigma+$ & est. & $\sigma-$ & $\sigma+$ & est. & $\sigma-$ & $\sigma+$ & est. & $\sigma-$ & $\sigma+$ \\ 
                \hline
S35 & 1.75 & 0.04 & 0.05 & 1.88 & 0.59 & 0.69 & 0.0269 & 0.0014 & 0.0014 & 7.79 & 0.94 & 0.98 \\ 
S35-9 & 1.73 & 0.04 & 0.04 & 2.14 & 0.54 & 0.55 & 0.0267 & 0.0013 & 0.0014 & 8.84 & 0.93 & 0.78 \\ 
S80 & 1.75 & 0.03 & 0.04 & 1.89 & 0.57 & 0.65 & 0.0269 & 0.0010 & 0.0010 & 7.72 & 0.85 & 0.88 \\ 
S80-9 & 1.73 & 0.03 & 0.03 & 2.06 & 0.46 & 0.55 & 0.0265 & 0.0009 & 0.0010 & 8.94 & 0.84 & 0.71 \\
\hline 
S35w & 1.74 & 0.03 & 0.03 & 1.99 & 0.31 & 0.30 & 0.0268 & 0.0012 & 0.0012 & 7.65 & 0.54 & 0.55 \\ 
S35-9w & 1.74 & 0.03 & 0.03 & 1.96 & 0.29 & 0.31 & 0.0268 & 0.0012 & 0.0012 & 9.10 & 0.59 & 0.50 \\ 
S80w & 1.74 & 0.01 & 0.02 & 1.98 & 0.15 & 0.10 & 0.0268 & 0.0006 & 0.0007 & 7.61 & 0.10 & 0.25 \\ 
S80-9w & 1.74 & 0.01 & 0.02 & 1.83 & 0.09 & 0.19 & 0.0267 & 0.0006 & 0.0006 & 9.28 & 0.38 & 0.08 \\ 
\hline
S35S & 1.74 & 0.13 & 0.15 & 2.00 & 0.58 & 0.67 & 0.0267 & 0.0055 & 0.0059 & 7.49 & 1.71 & 1.67 \\ 
S35-9S & 1.74 & 0.13 & 0.10 & 2.25 & 0.67 & 0.50 & 0.0267 & 0.0055 & 0.0069 & 8.57 & 1.92 & 0.82 \\ 
S80S & 1.74 & 0.10 & 0.15 & 2.00 & 0.59 & 0.67 & 0.0267 & 0.0055 & 0.0069 & 7.52 & 1.71 & 1.59 \\ 
S80-9S & 1.74 & 0.15 & 0.13 & 2.29 & 0.67 & 0.46 & 0.0267 & 0.0055 & 0.0069 & 8.57 & 1.97 & 0.88 \\ 
\hline
        \end{tabular}
\tablefoot{The algorithm adopted in the recovery is specified by the suffix in the scenario names. A suffix "w" indentifies the results obtained using the ML method, "S" identifies the resuls using the SCEPtER approach, while no suffix refers to the geometrical fit.}
\end{table*}

\begin{table*}[ht]
        \centering
        \caption{Mean and standard deviation of the stellar parameters ($\alpha_{\rm ml}$, $\Delta Y /\Delta Z$, $Z$, and age) recovered by direct likelihood integration.}
        \label{tab:directL}     
        \begin{tabular}{lcccc}
                \hline\hline
                & $\alpha_{\rm ml}$ & $\Delta Y /\Delta Z$ &          Z          &    age (Gyr)    \\ \hline
                S35D   &  $1.75 \pm 0.03$  &   $1.84 \pm 0.56$    & $0.0270 \pm 0.0011$ & $7.83 \pm 1.05$ \\
                S35-9D &  $1.74 \pm 0.02$  &   $2.07 \pm 0.46$    & $0.0267 \pm 0.0008$ & $8.93 \pm 0.78$ \\
                S80D   &  $1.75 \pm 0.03$  &   $1.81 \pm  0.56$   & $0.0269 \pm 0.0010$ & $7.86 \pm 0.97$ \\
                S80-9D &  $1.74 \pm 0.02$  &   $2.04 \pm 0.50$    & $0.0267 \pm 0.0008$ & $8.92 \pm 0.78$ \\ \hline
        \end{tabular}
\end{table*}

\subsection{Results from the ML fit}

\begin{figure*}
        \centering
        \includegraphics[height=6.0cm,angle=-90]{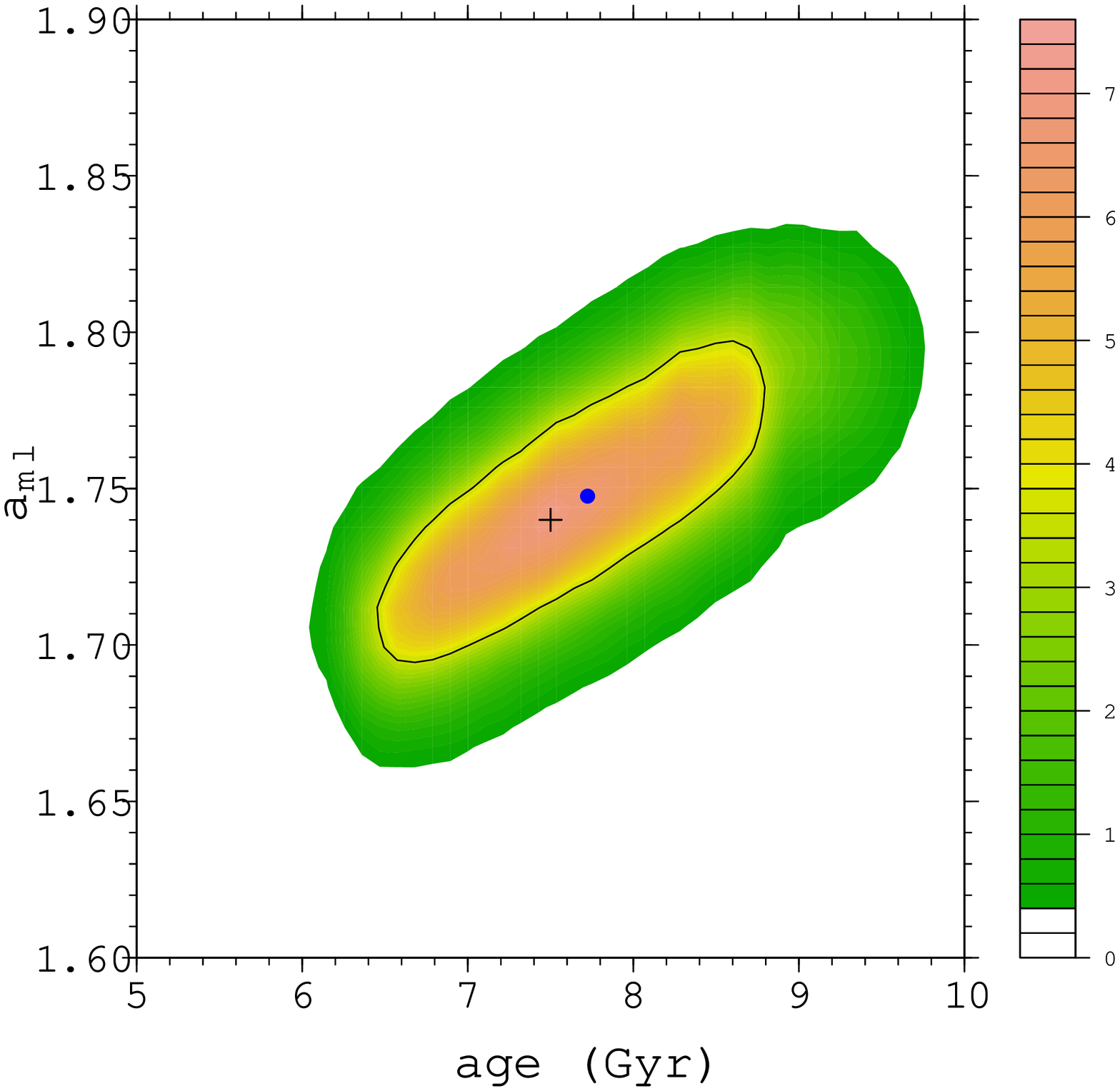}
        \includegraphics[height=6.0cm,angle=-90]{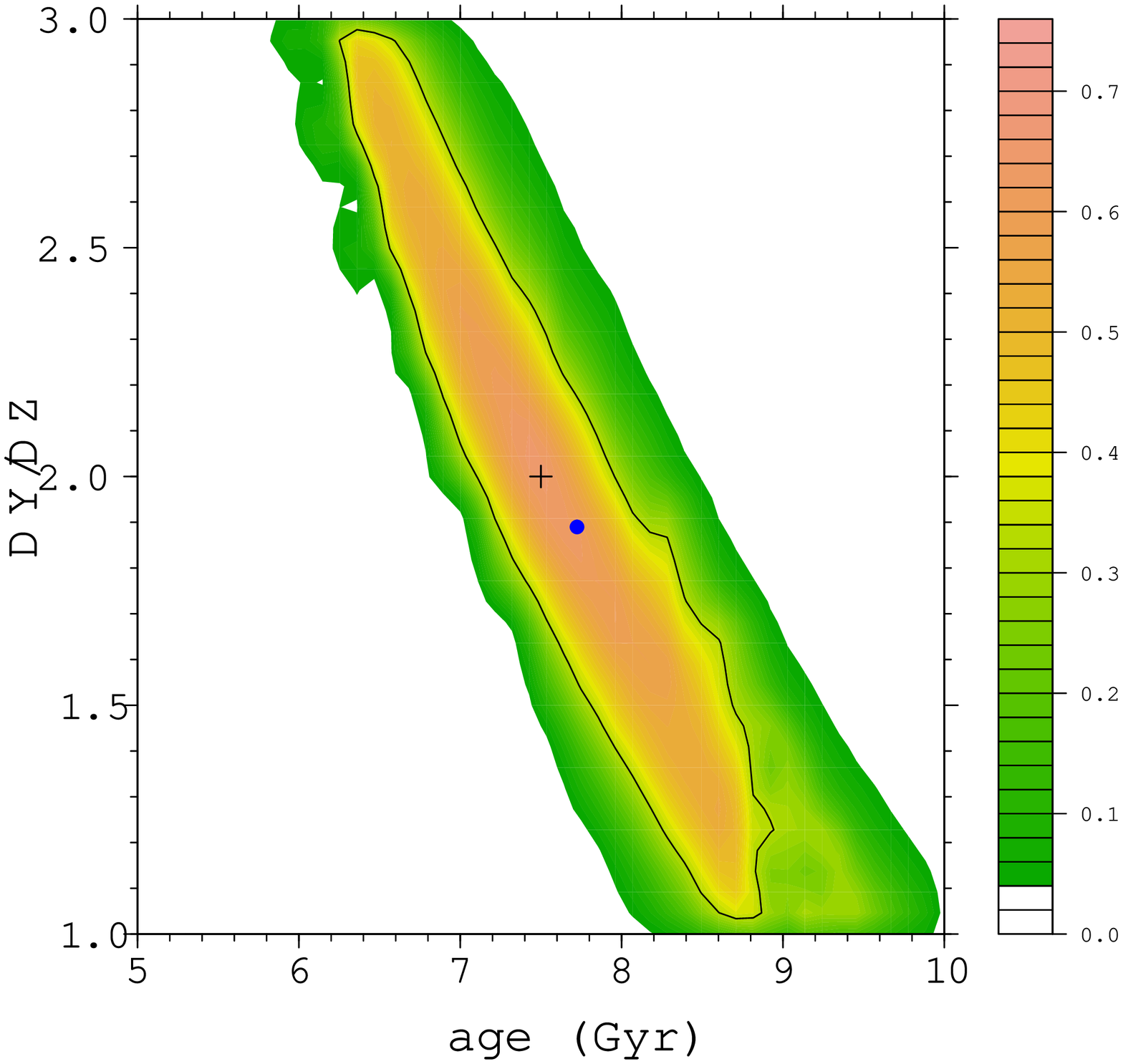}
        \includegraphics[height=6.0cm,angle=-90]{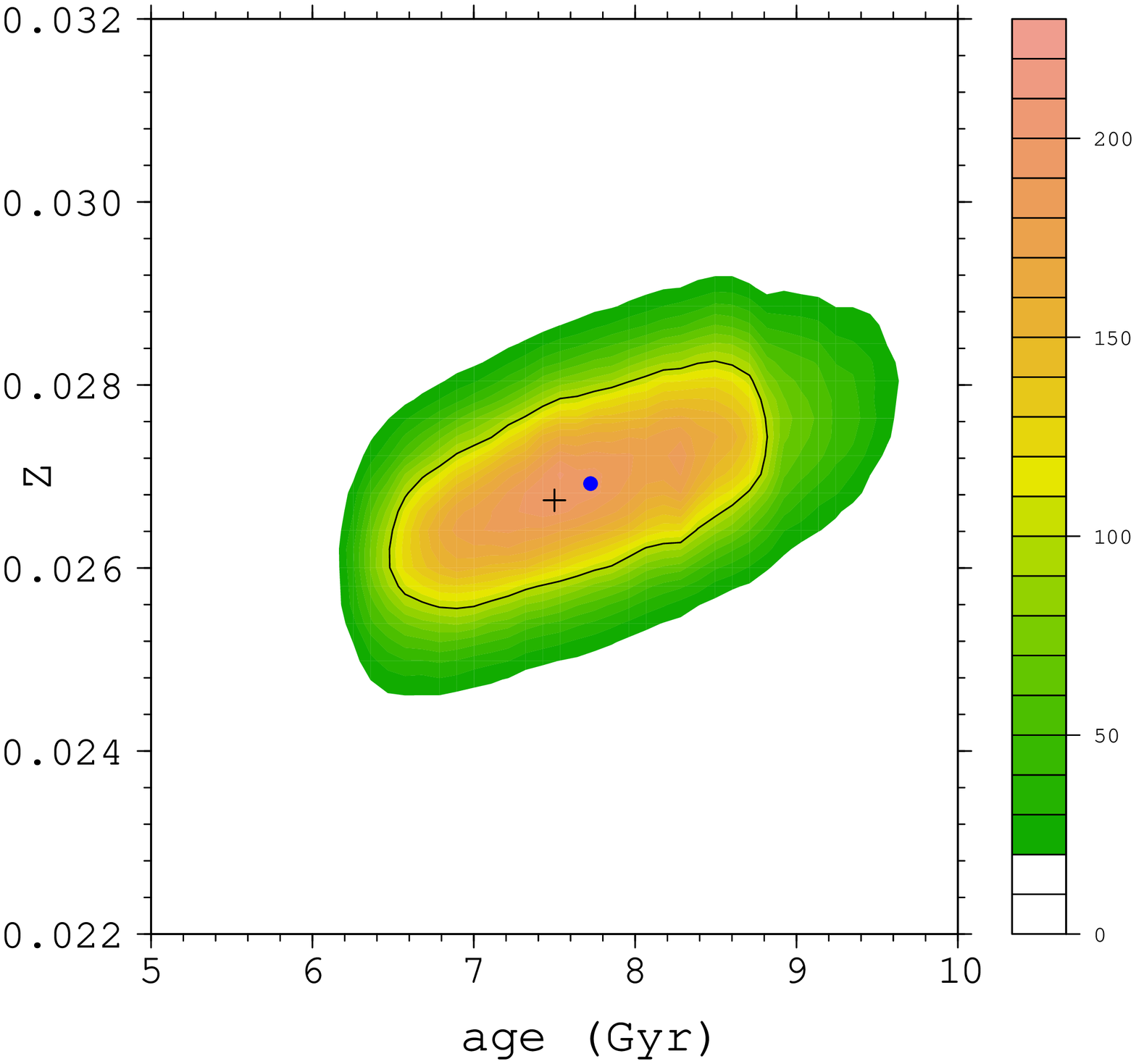}\\
        \includegraphics[height=6.0cm,angle=-90]{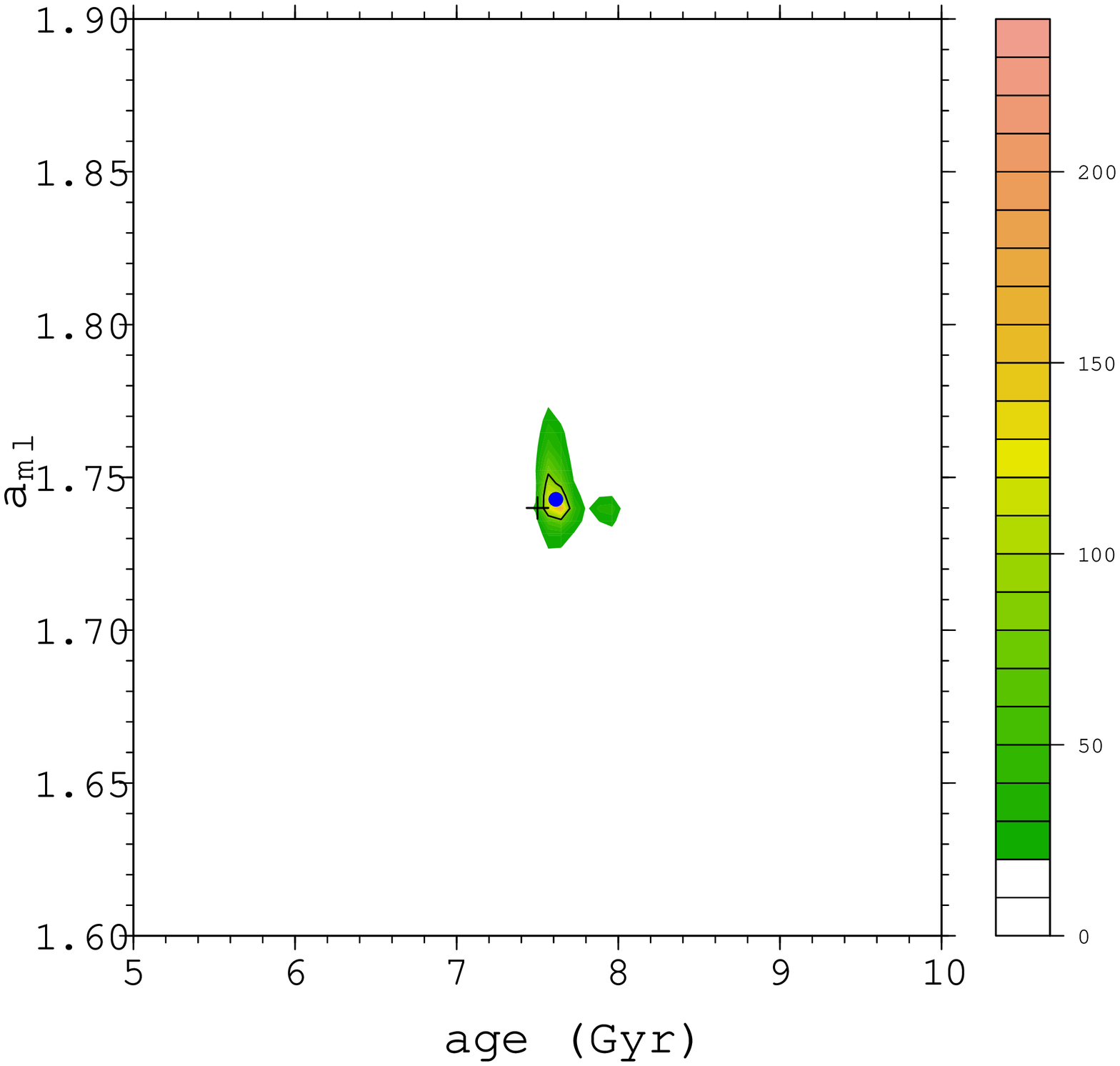}
        \includegraphics[height=6.0cm,angle=-90]{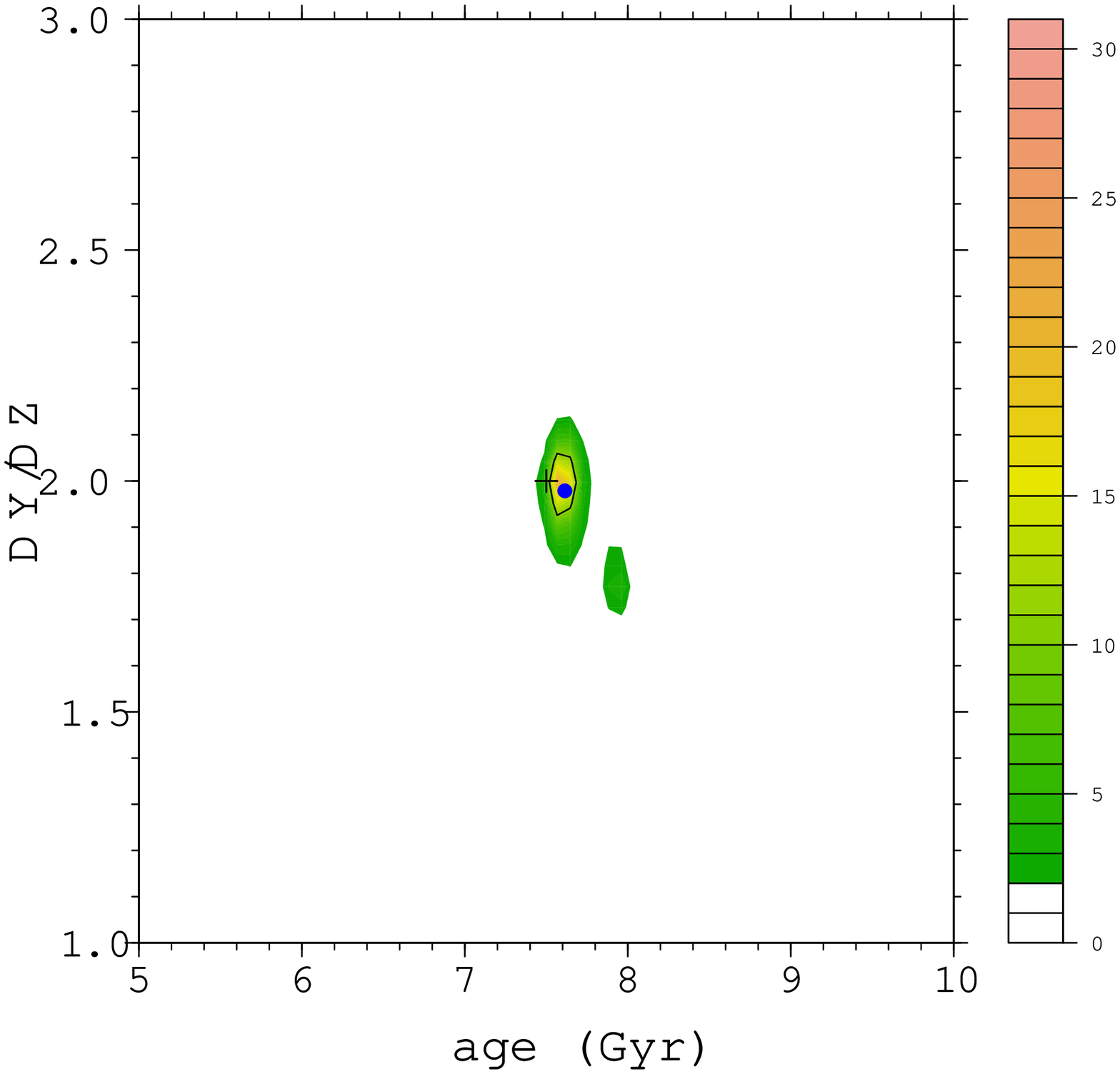}
        \includegraphics[height=6.0cm,angle=-90]{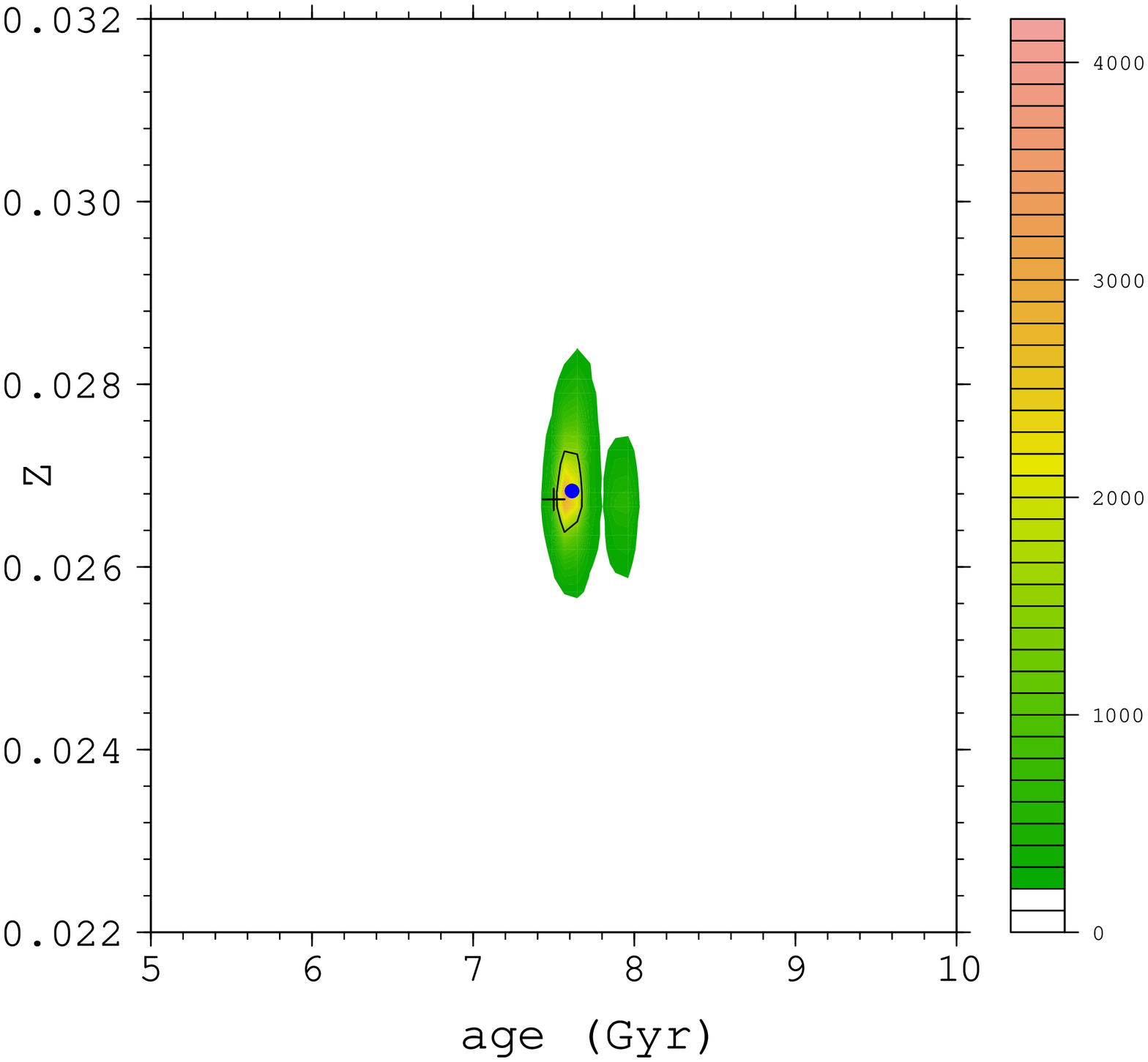}\\
        \includegraphics[height=6.0cm,angle=-90]{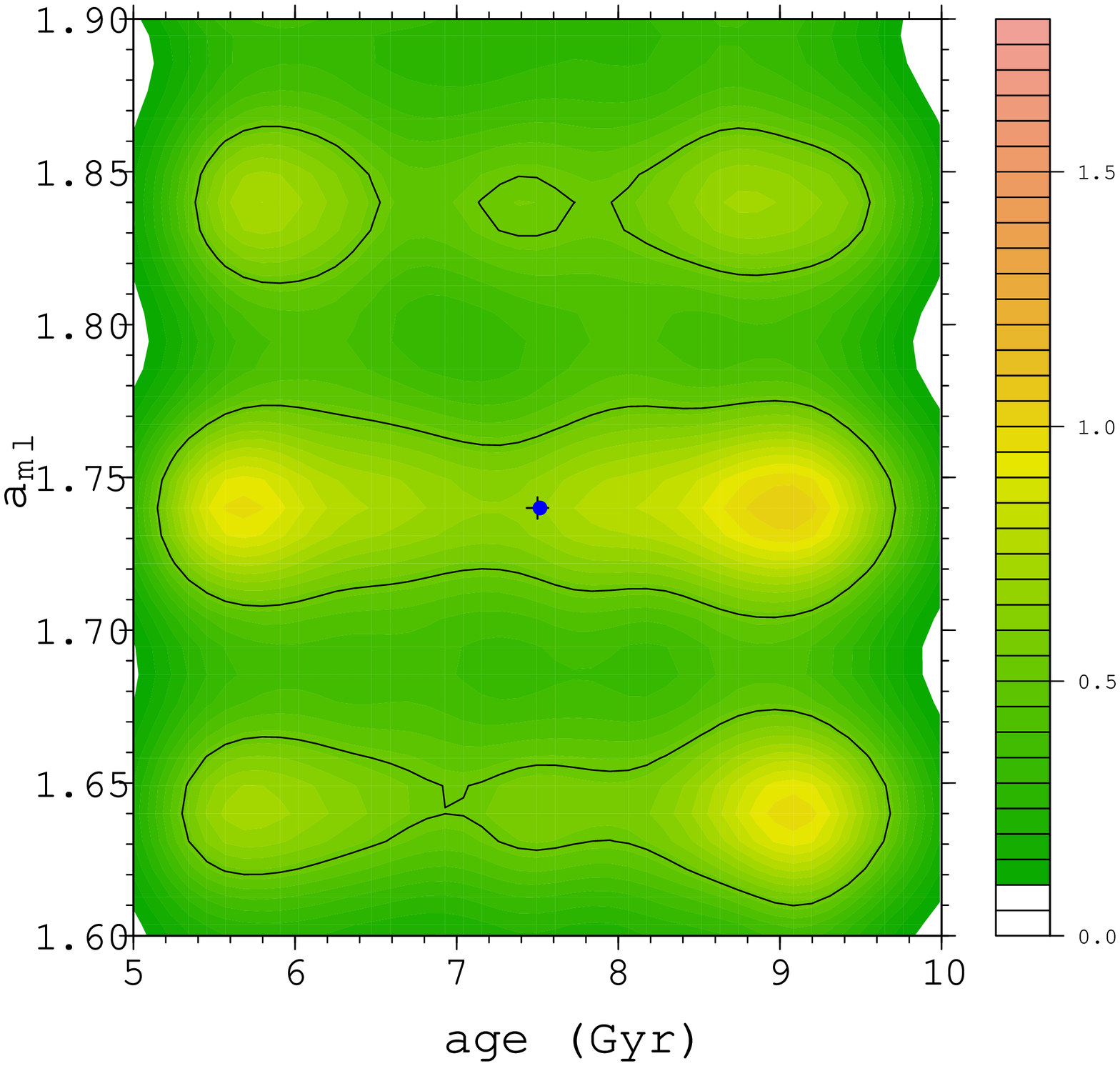}
        \includegraphics[height=6.0cm,angle=-90]{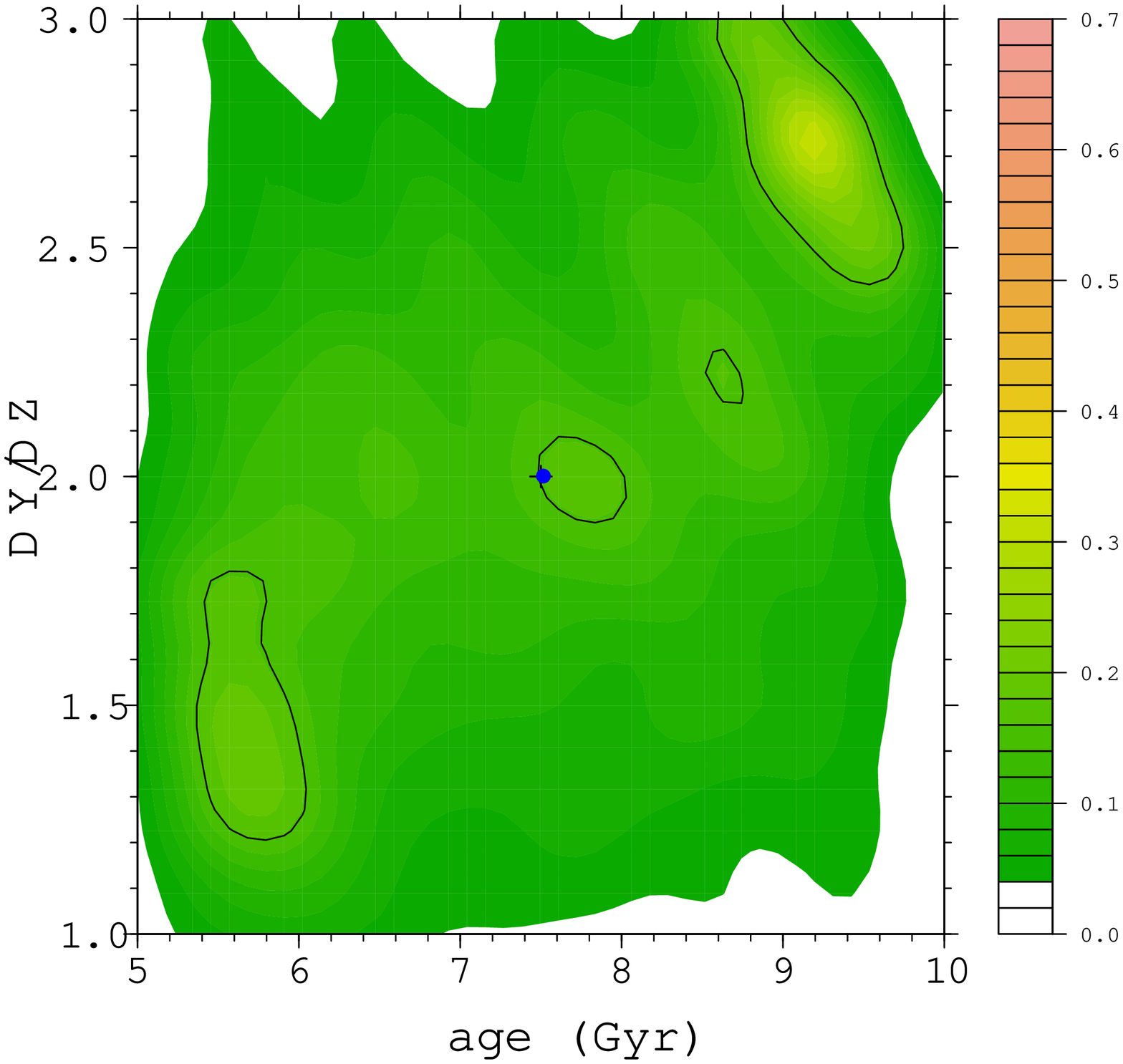}
        \includegraphics[height=6.0cm,angle=-90]{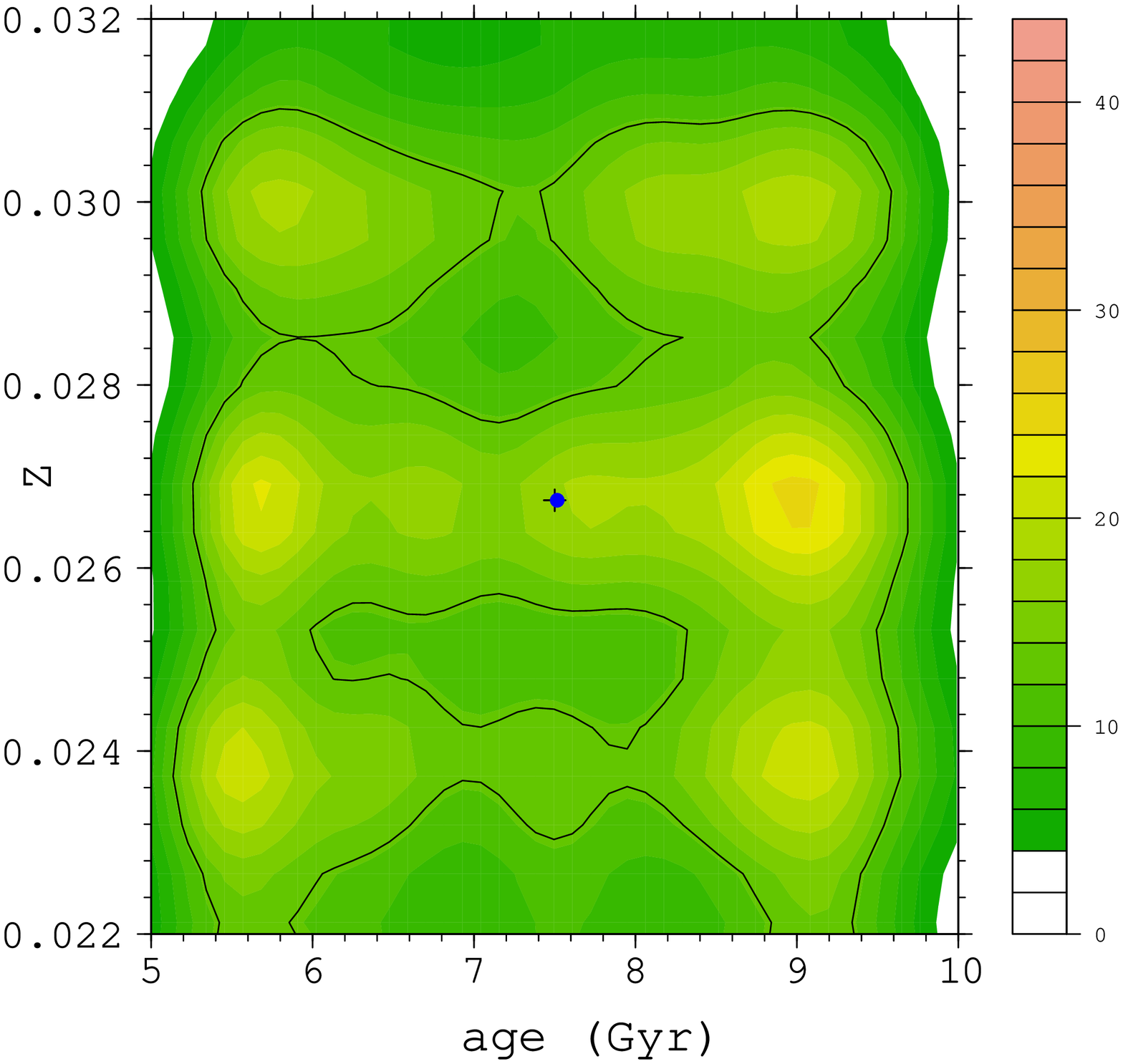}
        \caption{As in Fig.~\ref{fig:res7.5Gyr35}, but for samples of 80 RGB stars.  
        }
        \label{fig:res7.5Gyr80}
\end{figure*}

\begin{figure*}
        \centering
        \includegraphics[height=6.0cm,angle=-90]{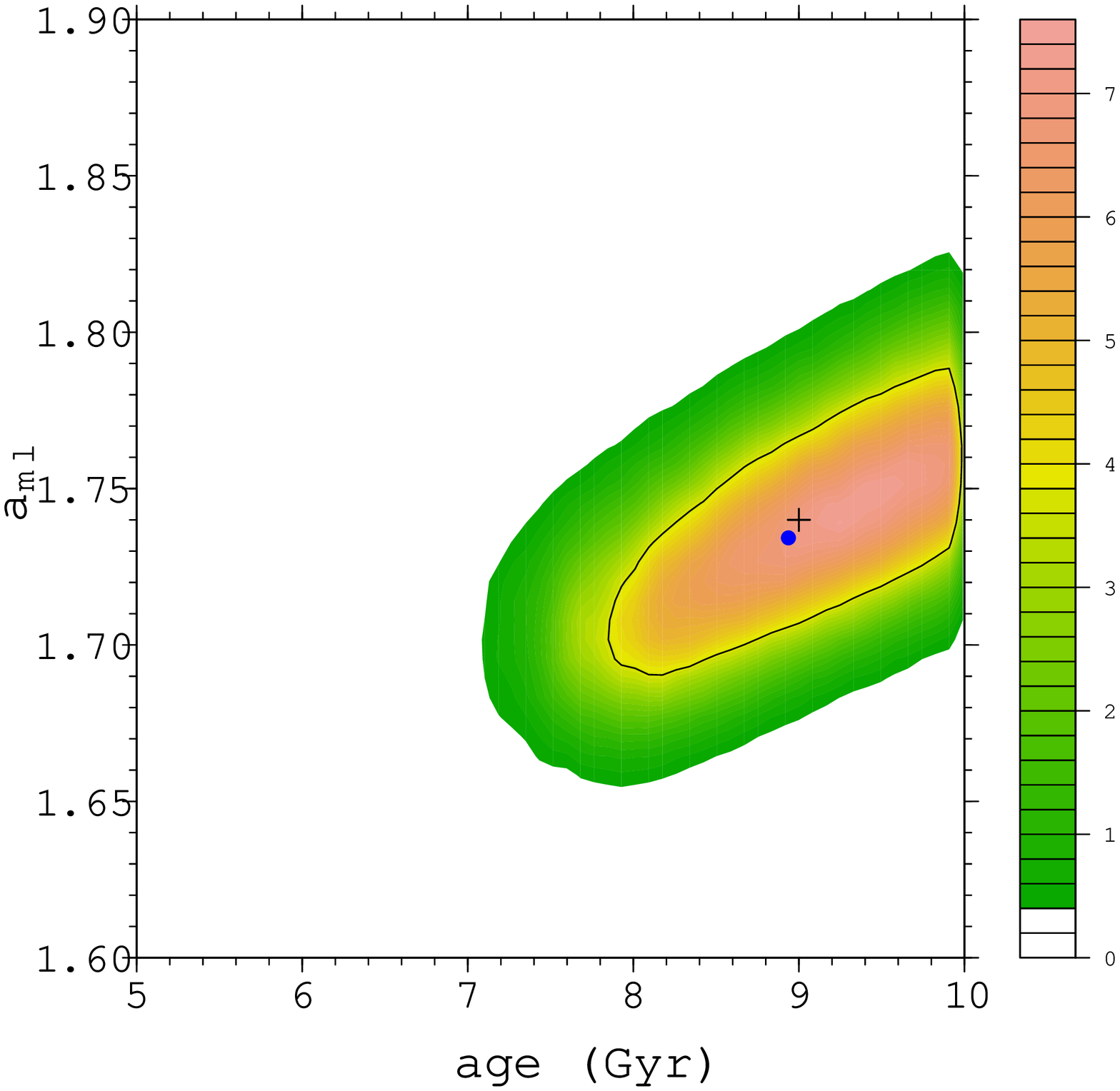}
        \includegraphics[height=6.0cm,angle=-90]{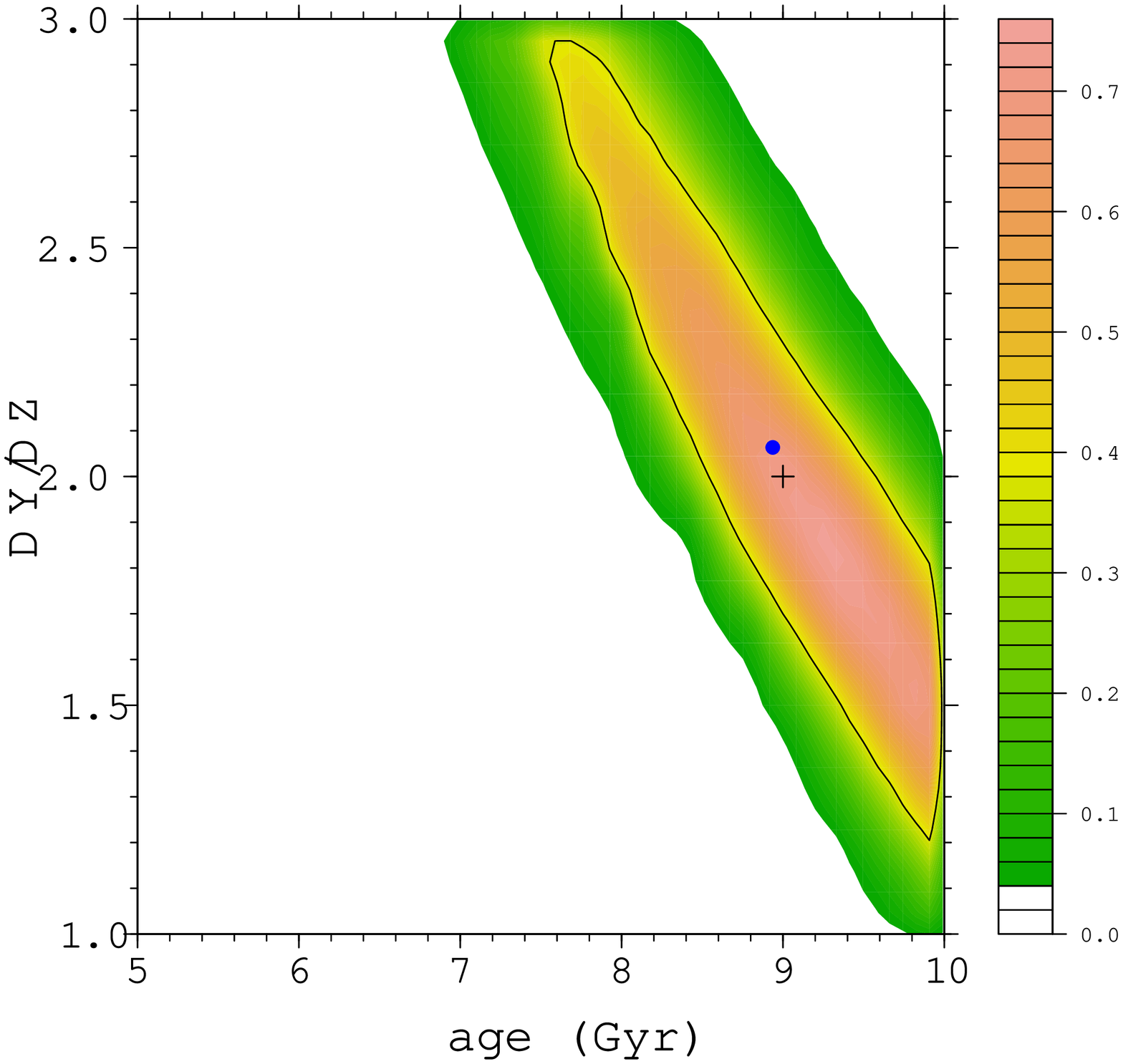}
        \includegraphics[height=6.0cm,angle=-90]{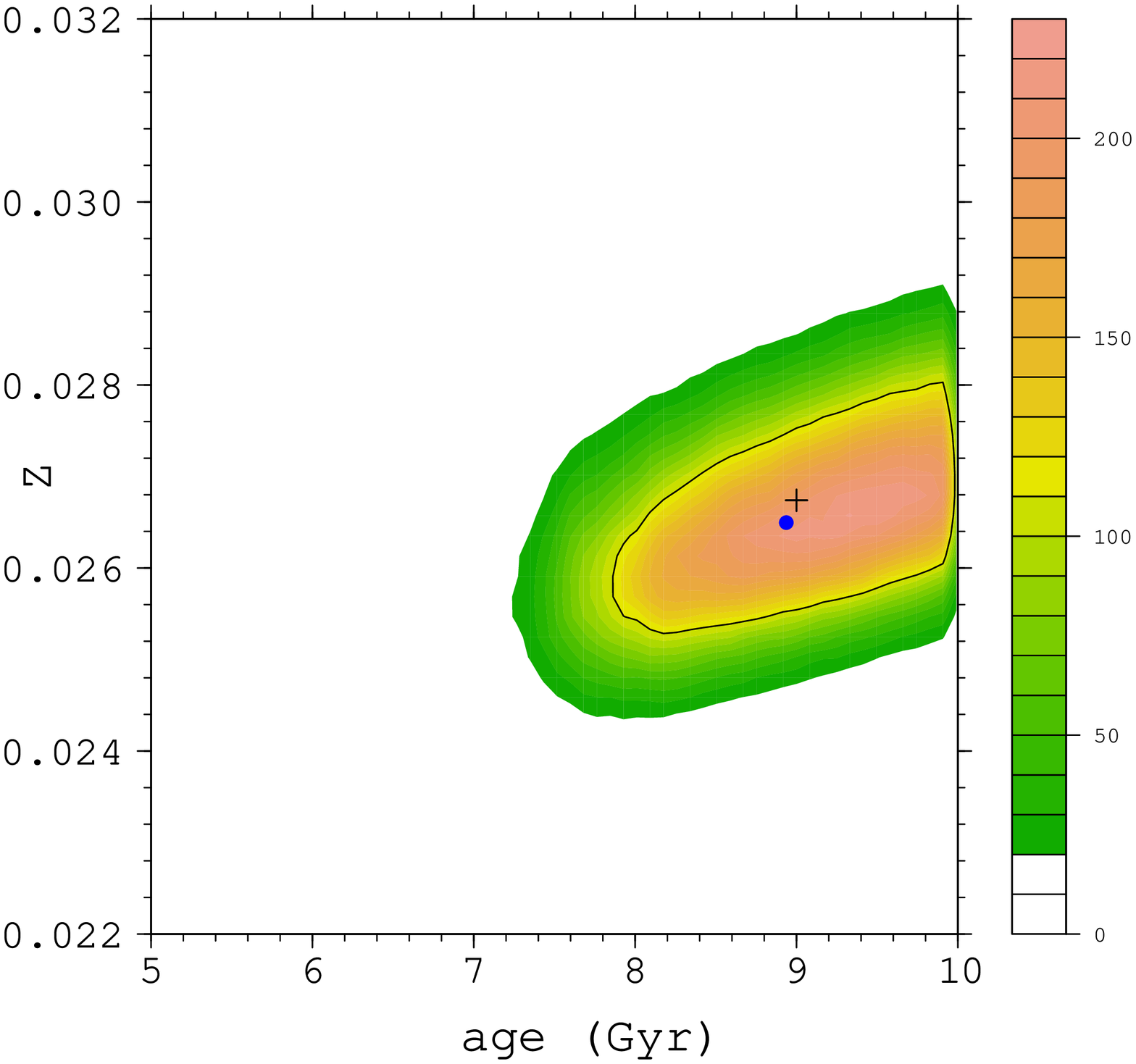}\\
        \includegraphics[height=6.0cm,angle=-90]{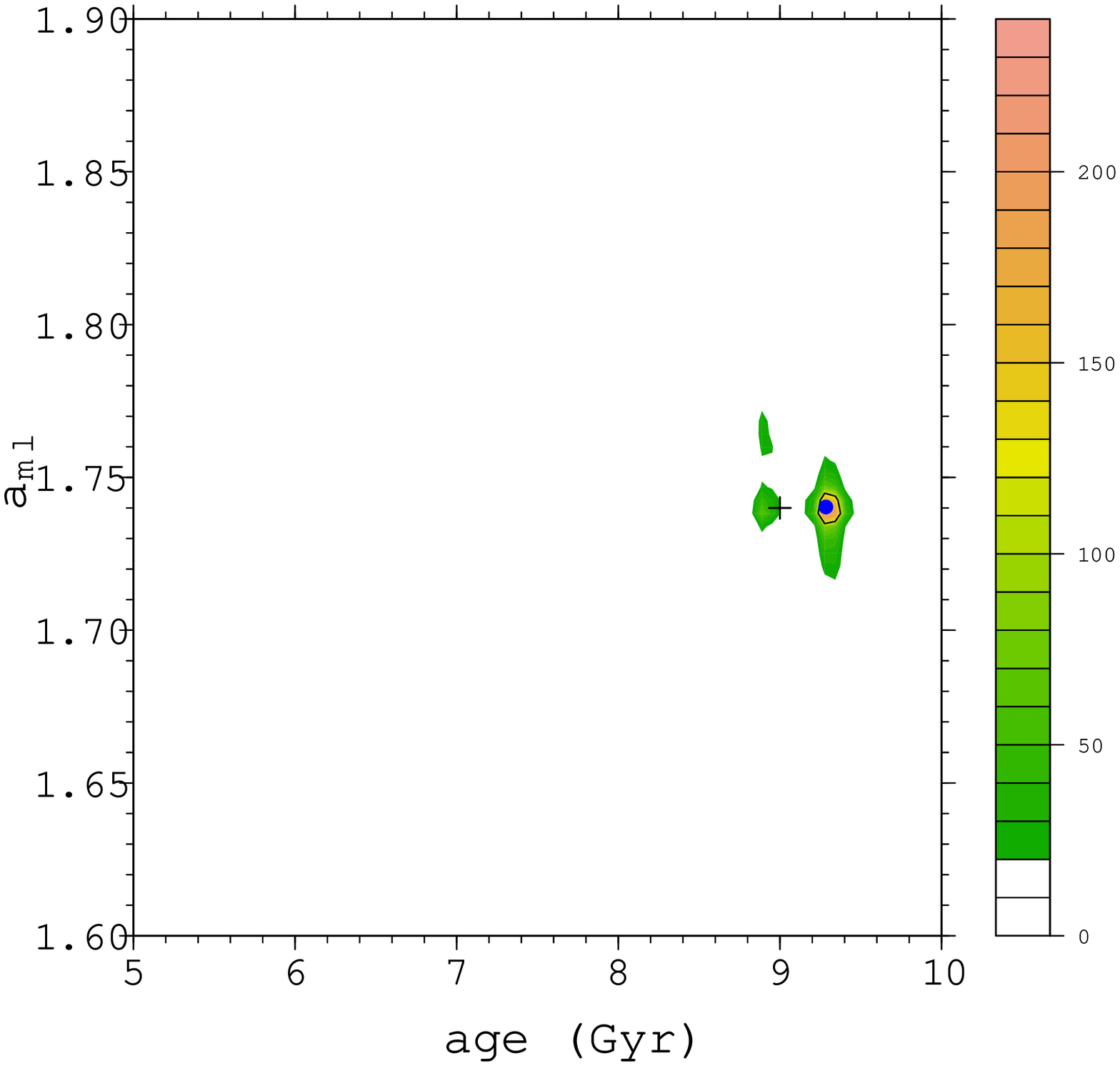}
        \includegraphics[height=6.0cm,angle=-90]{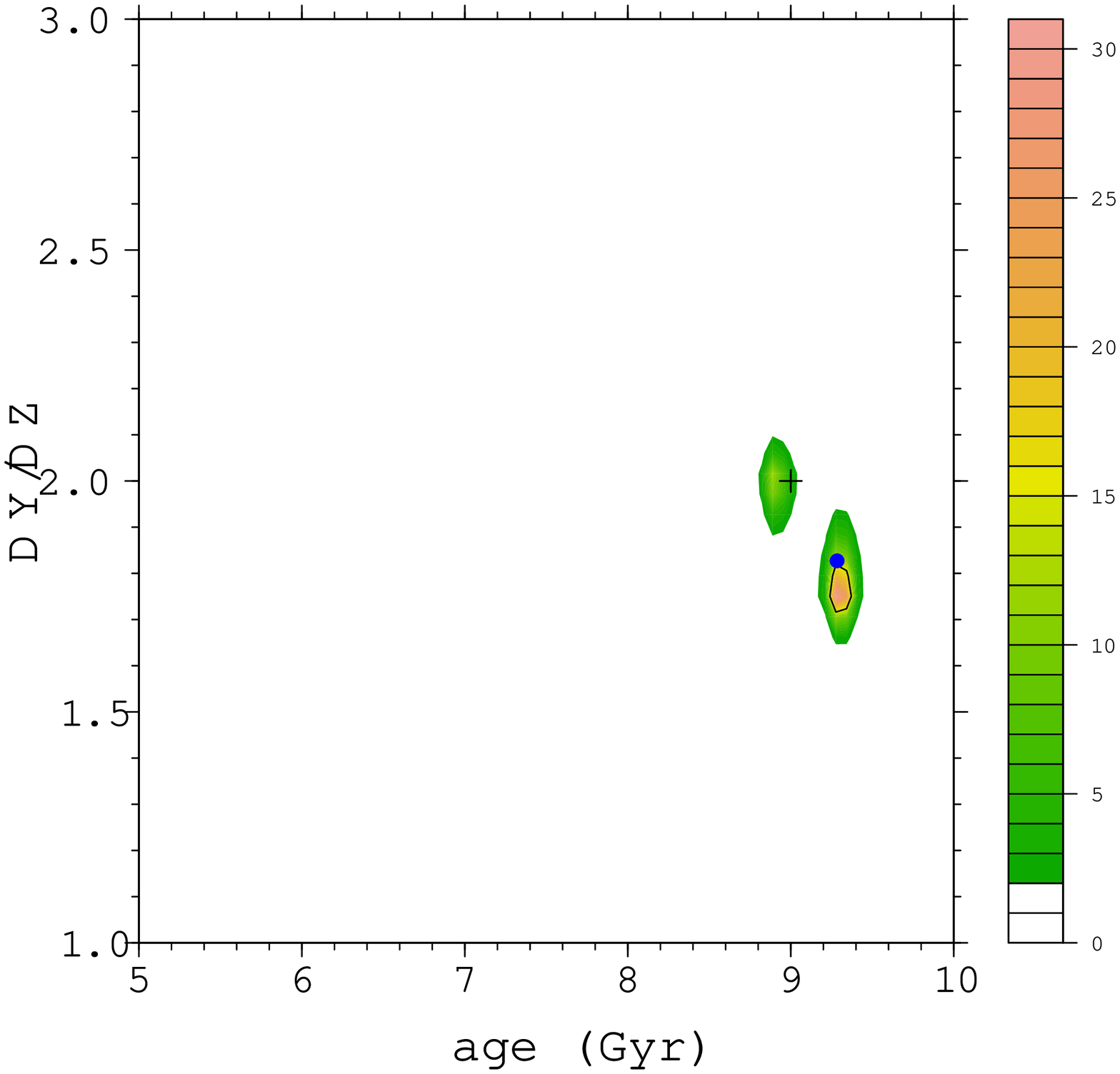}
        \includegraphics[height=6.0cm,angle=-90]{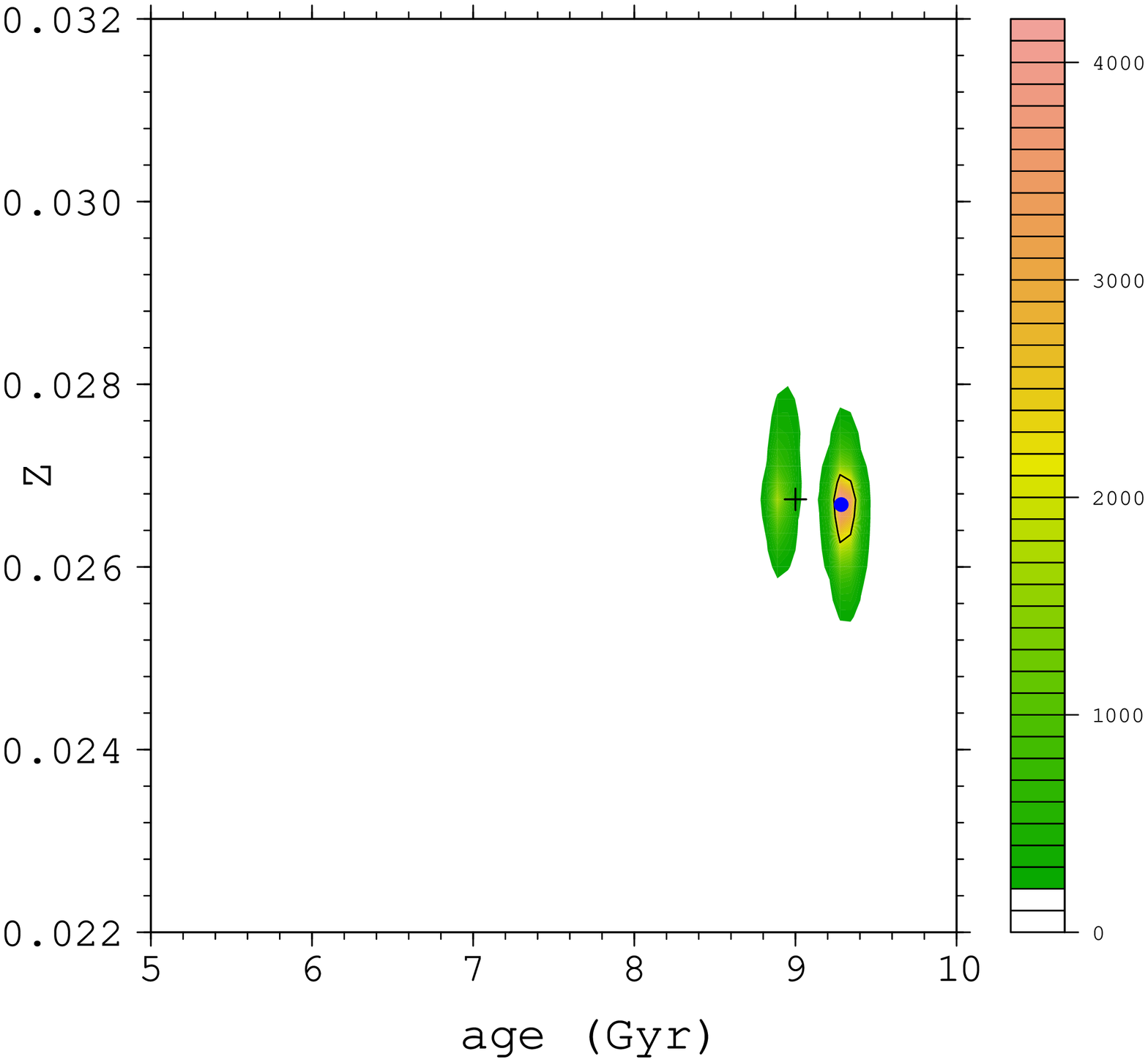}\\
        \includegraphics[height=6.0cm,angle=-90]{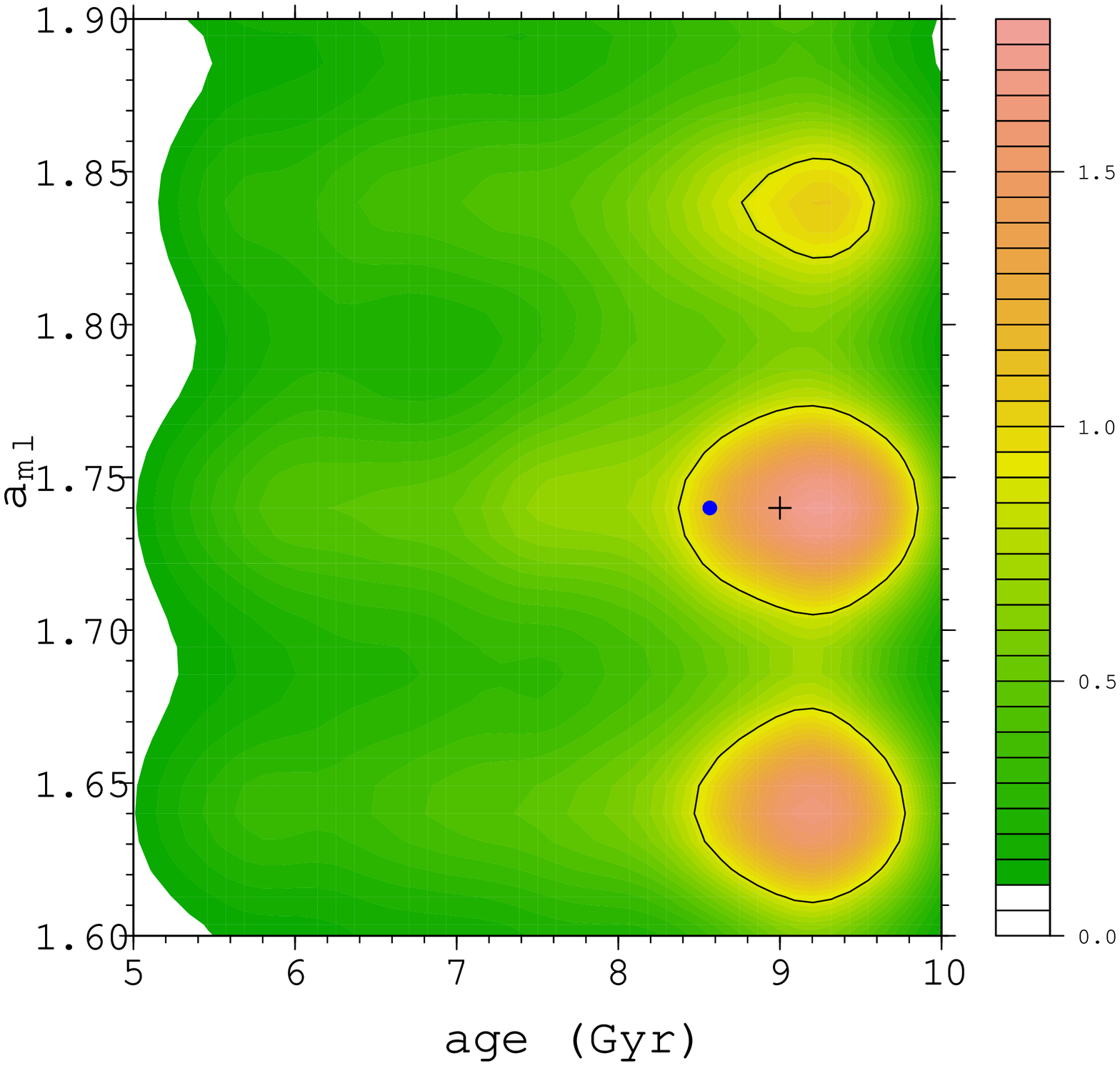}
        \includegraphics[height=6.0cm,angle=-90]{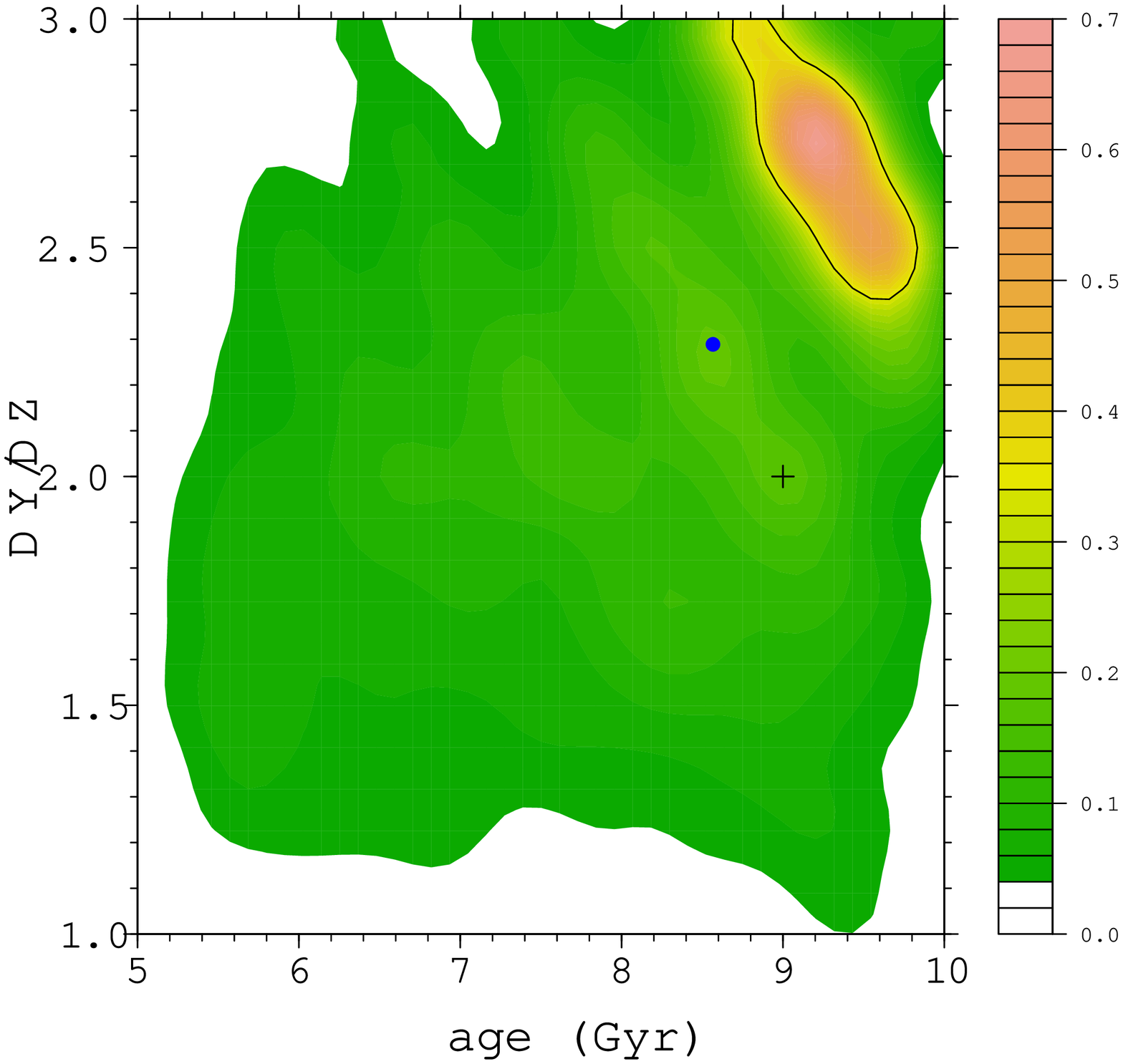}
        \includegraphics[height=6.0cm,angle=-90]{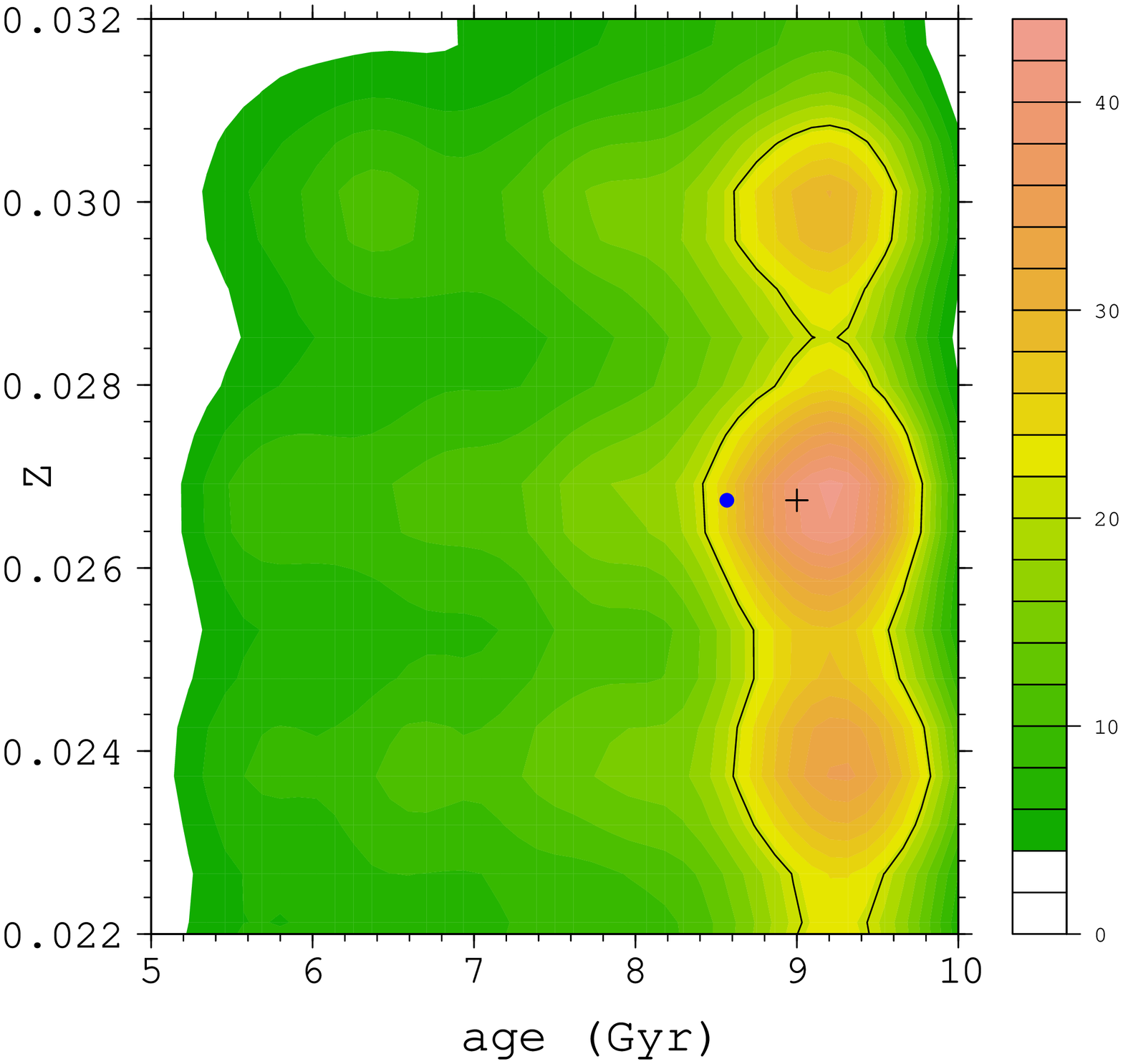}
        \caption{As in Fig.~\ref{fig:res7.5Gyr80}, but for a true age of 9.0 Gyr.  
        }
        \label{fig:res9.0Gyr80}
\end{figure*}

The method that adopts an ML fit (Sect.~\ref{sec:sub-ML}), taking into account the evolutionary timescale on the isochrone (which differentiates this method from the pure geometrical fit), provides similar results to the geometrical method, with some interesting differences. The biases on the recovered age are of about 0.15 and 0.1 Gyr for scenarios S35w and S80w, while for scenarios S35-9w and S80-9w these biases are about 0.10 and 0.3 Gyr. 
In these cases, the method provides a peak in density for values of the $\Delta Y/\Delta Z$ parameters of about 1.75 and ages of about 9.4 Gyr (central panel in the middle rows of Figs. \ref{fig:res9.0Gyr35} and \ref{fig:res9.0Gyr80}). This is the origin of the detected age biases. Moreover, no edge effects mitigate the overestimation because the variance of the posterior density is low and the distribution is not heavily truncated at 10 Gyr, that is, the edge of the grid.  
Figures~\ref{fig:res7.5Gyr80} and \ref{fig:res9.0Gyr80} show a drastic reduction of the error on the ML recovered parameters for larger sample sizes. In all the scenarios the random errors on the age are about one quarter of the corresponding quantities from the geometrical fitting. This reduction comes from the different approaches to the fit of the two techniques. For the geometrical method, two isochrones that pass at the same minimum distance from a reference point provide identical likelihood. This is not the case for the ML approach, which takes into account the whole evolution of the isochrone and assigns a higher likelihood to a curve that stays closer to the reference point for a larger evolutionary portion. Further details are discussed in Appendix~\ref{app:toy}.   

The biases in the initial helium content are almost negligible and the error on the recovered $\Delta Y/\Delta Z$ lower than about 0.3 and 0.15 for  scenarios with 35 and 80 stars, respectively. Indeed, these impressive results come from relatively small differences in the shape of the isochrones when the initial helium content varies. While the difference is so subtle that the isochrones are hardly distinguishable by eye, nevertheless it is more than enough for the ML algorithm to discriminate among them. As mentioned above, the $\Delta Y/\Delta Z$ shows a tendency for underestimation in the S80-9w scenario, leading to the value of 1.83, which is marginally consistent with the true value of 2.0 given the very small error of $+0.19$. 
The random errors on those estimates are significantly smaller, by about a half, than those from geometrical fitting.

The length of the chain needed to reach the convergence is another major difference between ML and geometrical estimates. Due to the shown unimodality, the geometrical fit reaches convergence much faster: usually the geometrical chains were stationary and well mixed after about one quarter of the lengths stated in Sect.~\ref{sec:schema}.         
For the 80-star scenarios the convergence and mixing tests of Gelman-Rubin and Geweke \citep{Gelman1992, Geweke1992} reported problems for five chains at 7.5 Gyr and seven at 9.0 Gyr. However, the removal of them from the global statistics leads to negligible differences in the final estimates of the parameters. Indeed the estimates from these problematic chains are indistinguishable from those without convergence issues. The problems reported by the tests arise due to the heavily peaked multimodal posterior distributions (see second row in Figs.~\ref{fig:res7.5Gyr80} and \ref{fig:res9.0Gyr80}); the proposed Gaussian distribution has more difficulty in properly mapping all the local maxima of the posteriors. This is a well-known difficulty in performing MCMC integration, which can be tackled by a change of the proposal distribution to a form that better mimics the posterior \citep[see e.g.][]{Haario2001,RobertCasella,Adaptive2011}, which is a difficult task in high-dimensional space.
Other possible remedies include adopting a longer chain or relying on more advanced sampling schema such as interacting MCMC \citep[][]{RobertCasella,Adaptive2011}. However, as stated above, these convergence problems minimally alter the present analysis and do not justify the adoption of these complex integration methods. 

In summary, the main difference between the geometrical and ML fitting methods is an appealing shrink of the random errors granted by the ML approach. As discussed in Sect.~\ref{sec:raneff}, when dealing with the reproducibility of the results from the various techniques in the presence of different artificial perturbations on the data, the increase in the precision of the ML estimates can lead  the method to a struggle to include the true parameter values in their estimated credible intervals.

\subsection{Results from the SCEPtER pipeline}

\begin{figure*}
        \centering
        \includegraphics[height=8.0cm,angle=-90]{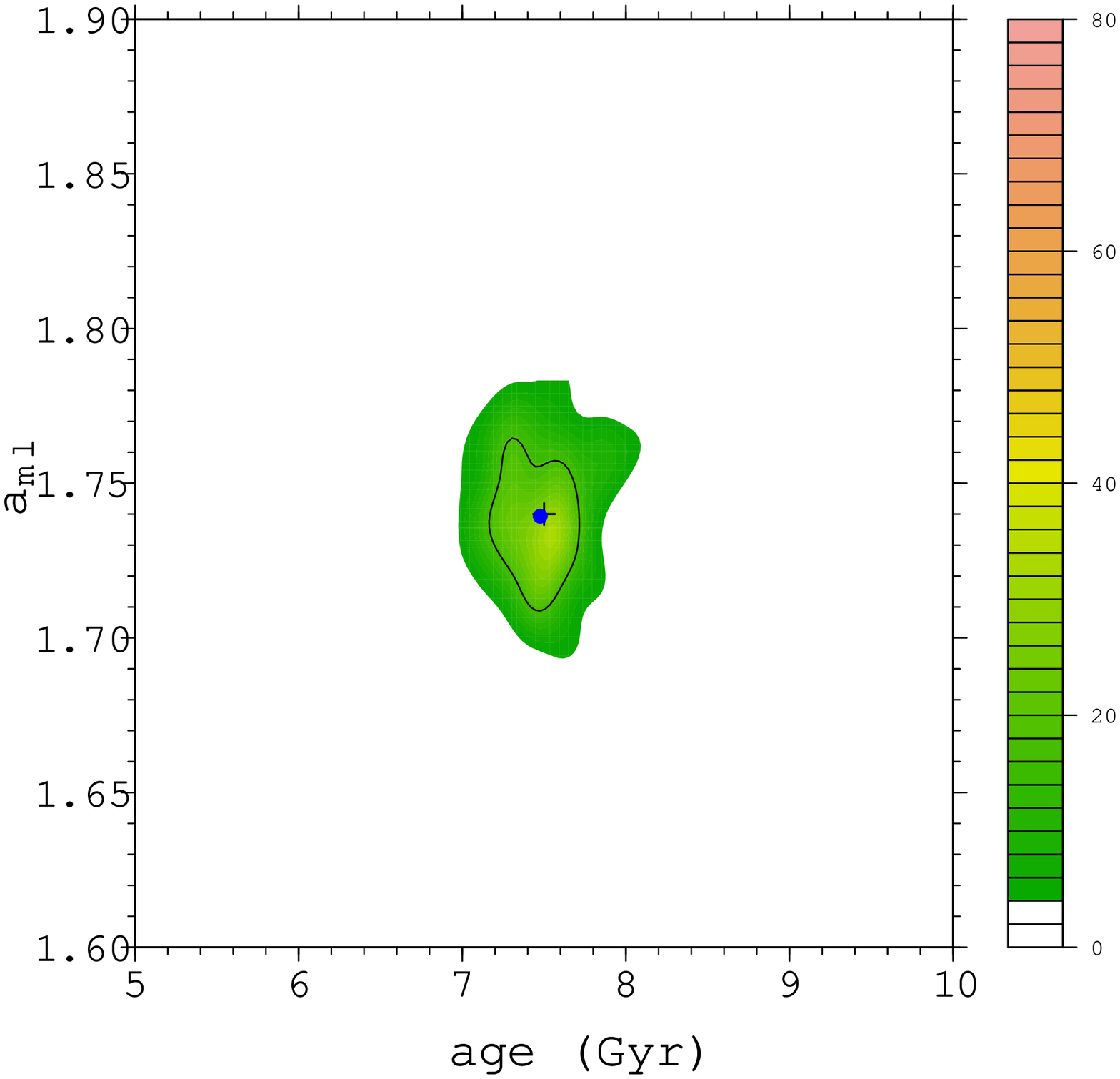}
        \includegraphics[height=8.0cm,angle=-90]{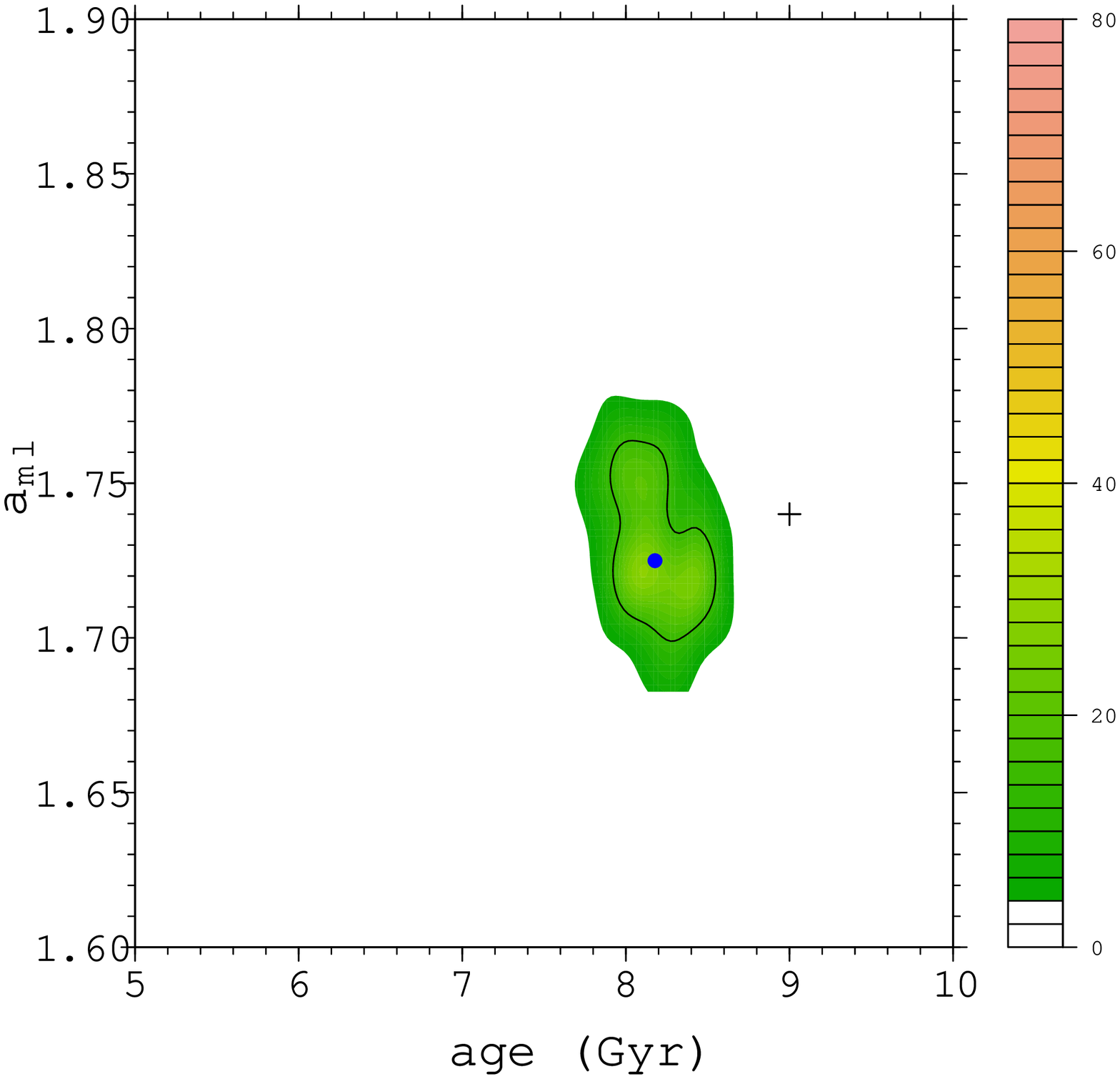}\\
        \includegraphics[height=8.0cm,angle=-90]{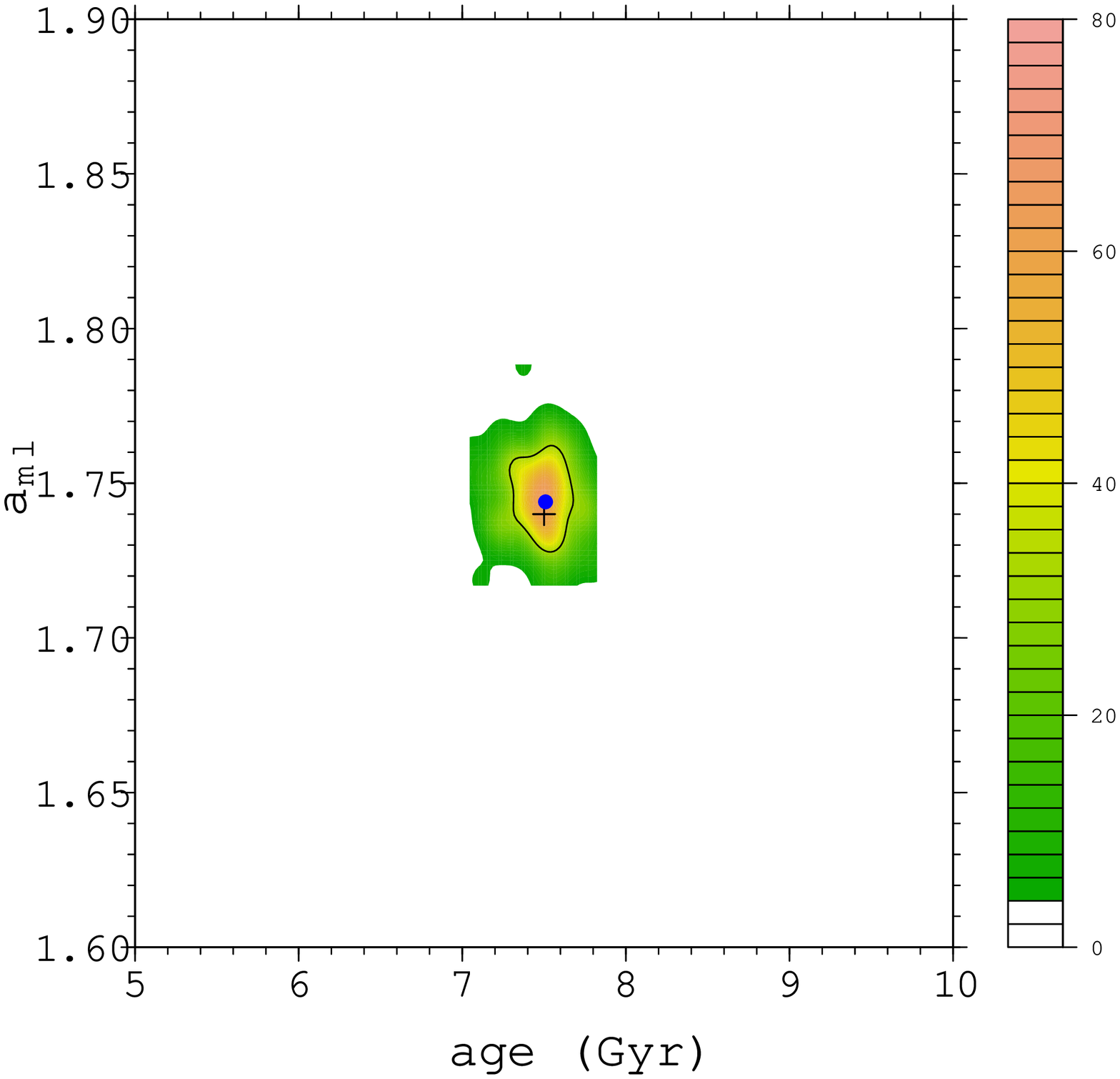}
        \includegraphics[height=8.0cm,angle=-90]{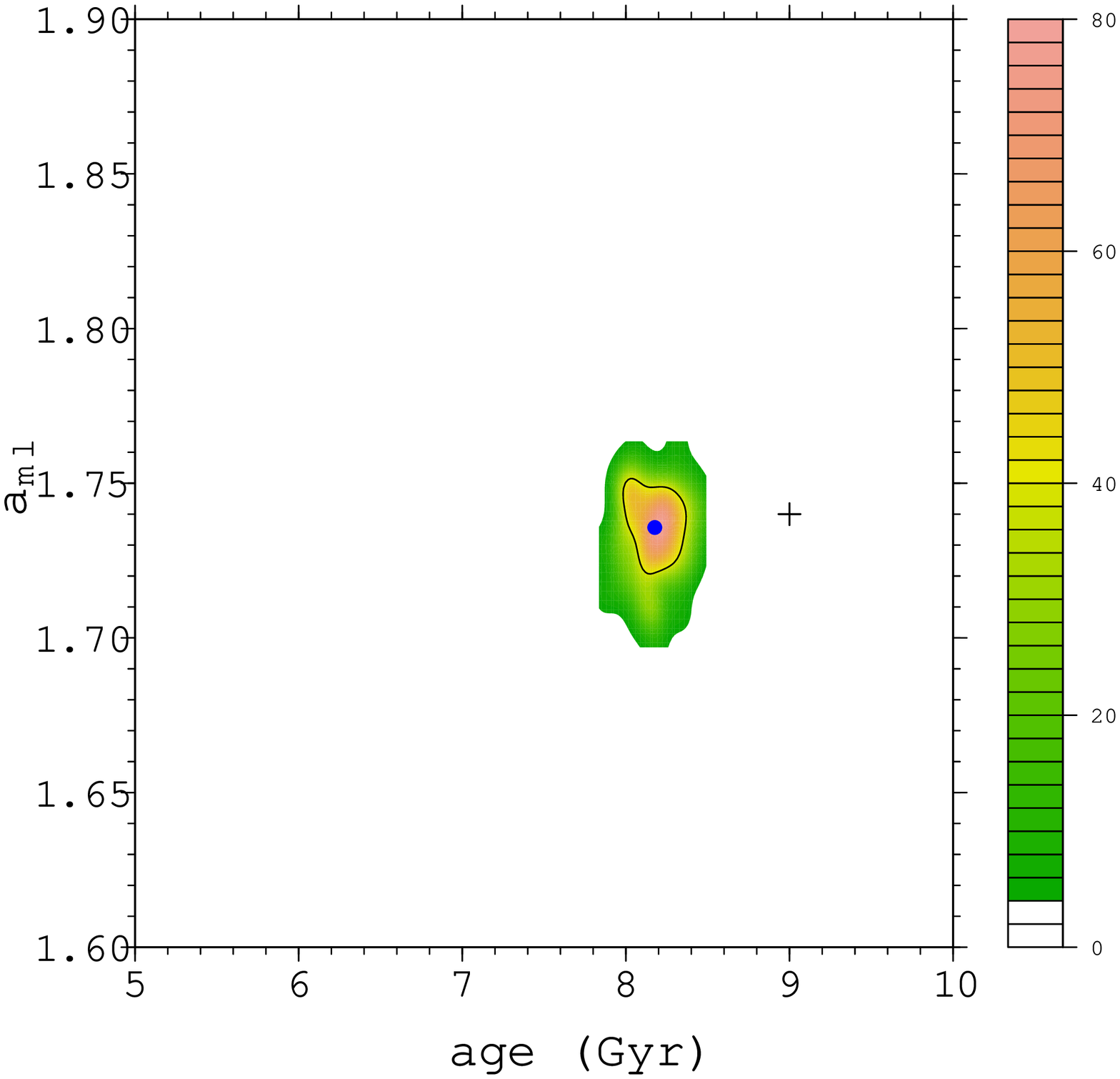}
        \caption{{\it Top row}, {\it left}: 2D density of probability in the age vs. $\alpha_{\rm ml}$ plane from samples of 35 RGB stars (true age 7.5 Gyr) from the mean of the SCEPtER fit of individual stars (see text). {\it Right}: Same as in the left panel but for a true age of 9.0 Gyr.  {\it Bottom row}: As in the top row, but for samples of 80 RGB stars.
        }
        \label{fig:resSCEPtER-mean}
\end{figure*}

The single and independent star fit (Sect.~\ref{sec:sub-scepter}) shows, as expected, the largest random error because it does not make use of the information of common parameters among the whole sample. 
While Figs.~\ref{fig:res7.5Gyr35} to \ref{fig:res9.0Gyr80} show the joint density from the  whole sample of stars, it is possible to adopt a different approach. For each of the Monte Carlo runs one can compute the mean of the stellar parameters over the sample and then adopt  these values to construct the joint density. By the law of large numbers it follows that the dispersion of the distribution of the means will shrink as the inverse of the square root of the sample size. Figure~\ref{fig:resSCEPtER-mean} shows, in the $\alpha_{\rm ml}$ versus age plane, the reconstructed densities of the mean. The results are unbiased for a true age of 7.5 Gyr, while a bias of almost $-0.9$ Gyr occurs for the 9.0 Gyr true age scenarios. The result for 7.5 Gyr
comes from the large, symmetric errors around the central value and is therefore of no practical relevance. The one for 9.0 Gyr shows a large bias, suggesting that this method has an accuracy that is clearly inferior to geometrical and maximum likelihood approaches.
The magnitude of the bias stems directly by the fact that the variability of the estimates is of the same order of magnitude or even larger than the half-width of the age interval spanned by the grid. Therefore, the proximity of the upper edge of the grid to the true solution at 9.0 Gyr limits the possibility to explore higher ages, thus inducing a hard boundary in the estimated ages, leading to an edge effect similar to those discussed in \citet{scepter1}. The same does not occur at 7.5 Gyr.

\section{Assessing the sample variability}\label{sec:raneff}

The results presented in the previous section obviously depend in some way on the samples adopted for their computation, and therefore depend on the random Gaussian perturbation added to the artificial stars  mimicking the observational uncertainties. Therefore it is worth exploring the relevance of this particular problem; in other words how stable are the fits owing to the random nature of observational errors?

To this purpose we can exploit the results of the simulation to directly evaluate the variability on the recovered ages, splitting the variance into two components: a first one  ($\sigma_g$) linked to the variability of the mean age among the Monte Carlo simulations, and a second one ($\sigma$) that accounts for the residual variance.  
Therefore $\sigma$ is directly linked to the effect of the observational uncertainties on the recovered values, taking the observed values as perfectly unbiased. This is the quantity that would affect a fit for real stellar objects. On the other hand, the $\sigma_g$ components, which cannot be assessed for real objects, are due to the fact that a bias (either random or systematic) in the observational values with respect to their true values would modify the median recovered parameters. This value is therefore linked to the dispersion of the medians.
A $\sigma_g$ that is negligible with respect to $\sigma$ implies that the mean values are recovered with a low spread with respect to the precision of the estimates. If this is the case, the hidden error source owing to the uncontrollable error on the observations can be safely neglected. On the other hand, when $\sigma_g$ is equal to or larger than $\sigma$, the spread in the mean recovered values is non-negligible. Therefore random observational errors can lead to different age estimates. Due to the impossibility for the observer to control this  source of uncertainty,     
a large ratio $\sigma_g/\sigma$ would pose serious problems, implying that the fitted age is prone to bias.

To assess the values of the two variance components, we adopted a very powerful statistical method: a random effect model fit \citep{venables2002modern,lme4}, which is briefly introduced in Appendix~\ref{app:raneff}.

\begin{table}[ht]
        \centering
        \caption{Age variance components from random models fit in the twelve considered scenarios (see text).}
        \label{tab:mixed}       
        \begin{tabular}{lcccc}
                \hline\hline
                & $\sigma$ & $\sigma_g$ & ratio & $q$\\ 
                \hline
S35 & 0.84 & 0.34 & 0.40 & 60\%\\ 
S35-9 & 0.74 & 0.26 & 0.35 & 60\% \\ 
S80 & 0.77 & 0.25 & 0.33 & 70\% \\ 
S80-9 & 0.68 & 0.16 & 0.23 & 60\%\\ 
\hline
S35w & 0.49 & 0.33 & 0.66 & 50\%\\ 
S35-9w & 0.48 & 0.27 & 0.57 & 55\%\\ 
S80w & 0.21 & 0.12 & 0.55 & 10\%\\ 
S80-9w & 0.23 & 0.14 & 0.60 & 15\%\\ 
\hline
S35S & 1.42 & 0.00 & 0.00 & 100\%\\ 
S35-9S & 1.26 & 0.04 & 0.04 & 100\%\\ 
S80S & 1.39 & 0.08 & 0.06 & 100\%\\ 
S80-9S & 1.30 & 0.00 & 0.00 & 100\%\\
\hline
\end{tabular}
\tablefoot{The columns contain: the residual standard error $\sigma$; the group standard error $\sigma_g$; their ratio; and the percentage $q$ of samples (rounded to the nearest 5\%) for which the true age is inside the $1 \sigma$ credible intervals.}
\end{table}

The standard error components in the twelve considered scenarios are reported in Table~\ref{tab:mixed} along with the percentage $q$ of the Monte Carlo experiments including the true values of all the parameters in the estimated $1 \sigma$ interval from the fit, that is, the error on the individual Monte Carlo reconstructions.
In this way it is possible to compare the $1 \sigma$ interval coverage with respect to  the prediction under the assumption that the parameters follow a Gaussian distributions with variance $\sigma^2$ (i.e. 68\%). 
The three methods show notably different performances. The single-star fit has the lower $\sigma_g$ variability both in absolute and in relative terms. Therefore the results from this technique can be considered robust against observational errors. Obviously this comes at the cost of a remarkably lower precision in the estimates. The geometrical fitting method has the second smallest  variance among realisations, that accounts for about 35\% of the random residual error. This contribution is not dominant but is not negligible either, so any sensible analysis of real cluster RGB stars should account for it. As shown in the last column of Table~\ref{tab:mixed}, about 30\% of the Monte Carlo perturbations lead to estimates of parameters whose $1 \sigma$ credible interval does not include the true value.
Although the contribution of random fluctuations in the observables to the overall variability in the estimates is usually neglected, the present results confirm the finding 
for detached binary systems discussed in \citet{BinTeo}. That work reports a $\sigma_g$ error as large as two thirds of the residual error $\sigma$. Ultimately, it seems that the role of random variability due to 
unavoidable measurement errors in the observational constraints is more important than commonly thought.

The largest $\sigma_g/\sigma$ ratio  comes from the ML fit that accounts for the evolutionary timescale. This result comes mainly from the reduction of the residual error on the age estimates discussed in Sect.~\ref{sec:results}. Overall the variance   among realisations is about 60\% of the random residual error, nearly double that resulting from geometrical fit. This leads to the risk of obtaining a best fit set of parameters that is inconsistent with the true ones. Indeed Table~\ref{tab:mixed} shows the dramatic low $q$ value of about 10\% for the 80-star scenarios, with clear evidence that the problem gets worse as the sample size increases. 
It is therefore clear that even when stellar tracks and synthetic data perfectly match -- apart from random variations -- the adoption of the estimated errors from the ML fit is overoptimistic, greatly increasing the risk of failing to include the true values of the estimated parameters in the obtained $1 \sigma$ error range.

In light of these results it seems the pure geometrical procedure may offer greater control of the random variability than the ML approach when fitting real objects.

\section{Technique comparison for single-star fits}\label{sec:cfr}

As discussed above, different methods are routinely adopted in the literature for age fitting \citep[see][and references therein]{Valls2014}. 
Whereas it is common to present stellar estimates obtained using methods relying upon different grids of stellar models, differences in the estimates obtained in this way, but adopting a different fit method, are rarely explored (see e.g. \citealt{Bazot2012} for a fit of $\alpha$ Cen A or \citealt{Jorgensen2005} for MS and SGB applications).  
It is therefore interesting to verify the relative performance of the techniques discussed above, but in the framework of single-star fits in the RGB phase. This section is devoted to the comparison of the age results from the three fitting methods described above.  

\begin{figure}
        \centering
        \includegraphics[height=8.0cm,angle=-90]{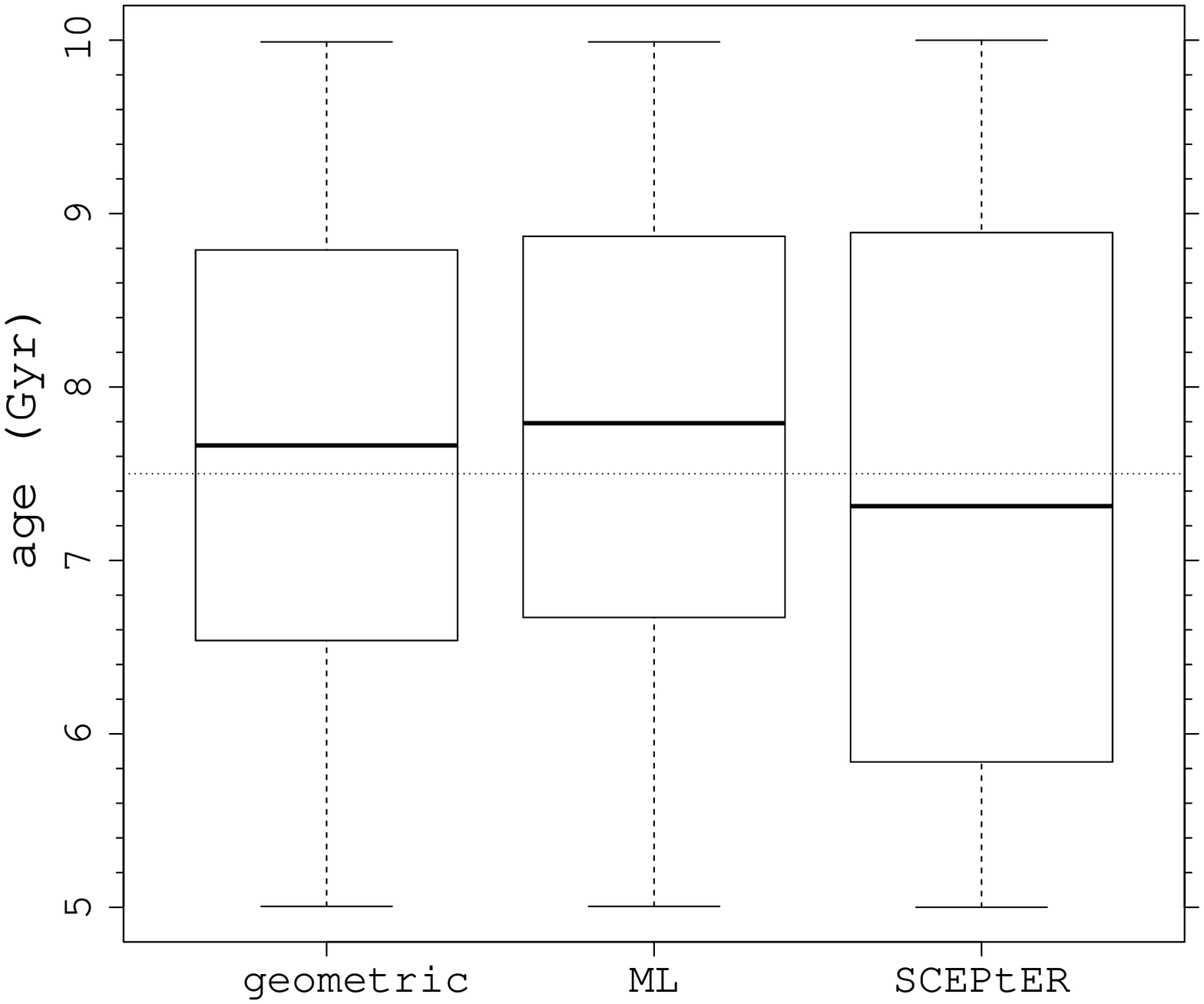}
        \caption{Box plot of the age estimates for single RGB star fits from the three adopted methods in the text.
        }
        \label{fig:boxplot}
\end{figure}

The comparison is performed selecting 20 parent artificial stars in the RGB phase, equally spaced in $\Delta \nu/\Delta \nu_{\sun}$ from 0.08 to 0.009 on the 7.5 Gyr isochrone ($Z = 0.02674$, $Y = 0.2752$), that is, centred over the grid described in Sect.~\ref{sec:grids}.
Each star was subjected to Gaussian perturbations of the observables, as described in Sect.~\ref{sec:method}. Each synthetic star was then individually fitted using one of the three techniques. The perturbation procedure was repeated 70 times for each star, to account for different observational errors. 
Alternatively to Sect.~\ref{sec:fittingML}, the SCEPtER method was employed to recover not only the central values but also for estimating their errors, with a Monte Carlo sampling that is fully described in \citep{eta,bulge}. 
Briefly, for each star, 500 secondary artificial stars were generated from the same Gaussian distribution adopted in the first perturbation step. All these stars were fitted. The median of the fitted values was usually adopted as the best estimate, and the 16th and 84th percentiles provide information about the errors. 

For the MCMC fit, eight parallel chains of 7\,000 iterations in length were evolved (after a burning-in chain of  3\,000 iterations) for each star, and then thinned to reduce the sample size to 500 points for computational reasons. The consistency of the estimates  from the thinned sample with those of the whole sample was verified.

In summary, the design of the simulation is fully crossed, that is, the experimental units (20 synthetic stars) are fitted by the three techniques, and the Monte Carlo simulation is repeated 70 times.
This approach allows one to identify the different sources of variability in the final estimate. It is clear that the individual differences among the various Monte Carlo realisations are not of direct interest by themselves. The main information coded in these data is the variability that can occur only by chance in the results due to errors in measurements carried out on the observed stars. Indeed, although the individual differences between SCEPtER and MCMC estimates are interesting \textit{per se} if assessed directly, they are much more informative if considered as representative of possible variability in the estimates owing to the choice of the fitting method. 
Obtaining unbiased and reliable estimates for these quantities from the bulk of the simulations is a relatively challenging task. As in Sect.~\ref{sec:raneff}, we exploit the power of a random-effect model fit (see Appendix~\ref{app:raneff}). This method was adopted to disentangle the residual variability in the age estimates $\sigma$ from $\sigma_a$, $\sigma_b$, and $\sigma_c$ those due to differences in the fitting methods, the choice of progenitor stars, and the individual Monte Carlo realizations, respectively.
A source of variability is relevant whenever it is large with respect to $\sigma$.

The fit of the model provided a mean $\mu = 7.59$ Gyr as the global age estimate, very close to the nominal value of 7.5 Gyr. Figure~\ref{fig:boxplot} shows a boxplot for the age estimates from the three methods. Pure geometrical and ML estimates gave nearly identical distributions of the ages, while the SCEPtER estimates show a larger variance. This is possibly due to the coarseness of the grid: while the first two methods can rely on isochrone interpolation to explore the posterior density, the SCEPtER method exploits only the existing grid points and therefore requires a much denser grid to offer equivalent accuracy. 
The global standard deviation was $\sigma = 1.41$ Gyr, revealing a large indetermination in the age fit. The other variance components are $\sigma_a = 0.19$ Gyr, $\sigma_b = 0.11$ Gyr, and $\sigma_c = 0.11$ Gyr. These results seems to point to a negligible variability due to either the position on the  RGB or random errors on the observables ($\sigma_b/\sigma = \sigma_c/\sigma \approx 8\%$); more interestingly, the variability among the three estimation techniques is also quite low ($\sigma_a/\sigma \approx 13\%$). We also verified that the larger variance affecting SCEPtER estimates does not pose a problem of heteroscedasticity (i.e. non-uniform variance) for the the mixed-effect model. To this purpose we repeated the analysis working with the rank of the age\footnote{Ranks are obtained by sorting the age into 
ascending order and replacing each value by its relative position in the ordered set.} and obtaining an equivalent result ($\sigma_a/\sigma \approx 10\%$).  

In light of these results, it seems that -- for single-star fits in the RGB phase -- the choice of method for age recovery has no severe effect on the final age bias. It is clear however that this result cannot be generalised to include other evolutionary phases; further theoretical investigation is needed to address the question of the actual impact of the choice of fitting method on the final stellar age estimate.

\section{Conclusions}\label{sec:conclusions}

We performed a theoretical investigation on the biases and random uncertainties affecting the   age estimates from RGB stars in clusters. We focussed on the age determination based on 
 a mixture of classical surface ($T_{\rm eff}$ and [Fe/H]) and asteroseismic ($\Delta \nu$ and $\nu_{\rm max}$) observables. We built a mock data set of artificial stars with properties that mimic the old galactic open cluster NGC 6791, with metallicity [Fe/H] = 0.3 and an initial helium abundance of $Y = 0.302$, corresponding to a helium-to-metal enrichment ratio of $\Delta Y/\Delta Z = 2.0$. 

By means of Monte Carlo simulations we studied the performance of OC age reconstruction
given a set of stars in the RGB evolutionary phase. We analysed clusters for two different ages, namely 7.5 Gyr and 9.0 Gyr, and for two hypothetical observational sample sizes, namely 35 and 80 stars. For each scenario, stars were sampled from the reference isochrones, and Gaussian perturbations were added to these  quantities, to account for the observational errors. We adopted as typical uncertainties 75 K in $T_{\rm eff}$, 0.1 dex in [Fe/H], 1\% in $\Delta \nu$ and 2.5\% in $\nu_{\rm max}$.

For each of the four aforementioned scenarios we performed the recovery with three different methods: a pure geometrical isochrone fitting, a Bayesian maximum-likelihood fit, and an independent fit for single stars by means of the SCEPtER pipeline \citep{eta}, the latter being adopted as a reference for the performance of the other techniques. The artificial stars were sampled from the same grid of stellar models used in the recovery. Therefore we worked in an ideal case where stellar models perfectly match observational data. This means that we evaluated the maximum possible performance of the fitting techniques, that is, the minimum biases and random errors. When real stars are used rather than artificial ones, larger errors and biases are to be expected.

Overall, the performances of the methods were found to be similar. The mixing-length value and 
the metallicity $Z$ are recovered accurately from all the scenarios by all the methods.
The geometrical method slightly overestimated the age by about 0.3 Gyr for the scenarios with a true age of 7.5 Gyr, and underestimated it by about 0.2 Gyr for the scenarios of 9.0 Gyr. 
The value of the initial helium content is underestimated for the scenarios of 7.5 Gyr and overestimated for those of 9.0 Gyr; however these values  are still consistent with the true value $\Delta Y/\Delta Z = 2.0$ since they are affected by large random errors. This is due to the fact that the initial helium content impacts very mildly on the effective temperature of the RGB, therefore different initial helium abundances result in nearly identical isochrones apart from the age. 

The ML technique  provided similar biases but with a much lower variance. The age is overestimated by about 0.1 Gyr for the scenarios of 7.5 Gyr and 0.2 Gyr for the scenarios of 9.0 Gyr. The initial helium content is accurately estimated, with a small error. These results highlight the benefit of considering the whole isochrone's path in the fit of an object and not only the distance to the closer point. This method provides random errors on the fitted quantities that are about one quarter of those returned by the pure geometrical fit. However, due to the multimodal nature of the posterior probability densities for the samples of 80 stars the method shows a much slower chain convergence in the MCMC, requiring about four times the number of iterations of the geometrical method.

The reference independent fit of single stars showed, as expected, a large variance. Taking into account only the mean values from each simulation -- thus exploiting the variance reduction thanks to the laws of large numbers --  we obtained an unbiased estimate for the scenario of 7.5 Gyr (not really informative, due to the symmetry of the grid around this value) and a bias of about $-0.4$ Gyr for the scenario of 9.0 Gyr.

The most important difference between the first two techniques comes from the robustness of the results against observational errors. We investigated how different perturbations of the synthetic data could lead to different age estimates. 
For the first fitting technique, we found that estimations starting from the same sample suffer from a $1 \sigma$ variability of about 0.3 Gyr from one Monte Carlo run to another. This value is not negligible because it is about 45\% of the intrinsic variability due to the observational error. The Bayesian fitting method showed a similar  variance between runs but owing to the reduction of the global random component, its impact on the final variability is about 65\%. This larger variability due to the random perturbation leads most simulations -- up to 90\% -- to fail to include the true parameter values in their estimated $1 \sigma$ credible interval.

Therefore the results obtained with the purely geometrical method are more resistant to 
observational errors than those obtained with an ML fit that account for the stellar evolutionary time scale.  
This is of particular relevance when the methods are adopted in practice to obtain age estimates from real data. As anticipated in Sect.~\ref{sec:method}, the uncertainty in the input physics neglected in the present work can play a relevant role. Given the very narrow credible intervals from ML estimates, the unexplored error sources can in principle provide a much wider error interval on the recovered parameters. As an example, a rigid percentage variation in the radiative opacity -- which is not, as discussed in Sect.~\ref{sec:method}, a very realistic method to account for this uncertainty for RGB stars -- would directly propagate to the estimated age \citep[see e.g.][]{Chaboyer1996, incertezze1}. This means that a $\pm 5\%$ uncertainty corresponds to about 0.4 Gyr for an isochrone at 7.5 Gyr. While the $1 \sigma$ error range from geometrical inference is larger than this value, this is not the case for ML estimates, leading to a false sense of accuracy.

While the inclusion of the information of the stellar evolutionary timescale in the fitting was advised for alleviating the problems linked to the age-metallicity degeneracy in the MS \citep[][and references therein]{Valls2014}, it seems that -- in agreement with the discussion in \citet{eta} -- it does not always lead to better results and the gain in the increased precision in the estimates is counterbalanced by the lower protection against random fluctuation than is provided by a  purely geometrical fit. Ultimately, the merits and the drawbacks of the methods have to be evaluated in light of the specific sample and evolutionary phases under investigation.  

Finally, we compared the performance of the three fitting methods in the framework of single RGB star age estimation. Owing to the fact that we adopt the same grid of stellar models in the procedure, this comparison sheds some light on the differences that could be ascribed to relying on different algorithmic approaches in the age fitting.
We obtained that the variability linked to the choice of the fitting method is minor, being about 15\% of the variability caused by the observational uncertainties. Therefore it seems that for RGB stars the choice of fitting scheme does not contribute significantly to the final age bias.
It is however clear that the validity of this statement is confined to the explored evolutionary phase. Further theoretical investigation is needed to explore the actual impact of the choices in the fitting method on the obtained results.

\begin{acknowledgements}
We thanks our anonymous referee for the useful comments and suggestions.
This work has been supported by PRA Universit\`{a} di Pisa 2018-2019 
(\emph{Le stelle come laboratori cosmici di Fisica fondamentale}, PI: S. Degl'Innocenti) and by INFN (\emph{Iniziativa specifica TAsP}).
\end{acknowledgements}

\bibliographystyle{aa}
\bibliography{biblio}

\begin{appendix}

\section{Details on the adopted Monte Carlo Markov chain}\label{app:MCMC}

The adopted MCMC method starts by establishing a good starting point in the $\theta$ space and evaluates the scale over which the likelihood function varies. Such estimates are obtained by  randomly sampling 7\,000 isochrones on the grid, and evaluating their likelihood. 
This step sets the initial value and the initial guess of the covariance matrix in the $\theta$ space for the jump function. Then a burning-in chain of 3\,000 points is generated in the following way.
We use $\theta_k$ to denote the value of the parameters after the step $k $ and $\Sigma$ the covariance matrix. A proposal point in the $\theta$ space is generated adopting a Gaussian jump function with mean $\theta_k$ and covariance $\Sigma$
\begin{equation}
\theta_{k+1} \sim N(\theta_k, \Sigma).\label{eq:MCMC-jump}
\end{equation} 
The choice of the jump function is usually critical in preventing the algorithm from being trapped in local minima (see \citealt{Haario2001}, and \citealp{Bazot2012} for an application to $\alpha$ Cen A modelling). In our particular case however the process was facilitated by the artificial objects being sampled from the exact grid adopted in the recovery. Therefore, apart from the random perturbations in the "observables", we do not have to face systematic discrepancies, that are unavoidable when attempting a fit for real stars. Therefore we found that the simple Gaussian jump function performs well against a mixture of Gaussian distributions with different variances or a mixture of Gaussian and uniform distributions. 

We use $P(\theta)_k$ and $P(\theta)_{k+1}$ to denote the likelihood of the solutions $\theta_k$  and 
$\theta_{k+1}$. We define $P_r = \min(\frac{P(\theta)_{k+1}}{P(\theta)_{k}},1)$. Then the following chain rule applies:
\begin{equation}
\theta_{k+1} = \begin{cases} \theta_{k+1}, & \mbox{with probability } P_r \\ \theta_{k}, & \mbox{with probability } 1-P_r \end{cases}
.\end{equation}
Due to the intrinsic degeneracy among the parameters $\theta$ the convergence speed of the chain is known to be sub-optimal (see \citealt{Haario2001}, and also \citealt{Kirkby-Kent2016} for a specific discussion in a detached binary system fit). Therefore after the burning-in phase the covariance matrix of the last 50\% of the proposed solutions was adopted to transform the $\theta$ variables to $\theta_\perp$ orthogonal ones. The step was performed by means of a principal-component analysis \citep{simar,Feigelson2012}, a statistical technique that computes a set of independent and orthogonal linear combinations $\theta_\perp$ of the original $\theta$ variables. The chain is then built in the newly computed space, achieving a better convergence speed. At the end the transformation was inverted and the results were remapped in the original $\theta$ space.

The burning-in chain is then discarded; the MCMC sampling for the geometrical fitting required  the evolution of eight parallel chains of 7,000 iterations in length for samples of 35 stars and 21\,000 iterations for samples of 80 stars. For the ML approach the burn-in chain length was 9\,000 points.
Indeed the chains from geometrical fitting showed a faster convergence and they are stationary after about one quarter to one half of the adopted length.
In fact, the ML posterior density is sparse within the parameter space, implying a greater difficulty of the chain to obtain an accurate map.
The lengths of the chains are sufficient to achieve a good convergence and mixing, according to the statistical tests of     
Gelman-Rubin and Geweke \citep{Gelman1992, Geweke1992}.

\section{A toy model for maximum likelihood to geometrical distance comparison}\label{app:toy}

In this section we present a simple toy model to highlight the differences in the geometrical versus ML approaches to the isochrone fitting.  

\begin{figure}
        \centering
        \includegraphics[height=8.5cm,angle=-90]{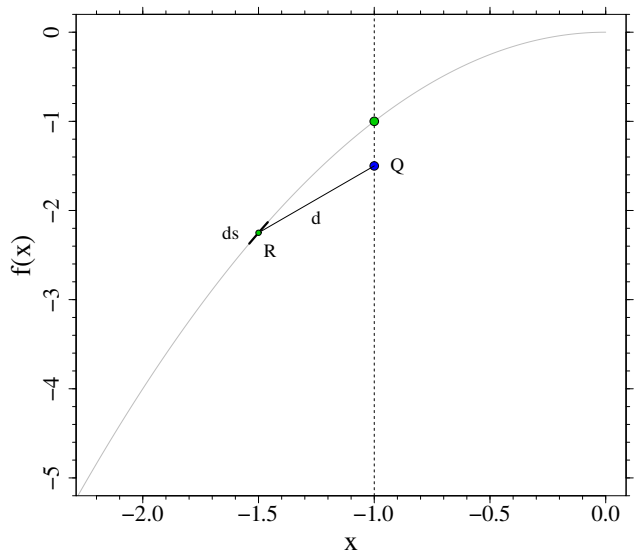}
        \caption{Toy model adopted for geometrical and ML fit comparison (see text). The dashed line identifies the possible positions of the point $Q$. }
        \label{fig:lik-toy}
\end{figure}

\begin{figure*}
        \centering
        \includegraphics[height=8.5cm,angle=-90]{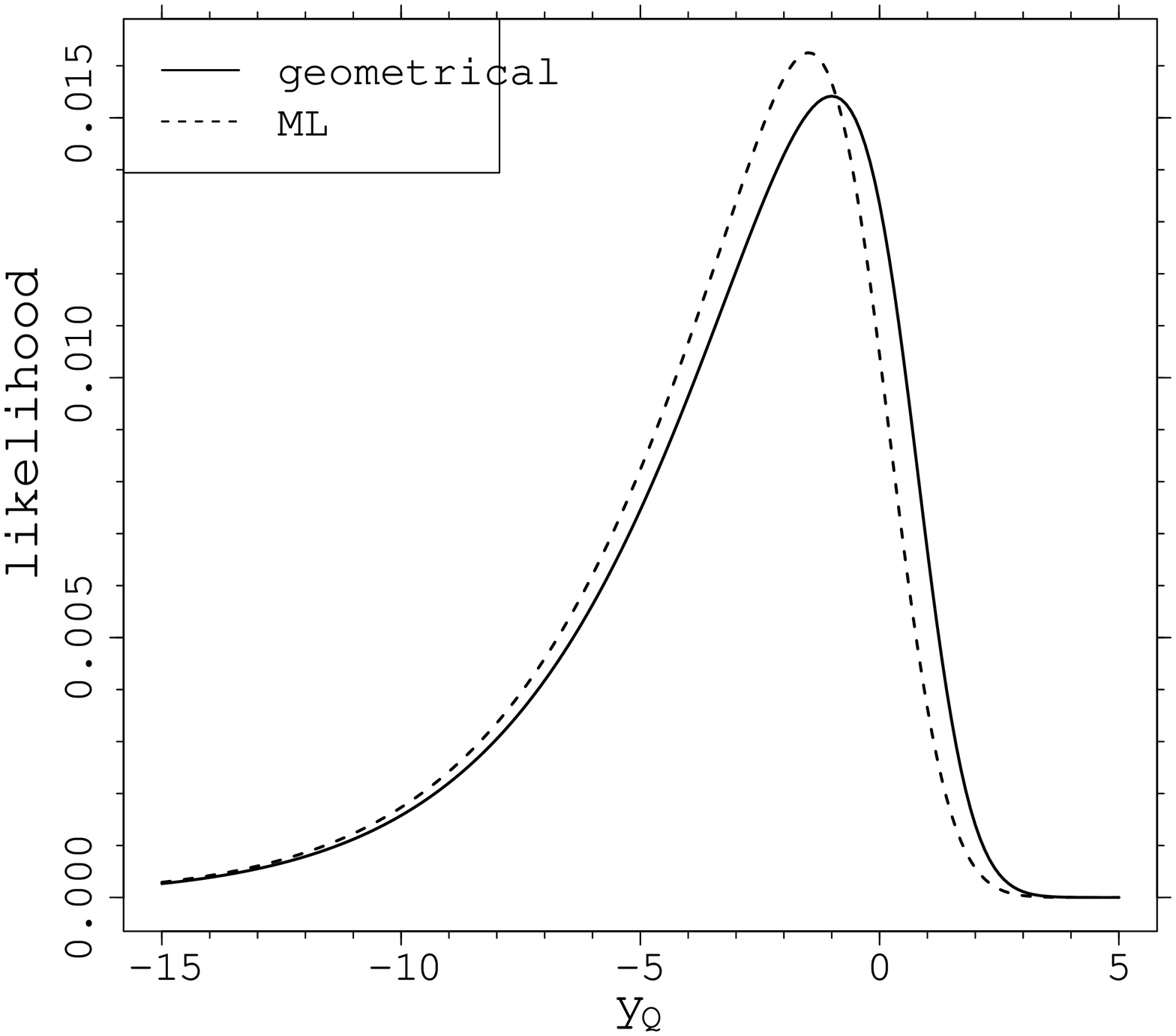}
        \includegraphics[height=8.5cm,angle=-90]{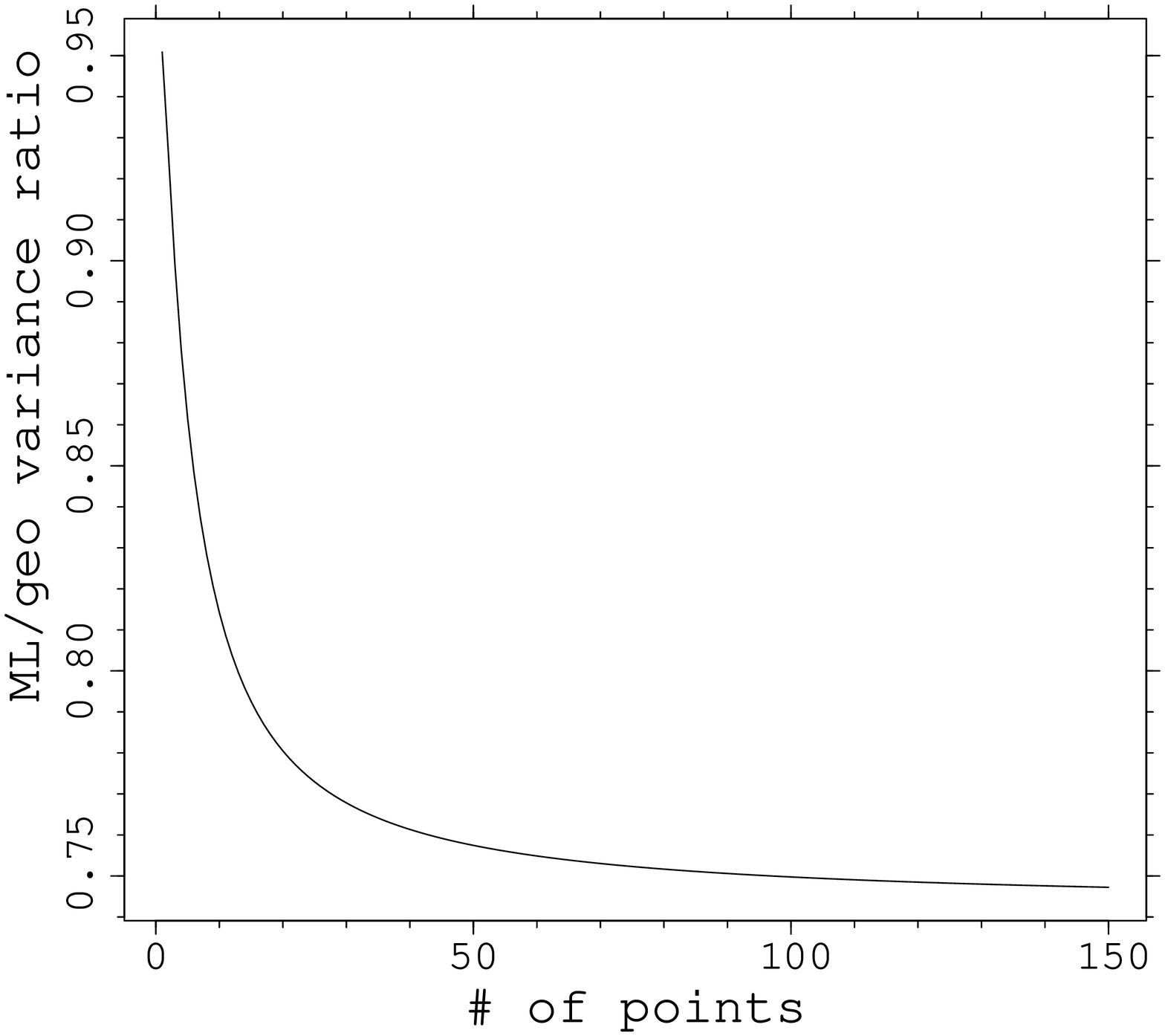}
        \caption{{\it Left}: Likelihood from the pure geometrical method (solid line) and for the ML method (dashed line) when varying the ordinate $y_Q$ of the observational data in the toy model (see text). {\it Right}: Ratio of the ML to geometrical variance versus the sample size. }
        \label{fig:lik-toy-res}
\end{figure*}

Let us assume that the fictional isochrone can be represented by the arc of parabola, as in  Fig.~\ref{fig:lik-toy}
\begin{equation}
f(x) = -x^2 \quad , \quad x \in [-5,0]. \label{eq:parabola}
\end{equation}
Let $Q = (-1, y_Q)$ be a point representing an "observation". Therefore, for $y_Q = -1,$ $Q$ lies on the arc of parabola. 
The squared distance $d^2$ from $Q$ to a point $R = (x, f(x))$ on the parabola is
\begin{equation}
d^2(x) = (-1-x)^2 + (y_Q + x^2) \label{eq:dist-toy}.
\end{equation} 
Using $d^2_{\rm min}$ to represent the minimum of $d^2$ over $x$
\begin{equation}
d^2_{\rm min} = \min_x d^2(x).
\end{equation}     
The geometrical fit would compute the likelihood of the arc of parabola given $Q$ as:
\begin{equation}
L_{\rm geo} = \exp(-d^2_{\rm min}/2)
.\end{equation}
Clearly, $L_{\rm geo}$ has a maximum for $d^2_{\rm min} = 0$, which occurs for $y_Q = -1$.

Differing from the above computation, the ML approach takes into account not only the nearest point but all the other points on the arc of parabola. 
We use $ds(x)$ to denote the the line element on the arc of parabola around the abscissa $x$ (see Fig.~\ref{fig:lik-toy}), and assume that the probability of all the points on the parabola are equal. Then the contribution of the point $R$ to the final likelihood is 
\begin{equation}
L_{\rm ML, R} = \exp(-d^2/2) \; ds = \exp(-d^2/2) \sqrt{1 + (2 x)^2} \;dx\label{eq:ML-toyR}
.\end{equation}
The full likelihood is obtained by integrating Eq.~(\ref{eq:ML-toyR}) over the arc of parabola
\begin{equation}
L_{\rm ML} = \int_{-5}^{0} \exp(-d^2/2) \sqrt{1 + (2 x)^2} \; dx
.\end{equation}

The left panel in Fig.~\ref{fig:lik-toy-res} shows the evolution of $L_{\rm geo}$ and $L_{\rm ML}$ for different values of $y_Q$. It is apparent that the ML estimates do not reach their maximum value for $y_Q = -1$, when the point $Q$ lies on the arch of parabola, but for a lower value, about $y_Q = -1.5$. In this configuration, thanks to the negative curvature of the arc of parabola, the point at abscissa $-1$ moves away from $Q$, but the distance of the neighbourhood region of the arc moves closer to $Q$ thus inducing the discrepancy between the two methods.         
For real isochrone fitting the problem is obviously complicated by the presence of different classes of 
isochrones, computed with varying metallicity and initial helium abundance. These parameters modify the shape of the isochrone and contribute in complex ways to the final bias on the estimates. 

Another key difference between the fitting methods is that the variance of ML estimates shrinks faster than those of the geometrical method when the sample size increases. This behaviour is caused by the fact that the former technique can exploit a greater amount of information than the latter: the ML approach not only uses the nearest point but also its neighbouring points to differentiate among the isochrones.
To simulate how the variances of the two techniques change with sample size we performed a simple exercise. We populate a sample set of observational points simply by considering the same point $Q$ multiple times.
The right panel in Fig.~\ref{fig:lik-toy-res} shows the trend of the variance ratio for different sizes of the observational sample. It is apparent that while the two variances are very close for a single observational point, the variance of the ML  method shrinks faster and is about three quarters of the variance from the geometrical method for a sample of size about 100.

\section{Random-effect models}\label{app:raneff}        

Random effect models were adopted in this work for disentangling the variance component 
owing to observational errors from the residual one (see Sect.~\ref{sec:raneff}), and to establish the relevance of different fitting methods, progenitor stars, and Monte Carlo realisations with respect to residual variance (see Sect.~\ref{sec:cfr}). We discuss the random effect model approach in the latter case; the adaptation to the former case is straightforward and is given at the end of this section. 

As described in Sect.~\ref{sec:method} a two-stage approach was adopted in the Monte Carlo simulations. First, $N_1$ artificial systems were generated and subjected to perturbation to account for the observational errors. Second, all these systems were fitted by the algorithms described in Sect.~\ref{sec:fittingML}, which adopts $N_2$ perturbed replicates of each system to evaluate the statistical errors of the estimated parameters.

We use $i$ to denote the method of fit ($i = 1, 3$), $j$ the progenitor artificial star ($j = 1, \ldots, 20$), $k$ the Monte Carlo run ($k = 1, \ldots, 70$), and $l$  the estimates from SCEPtER or MCMC methods ($l$ = 1, $\ldots$, 500). We use the dependent variable $Y$ to denote the age of the system.
A fixed-effect model for the age with respect to the previous variables would be specified as:
\begin{equation}
Y_{ijkl} = \mu + \alpha_i + \beta_j + \gamma_k + \varepsilon_{ijkl} 
,\end{equation}
where $\mu$ is the grand mean, $\alpha_i$ are the parameters (to be estimated from the model) for the difference in age among fitting methods, $\beta_j$ the parameters for the difference in age among progenitor stars, $\gamma_k$ the parameters for the difference among Monte Carlo replications,
and  $\varepsilon_{ijkl} \sim N(0, \sigma^2)$ is the error term.
This model corresponds to a classical three-way analysis of variance (ANOVA), and it is appropriate if the individual differences in the age among the exact artificially selected methods, progenitors, and replicates are of interest \citep[see e.g.][]{snedecor1989, Feigelson2012}. 

However, the studied stars are only a random sample of the possible ones than can be generated by the Monte Carlo procedure. Moreover, the two techniques employed in the analysis are possible choices among several others adopted in the literature.
Hence, it is interesting to estimate the {\it variability} in the mean age due to the random sampling and fitting methods. 
This goal is achieved by adopting a random effects model:
\begin{equation}
Y_{ijkl} = \mu + A_i + B_j + C_k + \varepsilon_{ijkl}\label{eq:raneff}
,\end{equation}
where $\varepsilon_{ij} \sim N(0, \sigma^2)$,  $A_i \sim N(0, \sigma_a^2)$, $B_j \sim N(0, \sigma_b^2)$, and $C_k \sim N(0, \sigma_c^2)$ are random variables, the latter three representing the difference between this model and the fixed effect one.
The estimates of $\sigma,$  the residual standard deviation, and $\sigma_a$, $\sigma_b$, and $\sigma_c$ , the standard deviation owing to the different source of variability, are the outcome of the model fitting.

The fit of the model in Eq.~(\ref{eq:raneff}) was performed using a restricted ML technique adopting the library {\it lme4} in R 3.4.3 \citep{lme4, R}.
Further details on the method and on its theoretical assumptions can be found in \citet{Laird1982, venables2002modern, lme4}.

For convenience we also report the model adopted for disentangling the variance component owing to observational errors from the residual one.
\begin{equation}
Y_{ij} = \mu + A_i + \varepsilon_{ij},
\end{equation}
with $\varepsilon_{ij} \sim N(0, \sigma^2)$ and $A_i \sim N(0, \sigma_g^2)$.

\end{appendix}
            
\end{document}